\documentclass[12pt]{article}
\usepackage{amssymb}
\usepackage{amsmath}
\usepackage{cite}
\usepackage[]{hyperref}
\input xy
\xyoption{all}

\usepackage{graphics,graphicx}
\usepackage{color}
\pdfoutput=1
\numberwithin{equation}{section}

\begin{document}
\begin{flushright}
  UWThPh2019-24
  \end{flushright}

\vspace*{0.5in}

\begin{center}

{\large\bf GLSMs, joins, and nonperturbatively-realized geometries}

\vspace{0.25in}

Johanna Knapp$^{1,2}$, Eric Sharpe$^3$

\vspace*{0.2in}

\begin{tabular}{cc}
{\begin{tabular}{l}
$^1$ Mathematical Physics Group \\
University of Vienna \\
Boltzmanngasse 5\\
1090 Vienna, Austria
\end{tabular}
} &

{\begin{tabular}{l}
$^2$ School of Mathematics and Statistics \\
University of Melbourne \\
Parkville 3010 VIC\\
Australia
  \end{tabular}}
  
\end{tabular}

\begin{tabular}{c}
  { \begin{tabular}{l}
$^3$ Department of Physics \\
Virginia Tech \\
850 West Campus Dr. \\
Blacksburg, VA 24061
\end{tabular} }
  \end{tabular}

{\tt johanna.knapp@unimelb.edu.au},
{\tt ersharpe@vt.edu}

$\,$

\end{center}

In this work we give a gauged linear sigma model (GLSM) realization of 
pairs of homologically projective dual Calabi-Yaus 
that have recently been constructed in the mathematics literature. Many of
the geometries can be realized mathematically in terms of joins. 
We discuss how joins can be described in terms of GLSMs and how the 
associated Calabi-Yaus arise as phases in the GLSMs. 
Due to strong-coupling phenomena in the GLSM, 
the geometries are realized via a mix of perturbative and non-perturbative effects. 
We apply two-dimensional gauge dualities to construct dual GLSMs. Geometries that are realized perturbatively in one GLSM, are realized non-perturbatively in the dual, and vice versa.

\begin{flushleft}
July 2019
\end{flushleft}

\newpage

\tableofcontents

\newpage

\section{Introduction}

Gauged linear sigma models (GLSMs) \cite{Witten:1993yc} 
provide a physical method to
construct and 
analyze stringy geometries and their moduli spaces. 
Over the last decade there has been a lot of progress in understanding
many aspects of GLSMs, including GLSMs with non-abelian gauge groups.

One of the advances in our understanding of geometry and GLSMs
has been to learn
that geometries can arise via nonperturbative effects,
in both nonabelian (see e.g. \cite{Hori:2006dk}) and
abelian (see e.g. \cite{Caldararu:2007tc}) GLSMs.
(See also \cite{Hellerman:2006zs}[section 12.2]).
In this context, dualities due to Hori \cite{Hori:2011pd} provide
a way to map non-perturbatively realized geometries to perturbatively 
realized ones in a dual theory.

Another advance of the last decade or so has been to understand that
geometric phases of the same GLSM are related
by homological projective duality \cite{kuz1,kuz2,kuz3},
which for Calabi-Yau GLSMs implies the phases are derived equivalent.
For example,
another aspect of the papers \cite{Hori:2006dk,Caldararu:2007tc} is that
they gave examples of GLSMs with non-birational geometric phases,
related by homological projective duality.
Since generic low-energy configurations of abelian and non-abelian GLSMs 
are actually non-geometric, those GLSMs which have more than one geometric 
phase are of particular interest.
Due to the rich structure of nonabelian theories, 
finding such examples in nonabelian GLSMs is a non-trivial task.

Recently, additional examples of homological projective duals were
described in the mathematics paper \cite{inoue}.  
One of the purposes of this paper is to
give GLSMs realizing those examples as geometric phases, and to explore
their properties and dualities.

Many of the examples of homological projective duality in
\cite{inoue}, as well as constructions of Calabi-Yau's in \cite{galkintalk},
involve a construction in algebraic geometry known as
a `join,' whose physical realization in GLSMs has not previously
been described. 
Thus, we begin in section~\ref{sect:joins:intro}
by giving
an introduction to joins and
their physical realization in some simple one-parameter GLSM examples.
At some level this section of our paper is also
a continuation of our efforts in
\cite{Caldararu:2017usq} 
to give GLSM-based realizations of other constructions in
algebraic geometry. 
In fact, as we shall see explicitly,
one of the models discussed in \cite{Caldararu:2017usq} 
fits into the framework of joins.
The one-parameter example we discuss there all have multiple non-abelian factors
in their gauge groups to which two-dimensional gauge dualities can be
applied, so we explore the dual GLSMs and verify that the phases of the
duals have the same geometric interpretation as in the original GLSM.
Phases that are realized perturbatively in one duality frame, as the critical
locus of a superpotential, are realized nonperturbatively in another,
and we also see examples in which geometry arises via a combination of
perturbative and nonperturbative effects.  We further propose an analogue
of joins for gauge theories, and also discuss a connection to Hadamard products and
Picard-Fuchs equations \cite{galkintalk} in this context.

Having described the basics of joins and their GLSM realizations,
in section~\ref{sect:multiparameter} 
we turn to the physical realization of the
homological
projective duals discussed in \cite{inoue}.
We give GLSMs in which those homological projective duals arise as different
phases.  We also apply gauge duality to those GLSMs, and check that one
recovers the same geometries in the same phases of dual GLSMs.

In the models discussed in \cite{inoue}, Calabi-Yau conditions are
stated which utilize relations amongst the divisors, and which do not
lift to the ambient space.  We discuss how those Calabi-Yau conditions
can be seen in GLSMs, a topic we elaborate upon in
appendix~\ref{app:nonambientcy}.


\section{Joins: introduction and one-parameter examples}
\label{sect:joins:intro}

\subsection{Overview of joins}

A join in algebraic geometry is a close analogue of the notion of
join in algebraic topology, where a join of $X$ and $Y$ is a quotient
of $X \times Y \times [0,1]$ in which $X$ is shrunk to a point at $0$ and
$Y$ is shrunk to a point at $1$.  In algebraic geometry,
given two algebraic varieties $M_1$,
$M_2$, each with a projective embedding
\begin{equation}
M_i \: \stackrel{ {\cal O}(H_i) }{\longrightarrow} \: 
{\mathbb P}(V_i),
\end{equation}
for vector spaces $V_1, V_2$,
one can define Join$(M_1,M_2)$ to be a union of the lines spanned
by the points $(x_1,0)$ and $(0,x_2)$ in 
${\mathbb P}(V_1 \oplus V_2)$.  (See for example
\cite{galkintalk} for an excellent introduction.)
Joins are typically singular, so one needs to either resolve or smooth.

We will work with a resolution,
the resolved join, which is defined to be
\begin{displaymath}
{\mathbb P}_{ M_1 \times M_2 } \left( {\cal O}(-H_1) \oplus
{\cal O}(-H_2) \right),
\end{displaymath}
where $H_{1,2}$ are the hyperplane classes of the two projective
embeddings.

For a simple example, suppose $M_1 = M_2 = {\mathbb P}^1$,
with the trivial embedding into ${\mathbb P}^1$ itself.
Then, the resolved join is
\begin{displaymath}
{\mathbb P}_{ {\mathbb P}^1 \times {\mathbb P}^1 } \left(
{\cal O}(-1,0) \oplus {\cal O}(0,-1) \right),
\end{displaymath}
which can be described by a GLSM with fields
$x_{1,2}$, $y_{1,2}$, $z_{1,2}$ with $U(1)^3$ charges
\begin{center}
\begin{tabular}{c|ccccrr}
$U(1)$ & $x_1$ & $x_2$ & $y_1$ & $y_2$ & $z_1$ & $z_2$ \\ \hline
$\lambda$ & $1$ & $1$ & $0$ & $0$ & $-1$ & $0$ \\
$\mu$ & $0$ & $0$ & $1$ & $1$ & $0$ & $-1$ \\
$\nu$ & $0$ & $0$ & $0$ & $0$ & $1$ & $1$
\end{tabular}
\end{center}

This can be projected to the classical join, which lives in
${\mathbb P}^3$, by taking as homogeneous coordinates $x_{1,2} z_1$,
$y_{1,2} z_2$.  These are all neutral under $\lambda$ and $\mu$, but
have charge $1$ under $\nu$, and so define homogeneous coordinates on
a ${\mathbb P}^3$.  

More intuitively, the classical join describes a line
between any two points of $M_1$ and $M_2$:  the ${\mathbb P}^1$ bundle
over $M_1 \times M_2$ certainly describes a line over each pair of
points in $M_1$ and $M_2$, satisfying that intuition, and resolves
singularities that arise
when points of $M_1$ and $M_2$ collide inside their projective embedding.
As a more primitive consistency check, the join of $M_1$ and $M_2$
should have dimension
\begin{displaymath}
\dim \, M_1 \: + \: \dim M_2 \: + \: 1,
\end{displaymath}
and the construction of the resolved join (as a ${\mathbb P}^1$ bundle
over the ordinary product $M_1 \times M_2$) certainly has that property.

The Calabi-Yau condition is straightforward to derive for a resolved
join.  Let $J$ denote the resolved join.  Then, for
\begin{displaymath}
\pi: \: J \: \longrightarrow \: M_1 \times M_2,
\end{displaymath}
we have the sequence
\begin{displaymath}
0 \: \longrightarrow \: {\cal O} \: \longrightarrow \:
{\cal O}(1) \otimes
\pi^* \left( {\cal O}(-H_1) \oplus {\cal O}(-H_2) \right) \:
\longrightarrow \: T_{\pi} \: \longrightarrow \: 0,
\end{displaymath}
hence
\begin{eqnarray*}
\det T_{\pi} & = & {\cal O}(2) \otimes 
\pi^* {\cal O}(- H_1 - H_2), 
\\
& = & \left( K_J \otimes \pi^* K_{M_1 \times M_2}^{-1} \right)^{-1},
\end{eqnarray*}
from which we derive
\begin{displaymath}
K_J \: = \: {\cal O}(-2) \otimes \pi^* \left(
K_{M_1 \times M_2} \otimes {\cal O}(+ H_1 + H_2) \right).
\end{displaymath}

\subsection{Joins and Hadamard products}

           In \cite{galkintalk} a connection between joins and Hadamard products of associated Picard-Fuchs differential operators was discussed. We will see this connection in the one-parameter examples discussed below, and so here we briefly review the relevant definitions.

           For this purpose, let us first recall the definition of Hadamard products. Consider two power series $u$ and $v$, satisfying Picard-Fuchs type differential equations (to be precise, the power series have to be {\em D-finite} \cite{vanstraten1}):
\begin{equation}
  u=\sum_{n=0}^{\infty}b_nz^n, \qquad v=\sum_{n=0}^{\infty}c_nz^n,
\end{equation}
satisfying $D_uu=0$ and $D_vv=0$. The Hadamard product of the power series is
\begin{equation}
  y=\sum_{n=0}^{\infty}a_nz^n=u\ast v=\sum_{n=0}^{\infty}b_nc_nz^n.
\end{equation}
These satisfy the differential equation $(D_u\ast D_v)y=0$. Note that it is not straightforward to determine  $(D_u\ast D_v)$. What one does in practice is to compute $y$ and determine the differential operator annihilating it via an ansatz.

One way applied in \cite{vanstraten1,vanstraten2} to construct fourth-order Picard-Fuchs operators is via Hadamard products of second order operators associated to elliptic curves. This yields differential operators that can be associated to one-parameter Calabi-Yau threefolds. In this work we will mainly be interested in models of non-abelian GLSMs realizing homological projective duality. 
We will construct Hadamard products explicitly in examples of 
elliptic curves constructed via GLSMs with gauge groups $U(2)$ and $(U(1)\times O(2)_+)/\mathbb{Z}_2$. These have been discussed in \cite{Hori:2013gga}. Concrete examples will be discussed in sections \ref{sect:ex:u2} and \ref{sect:ex:o+2}.
(See also \cite{dm} for a discussion of Hadamard products in a slightly
different context.)

\subsection{Example:  join of two hypersurfaces}

As a warm-up example, we will first describe the join of two
hypersurfaces in projective spaces.  This is an elementary example,
whose details will hopefully help illuminate the notion of joins for
readers.

Consider a degree $d_1$ hypersurface $X = \{f(x) = 0\}$ in ${\mathbb P}^n$,
described with homogeneous coordinates $x_0, \cdots, x_n$,
and the degree $d_2$ hypersurface $Y = \{ g(y) = 0 \}$ in ${\mathbb P}^m$,
described with homogeneous coordinates $y_0, \cdots, y_m$.
Then, Join($X,Y$) is described by the complete intersection
\begin{equation}
f(x) \: = \: 0, 
\: \: \:
g(y) \: = \: 0,
\end{equation}
in ${\mathbb P}^{n+m+1}$, with homogeneous coordinates
\begin{equation}
x_0, \cdots, x_n, y_0, \cdots, y_m.
\end{equation}
As a quick consistency check, note that the dimension of the join claimed
above is 
\begin{displaymath}
n+m+1 \: - \: 2 \: = \: n+m-1,
\end{displaymath}
which matches
\begin{displaymath}
\dim X \: + \: \dim Y \: + \: 1,
\end{displaymath}
as expected.  Furthermore, if the hypersurfaces $X$ and $Y$ are both 
Calabi-Yau, meaning $d_1 = n+1$ and $d_2 = m+1$, then the classical
join is another Calabi-Yau, as $d_1 + d_2 = (n+m+1)+1$.

Also note in passing that the classical Join($X,Y$) is singular:
it contains the locus where $f(x)$ vanishes because all the $x$ vanish
(essentially, a contraction of $X$ to a point),
as well as the locus where $g(y)$ vanishes because all the $y$ vanish
(similarly, a contraction of $Y$ to a point).
Only the intersection of these two loci is omitted. 
This structure is precisely analogous to the structure of the join in
algebraic topology, where one also contracts the two spaces to points,
at either end of the interval.

Furthermore, we shall see explicitly later that by rescaling different
terms so as to maintain the D-term constraint in the corresponding GLSM, we can realize 
one-parameter families connecting points on $X$ to points on $Y$ --
lines connecting points on either space, in other words.

What is going to make joins interesting in other cases is that the
embeddings into projective spaces may be considerably more complicated
than for the easy case of a hypersurface, so it may not be possible
to easily `eyeball' the answer by inspection as we have done above.
Our procedure in other cases, therefore, is to first write down the
resolved join, which can be done straightforwardly, and then blowdown
to recover the original join.

To make this clear, we shall illustrate the resolved join and its
blowdown next.  The resolved join in the case above is described by a GLSM
with gauge group $U(1)^3$ and fields
\begin{center}
\begin{tabular}{c|ccccrr}
& $x_i$ & $y_j$ & $p_1$ & $p_2$ & $z_1$ & $z_2$ \\ \hline
$U(1)_1$ & $1$ & $0$ & $-d_1$ & $0$ & $-1$ & $0$ \\
$U(1)_2$ & $0$ & $1$ & $0$ & $-d_2$ & $0$ & $-1$ \\
$U(1)_3$ & $0$ & $0$ & $0$ & $0$ & $1$ & $1$
\end{tabular}
\end{center}
with superpotential
\begin{equation}
W \: = \: p_1 f(x) \: + \: p_2 g(y).
\end{equation}
This describes a ${\mathbb P}^1$ bundle over the product
${\mathbb P}^n \times {\mathbb P}^m$.  The $z_i$ are homogeneous
coordinates on the fibers of that ${\mathbb P}^1$ bundle,
and $x_i$, $y_j$ are homogeneous coordinates on
${\mathbb P}^n$ and ${\mathbb P}^m$, respectively.
The embedding of a hypersurface
into its ambient projective space corresponds technically to
an embedding generated by an ample line bundle ${\cal O}(1)$,
so in the notation of \cite{inoue}, we take 
\begin{equation}
{\cal O}(-H_1) \: = \: {\cal O}_{ {\mathbb P}^n }(-1),
\: \: \:
{\cal O}(-H_2) \: = \: {\cal O}_{ {\mathbb P}^m }(-1),
\end{equation}
so $z_1$ and $z_2$ each have weight $-1$ under the $U(1)$s building
each of the two projective spaces in the base.

Next, we shall blowdown the resolution, to relate this more simply
to the classical join described above.  First, we blowdown the divisor
$\{ z_1 = 0 \}$, and eliminate $U(1)_3$.  This yields a $U(1)^2$
gauge theory with fields
\begin{center}
\begin{tabular}{c|ccccc}
& $x_i$ & $y_j$ & $p_1$ & $p_2$ & $z_2$ \\ \hline
$U(1)_1$ & $1$ & $0$ & $-d_1$ & $0$ & $-1$  \\
$U(1)_2$ & $0$ & $1$ & $0$ & $-d_2$ & $-1$ 
\end{tabular}
\end{center}

Next, we blowdown the divisor $\{ z_2 = 0 \}$ and eliminate $U(1)_2$.
This yields the $U(1)$ gauge theory with fields
\begin{center}
\begin{tabular}{c|cccc}
& $x_i$ & $y_j$ & $p_1$ & $p_2$ \\ \hline
$U(1)_1$ & $1$ & $1$ & $-d_1$ & $-d_2$
\end{tabular}
\end{center}
which, when combined with the superpotential
\begin{equation}
W \: = \: p_1 f(x) \: + \: p_2 g(y),
\end{equation}
the reader will recognize as the GLSM for the classical join described
initially.

The reader should note that, as expected, this GLSM is singular at points
where $X$ and $Y$ separately contract to points, in other words at points
where all the $x_i$ vanish, or points where all the $y_i$ vanish.
Furthermore, we have a line of points connecting $X$ and $Y$, related
by relative rescalings.  In more detail, the D-term constraint is
\begin{equation}
\sum_i |x_i |^2 \: + \: \sum_j |y_j|^2 \: = \: r.
\end{equation}
As we increase
\begin{displaymath}
\sum_i |x_i |^2,
\end{displaymath}
we decrease
\begin{displaymath}
\sum_j |y_j|^2,
\end{displaymath}
so as to keep the sum constant, which results in a one-parameter family 
with the two solutions
\begin{equation}
\sum_i |x_i |^2 \: = \: r,
\: \: \:
\sum_j |y_j|^2 \: = \: r,
\end{equation}
as endpoints.  Thus, we have lines connecting points on $X$ to points on $Y$.
This is more explicitly how this algebro-geometric join
is analogous to the join of algebraic topology.

Now, this singular GLSM admits a natural deformation, by letting
$f(x)$ also depend on $y_j$, and $g(y)$ also depend upon $x_i$:
\begin{equation}
W \: = \: p_1 \, f(x,y) \: + \: p_2 \, g(x,y).
\end{equation}
This now defines a complete intersection
${\mathbb P}^{n+m+1}[d_1,d_2]$, which for generic $f$ and $g$ is smooth.
(Mathematically, the singularities where $X$ and $Y$ contract to points have
high codimension, and so will not intersect a generic hypersurface.)
Such deformations will play an important role in our later examples.

This is the pattern we will follow in other examples -- we will first
write down the resolved join, blowdown to uncover the classical join,
and then analyze the resulting GLSM.  In some cases this may be overkill,
but it should provide a systematic procedure to understand these
constructions.

\subsection{Example:  Join($G(2,5),G(2,5)$) }
\label{sect:ex:u2}
\subsubsection{Ambient join}

To begin, we shall describe the Join of $G(2,5)$ with itself that was also discussed in \cite{inoue}. 
We will see that the result
is related to the intersection $G(2,5) \cap G(2,5)$ described physically in
\cite{Caldararu:2017usq}, following \cite{galkintalk}.
(See also \cite{Kapustka:2017jyt} for a related theory.)

As first discussed in \cite{Witten:1993xi},
a GLSM for the Grassmannian $G(k,n)$ is given by 
a $U(k)$ gauge theory with $n$ chiral superfields in the fundamental
representation.  To describe the resolution of the join of
$G(2,5)$ to another copy of $G(2,5)$, the total space of
a ${\mathbb P}^1$ bundle over their product,
we will use an $U(2) \times U(2) \times U(1)$ gauge theory with
fields $\phi^i_a$, $\tilde{\phi}^i_a$, $z_{1,2}$, 
$a \in \{1, 2\}$, $i \in \{1, \cdots, 5\}$ charged as follows:
\begin{center}
\begin{tabular}{c|cccc}
 & $\phi^i_a$ & $\tilde{\phi}^i_a$ & $z_1$ & $z_2$ \\ \hline
$U(2)$ & $\Box$ & ${\bf 1}$ & det$^{-1}$ & ${\bf 1}$ \\
$U(2)$ & ${\bf 1}$ & $\Box$ & ${\bf 1}$ & det$^{-1}$ \\
$U(1)_3$ & $0$ & $0$ & $1$ & $1$ 
\end{tabular}
\end{center}

Next, to compare this to other expressions for joins, we will blow
down the divisors $\{z_1 = 0 \}$ and $\{z_2=0\}$.  To describe this
more efficiently, we will use the fact that
\begin{equation}
U(2) \: = \: \frac{ SU(2) \times U(1) }{ {\mathbb Z}_2 },
\end{equation}
and describe gauge charges of the covering gauge group, bearing in mind that
we will take an orbifold at the end.
The charge and representation table can then be
rewritten as follows:
\begin{center}
\begin{tabular}{c|ccrr}
 & $\phi^i_a$ & $\tilde{\phi}^i_a$ & $z_1$ & $z_2$ \\ \hline
$SU(2)$ &  $\Box$ & ${\bf 1}$ & ${\bf 1}$ & ${\bf 1}$ \\
$U(1)_1$ & $1$ & $0$ & $-2$ & $0$ \\
$SU(2)$ &  ${\bf 1}$ & $\Box$ & ${\bf 1}$ & ${\bf 1}$ \\
$U(1)_2$ & $0$ & $1$ & $0$ & $-2$ \\
$U(1)_3$ & $0$ & $0$ & $1$ & $1$
\end{tabular}
\end{center}
where the gauge group is
\begin{equation}
\frac{ 
SU(2) \times U(1)_1 \times SU(2) \times U(1)_2
}{
{\mathbb Z}_2 \times {\mathbb Z}_2
} \times U(1)_3.
\end{equation}
Note that, schematically, in terms of the divisors corresponding to the
$U(1)$ factors,
$D_3 \sim 2 D_2 \sim 2 D_1$, where the $D_i$ are Picard group elements
nominally associated with the three $U(1)$s.

Blowing down the divisor $\{ z_1 = 0 \}$, and eliminating $U(1)_3$,
we get
\begin{center}
\begin{tabular}{c|ccr}
 & $\phi^i_a$ & $\tilde{\phi}^i_a$ & $z_2$ \\ \hline
$SU(2)$ &  $\Box$ & ${\bf 1}$ & ${\bf 1}$ \\
$U(1)_1$ & $1$ & $0$ & $-2$ \\
$SU(2)$ &  ${\bf 1}$ & $\Box$ & ${\bf 1}$ \\
$U(1)_2$ & $0$ & $1$ & $-2$ 
\end{tabular}
\end{center}

Blowing down the divisor $\{ z_2 = 0 \}$ and eliminating $U(1)_2$, 
we get
\begin{center}
\begin{tabular}{c|cc}
 & $\phi^i_a$ & $\tilde{\phi}^i_a$ \\ \hline
$SU(2)$ &  $\Box$ & ${\bf 1}$ \\
$U(1)_1$ & $1$ & $1$ \\
$SU(2)$ &  ${\bf 1}$ & $\Box$
\end{tabular}
\end{center}
To be clear, this is the matter content in a gauge theory with
gauge group
\begin{equation}
\frac{
SU(2) \times U(1)_1 \times SU(2)
}{
{\mathbb Z}_2 \times {\mathbb Z}_2
}.
\end{equation}

The two blowdowns we have performed are a prototype for analogous
manipulations on resolved joins we shall perform throughout this paper.

In passing, we observe that the structure above is part of the
structure that describes $G(2,5) \cap G(2,5)$ in \cite{Caldararu:2017usq}.
There, the intersection $G(2,5) \cap G(2,5)$ was described
in terms of a 
\begin{equation}
\frac{
U(1) \times SU(2) \times SU(2) 
}{
{\mathbb Z}_2 \times {\mathbb Z}_2
}
\end{equation}
gauge theory with matter $\phi^i_a$, $\tilde{\phi}^i_a$, and $p_{ij}$,
in the representations
\begin{center}
\begin{tabular}{c|ccr}
 & $\phi^i_a$ & $\tilde{\phi}^i_a$ & $p_{ij}$ \\ \hline
$SU(2)$ & $\Box$ & ${\bf 1}$ & ${\bf 1}$ \\
$U(1)$ & $1$ & $1$ & $-2$ \\
$SU(2)$ & ${\bf 1}$ & $\Box$ & ${\bf 1}$
\end{tabular}
\end{center}
with superpotential
\begin{equation}
W \: = \: p_{ij} \left( f^{ij}(B) \: - \: \tilde{B}^{ij} \right),
\end{equation}
where
\begin{equation}
B^{ij} \: = \: \epsilon^{ab} \phi^i_a \phi^j_b,
\: \: \:
\tilde{B}^{ij} \: = \: \epsilon^{ab} \tilde{\phi}^i_a \tilde{\phi}^j_b,
\end{equation}
are the baryons (Pl\"ucker coordinates), and $f^{ij}$ encodes a linear
flavor rotation in $GL(5)$.

In effect, the GLSM for the self-intersection $G(2,5) \cap G(2,5)$ is
encoding a subvariety within the space defined by the GLSM for the
(blowdown of the resolved) join.  This is in agreement with statements
in \cite{galkintalk}, which indicated that $G(2,5) \cap G(2,5)$ is
a subvariety of Join($G(2,5),G(2,5)$); we now see that relationship
physically in GLSMs.  We will further elaborate on this
relationship in section~\ref{sect:g25:deftoint}, where we will argue that a
Calabi-Yau complete intersection in the ambient Join($G(2,5),G(2,5)$)
can be deformed to the Calabi-Yau $G(2,5) \cap G(2,5)$.

\subsubsection{Calabi-Yau complete intersection}
\label{sect:g25:2:cyci}

Next, we will consider a Calabi-Yau complete intersection
in Join($G(2,5),G(2,5)$).  First, recall that a complete intersection
of five hyperplanes in a single $G(2,5)$ is Calabi-Yau, and in fact
is an elliptic curve.  In this section we will consider an analogous
complete intersection in either factor of Join($G(2,5),G(2,5)$).

First, recall that the resolution of the join was defined by
a $U(2) \times U(2) \times U(1)$ gauge theory with matter content
\begin{center}
\begin{tabular}{c|cccc}
 & $\phi^i_a$ & $\tilde{\phi}^i_a$ & $z_1$ & $z_2$ \\ \hline
$U(2)$ & $\Box$ & ${\bf 1}$ & det$^{-1}$ & ${\bf 1}$ \\
$U(2)$ & ${\bf 1}$ & $\Box$ & ${\bf 1}$ & det$^{-1}$ \\
$U(1)_3$ & $0$ & $0$ & $1$ & $1$
\end{tabular}
\end{center}

Now, on the face of it, to describe a Calabi-Yau complete intersection
in the space defined by the GLSM above, we would need the sum of the
$U(1)_3$ charges to be $2$, which would highly constrain possible
complete intersections.  

However, more general possibilities exist, which use relations amongst the
divisors, and so yield Calabi-Yau conditions which do not extend
to the entire ambient space.  We discuss in appendix~\ref{app:nonambientcy} 
how to understand Calabi-Yau conditions
that utilize relations along the intersection that are not inherited from
the ambient space, as well as GLSM constructions to make the
pertinent Calabi-Yau condition more clear.  In the present case,
briefly, we can rewrite the GLSM (for generic intersections) 
by performing
blowdowns which eliminate some
$U(1)$ factors, analogues of the blowdowns discussed in the
previous section.

Describing the gauge group as
\begin{equation}
\frac{
SU(2) \times U(1)_1 \times SU(2) \times U(1)_2
}{
{\mathbb Z}_2 \times {\mathbb Z}_2
} \times U(1)_3
\end{equation}
we see that the GLSM describing a complete intersection of 
five hyperplanes $G^{\alpha}(B^{ij})$ in the first $U(2)$
factor and five hyperplanes $\tilde{G}^{\beta}(\tilde{B}^{ij})$
in the second $U(2)$ factor has fields
\begin{center}
\begin{tabular}{c|ccrrrr}
& $\phi^i_a$ & $\tilde{\phi}^i_a$ & $z_1$ & $z_2$ & $p_{\alpha}$
& $\tilde{p}_{\beta}$ \\ \hline
$SU(2)$ & $\Box$ & ${\bf 1}$ & ${\bf 1}$ & ${\bf 1}$ &
${\bf 1}$ & ${\bf 1}$ \\
$U(1)_1$ & $1$ & $0$ & $-2$ & $0$ & $-2$ & $0$ \\
$SU(2)$ & ${\bf 1}$ & $\Box$ & ${\bf 1}$ & ${\bf 1}$ &
${\bf 1}$ & ${\bf 1}$ \\
$U(1)_2$ & $0$ & $1$ & $0$ & $-2$ & $0$ & $-2$ \\
$U(1)_3$ & $0$ & $0$ & $1$ & $1$ & $0$ & $0$
\end{tabular}
\end{center}
We also take this model to have the superpotential
\begin{equation}
W \: = \: \sum_{\alpha} p_{\alpha} G^{\alpha}(B^{ij})
\: + \: \sum_{\beta} \tilde{p}_{\beta} \tilde{G}^{\beta}(
\tilde{B}^{ij}),
\end{equation}
where
\begin{equation}
B^{ij} \: = \: \epsilon^{ab} \phi^i_a \phi^j_b,
\: \: \:
\tilde{B}^{ij} \: = \: \epsilon^{ab} \tilde{\phi}^i_a
\tilde{\phi}^j_b
\end{equation}
are the baryons (Pl\"ucker coordinates) in each $SU(2)$ factor.
This theory does not satisfy the usual Calabi-Yau condition for a GLSM;
the sum of the $U(1)_3$ charges is nonzero, for example.  However,
we shall see that after successive blowdowns, the resulting space
will satisfy the GLSM Calabi-Yau condition.

Blowing down the divisor $\{ z_1 = 0 \}$ and eliminating $U(1)_3$,
the table above becomes
\begin{center}
\begin{tabular}{c|ccrrr}
& $\phi^i_a$ & $\tilde{\phi}^i_a$ & $z_2$ & $p_{\alpha}$
& $\tilde{p}_{\beta}$ \\ \hline
$SU(2)$ & $\Box$ & ${\bf 1}$ &  ${\bf 1}$ &
${\bf 1}$ & ${\bf 1}$ \\
$U(1)_1$ & $1$ & $0$ & $+2$ & $-2$ & $0$ \\
$SU(2)$ & ${\bf 1}$ & $\Box$ &  ${\bf 1}$ &
${\bf 1}$ & ${\bf 1}$ \\
$U(1)_2$ & $0$ & $1$ &  $-2$ & $0$ & $-2$ 
\end{tabular}
\end{center}

Blowing down the divisor $\{ z_2 = 0 \}$ and eliminating $U(1)_2$,
the table becomes
\begin{center}
\begin{tabular}{c|ccrr}
& $\phi^i_a$ & $\tilde{\phi}^i_a$ & $p_{\alpha}$
& $\tilde{p}_{\beta}$ \\ \hline
$SU(2)$ & $\Box$ & ${\bf 1}$ & ${\bf 1}$ & ${\bf 1}$ \\
$U(1)_1$ & $1$ & $1$ & $-2$ & $-2$ \\
$SU(2)$ & ${\bf 1}$ & $\Box$ & ${\bf 1}$ & ${\bf 1}$
\end{tabular}
\end{center}
To be clear, the gauge group in this GLSM is now taken to be
\begin{equation}
\frac{
SU(2) \times U(1) \times SU(2)
}{
{\mathbb Z}_2 \times {\mathbb Z}_2
},
\end{equation}
and this GLSM has the superpotential
\begin{equation}
W \: = \: \sum_{\alpha} p_{\alpha} G^{\alpha}(B^{ij})
\: + \: \sum_{\beta} \tilde{p}_{\beta} \tilde{G}^{\beta}(
\tilde{B}^{ij}).
\end{equation}

It is straightforward to check that, since there are five
$p_{\alpha}$ and five $\tilde{p}_{\beta}$, the sum of the $U(1)_1$
charges now vanishes, and so this GLSM describes a Calabi-Yau complete
intersection in the join of $G(2,5)$ with itself. 

For completeness, let us also walk through the phases of this theory. 

Let $r$ denote the Fayet-Iliopoulos parameter associated to $U(1)_1$.
More or less by construction, for $r \gg 0$, one has a (singular)
geometric phase
describing the complete intersection in the (blowdown of the
resolution of the) join of $G(2,5)$ with itself.
In the notation of \cite{inoue}, if we let
\begin{equation}
J \: = \: {\rm Join}(G(2,5), G(2,5)),
\end{equation}
$V_5$ be a five-dimensional vector space,
$W \subset \wedge^2 V_5$ be the vector subspace defined by the
first five hyperplanes $\{ G^{\alpha} \}$
in (the Pl\"ucker embedding of) $G(2,5)$,
and $\tilde{W} \subset \wedge^2 V_5$ be the vector subspace defined
by the second five hyperplanes $\{ \tilde{G}^{\beta} \}$ in 
(the Pl\"ucker embedding of) the second $G(2,5)$, then this complete
intersection in the join could be described as
\begin{equation}
J \times_{ {\mathbb P}( \wedge^2 V_5 \oplus \wedge^2 V_5) }
{\mathbb P} W \times_{ {\mathbb P}( \wedge^2 V_5 \oplus \wedge^2 V_5)}
{\mathbb P} \tilde{W}.
\end{equation}

To be clear, the join above is singular mathematically, and that will
be reflected in additional noncompact directions in the GLSM.  In the
next section, we will smooth this model by considering generic
superpotential deformations.

The $r \ll 0$ phase is more interesting.
In this phase, D-terms imply that
the $\{ p_{\alpha}, \tilde{p}_{\beta} \}$ do not
all vanish.
Here, the superpotential can be interpreted as a mass matrix
with entries linear in $p_{\alpha}$, $\tilde{p}_{\beta}$:
\begin{equation}
W \: = \: \phi^i_a \phi^j_b \left( \epsilon^{ab} f^{\alpha}_{ij} p_{\alpha}
\right) \: + \:
\tilde{\phi}^i_a \tilde{\phi}^j_b \left( \epsilon^{ab} \tilde{f}^{\beta}_{ij}
\tilde{p}_{\beta} \right),
\end{equation}
where
\begin{equation}
G^{\alpha}(B) \: = \: f^{\alpha}_{ij} \epsilon^{ab} \phi^i_a \phi^j_b,
\: \: \:
\tilde{G}^{\beta} \: = \: \tilde{f}^{\beta}_{ij} \epsilon^{ab} \tilde{\phi}^i_a
\tilde{\phi}^j_b.
\end{equation}
The $f^{\alpha} p_{\alpha}$, $\tilde{f}^{\beta} \tilde{p}_{\beta}$
each define an antisymmetric $5 \times 5$ matrix, and as each is antisymmetric,
their possible ranks are $4$, $2$, and $0$.

Our analysis of these terms then closely follows the analysis
of the R{\o}dland example in \cite{Hori:2006dk}.  Over loci where an
antisymmetric matrix has rank $4$, there is only one massless doublet
of the corresponding $SU(2)$, so from the analysis of
\cite{Hori:2006dk}, there are no vacua.  Over loci where an antisymmetric
matrix has rank $2$, on the other hand, there are three massless
doublets of the corresponding $SU(2)$, which corresponds to a unique
supersymmetric vacuum in the physics of that $SU(2)$.

Thus, we see that in the space of $p_{\alpha}$, $\tilde{p}_{\beta}$,
the theory flows to the intersection of the loci where the two
mass matrices $f^{\alpha}_{ij} p_{\alpha}$, $\tilde{f}^{\beta}_{ij}
\tilde{p}_{\beta}$ each have rank two.  
Following \cite{Donagi:2007hi}[section 4.2.2], 
this means that they are in the image of the Pl\"ucker embeddings of
each copy of $G(2,5)$.  As a result,
those two quantities -- $f^{\alpha}_{ij} p_{\alpha}$,
$\tilde{f}^{\beta}_{ij} \tilde{p}_{\beta}$ -- define the vector subspaces
$W^{\perp}$, $\tilde{W}^{\perp}$, for the $W, \tilde{W} \subset
\wedge^2 V_5$ defined earlier, and in the notation of \cite{inoue},
this phase describes the space 
\begin{equation}
J \times_{ {\mathbb P}( \wedge^2 V_5^* \oplus \wedge^2 V_5^*) }
{\mathbb P} W^{\perp} \times_{ {\mathbb P}( \wedge^2 V_5^* \oplus 
\wedge^2 V_5^*)}
{\mathbb P} \tilde{W}^{\perp},
\end{equation}
for $J$ the Join($G(2,5),G(2,5)$).

So far, we have discussed the phases of this one-parameter GLSM.
In principle, one can dualize in one or both of the $SU(2)$
factors, following \cite{Hori:2011pd}[section 5.6],
and when one does so, one finds that dual descriptions realize
the same geometry, but differently:  phases that are realized
perturbatively (as the critical locus of a superpotential) here,
are there realized via nonperturbative effects, ala
\cite{Hori:2006dk}, and vice-versa.  

We will see analogous analyses for other models later in this paper.
We omit that analysis here for two reasons:
\begin{itemize}
\item The geometries described by these GLSMs, joins of 
complete intersections in $G(2,5)$ are extremely singular, as are all
classical joins.  They have singularities where either factor shrinks to
zero size.
\item We can smooth the singularities by deforming the superpotential,
but when we do so, we recover a model that was extensively
analyzed in \cite{Caldararu:2017usq}, including how the phases of
dual GLSMs are related to one another.
\end{itemize}
We shall elaborate on the last point, the deformation of this theory
to an example studied in \cite{Caldararu:2017usq}, in the next
subsection.

\subsubsection{Deformation to $G(2,5)\cap G(2,5)$}
\label{sect:g25:deftoint}

So far in this section, we have discussed GLSMs for 
Calabi-Yau complete intersections in Join($G(2,5),G(2,5)$).  However,
the reader may have noticed that the superpotential in this model is
not generic -- more general superpotential terms are certainly consistent
with the symmetries of the theory.  In fact, by adding more generic terms,
one can deform such complete intersections in Join($G(2,5),G(2,5)$) 
into the intersection $G(2,5) \cap G(2,5)$, another
Calabi-Yau threefold whose GLSM was discussed in \cite{Caldararu:2017usq}.

We can see this deformation explicitly as follows.  Begin with the model
of section~\ref{sect:g25:2:cyci}, the complete intersection in the join
described in the first duality frame.
Note that we can write the superpotential as
\begin{equation}
  W=B^{ij}(f_{ij}^{\alpha}p_{\alpha}+g_{ij}^{\beta}\tilde{p}_{\beta})+\tilde{B}^{ij}(\tilde{f}_{ij}^{\alpha}p_{\alpha}+\tilde{g}_{ij}^{\beta}\tilde{p}_{\beta}).
  \end{equation}
Now we can define
\begin{equation}
  p_{ij}(p_{\alpha},\tilde{p}_{\beta})=f_{ij}^{\alpha}p_{\alpha}+g_{ij}^{\beta}\tilde{p}_{\beta}.
\end{equation}
Due to antisymmetry of the $B^{ij}$ we have $p_{ij}=-p_{ji}$. So the $10$ degrees of freedom from $(p_{\alpha},\tilde{p}_{\beta})$ can be recombined into $10$ antisymmetric singlets $p_{ij}$ of the same charges. The coefficient of $\tilde{B}^{ij}$ in the superpotential is nothing but a linear transformation of the $p_{ij}$, which we can also interpret as a linear rotation in the $\tilde{B}^{ij}$. Therefore we can write the superpotential as
\begin{equation}
  W=p_{ij}(B^{ij}-f^{ij}(\tilde{B})).
\end{equation}
Hence, we have recovered the model of \cite{Caldararu:2017usq}. As a further consistency check we note that the effective potential on the Coulomb branch is the same for the models  of \cite{Caldararu:2017usq} and section \ref{sect:g25:2:cyci}.

\subsubsection{Picard-Fuchs operator and Hadamard products}
              
The Picard-Fuchs operator associated to this model has already been identified in \cite{Caldararu:2017usq}. Indeed, it is a Hadamard product of the operators associated to an elliptic curve associated to a GLSM with gauge group $U(2)$ discussed in \cite{Hori:2013gga}. 

             Let us briefly recall the discussion of \cite{Hori:2013gga}. One can construct a GLSM associated to an elliptic curve by choosing $G=U(2)$ with five fundamentals $x_i$ and and five singlets $p^k$ transforming in the inverse determinantal representation. The superpotential is
\begin{equation}
  W=\sum_{i,j=1}^5A^{ij}(p)x_i^ax_j^b\epsilon_{ab}, \qquad A^{ij}(p)=\sum_{k=1}^5A^{ij}_kp^k
\end{equation}
The $r\gg0$ phase is a complete intersection of codimension $5$ in $G(2,5)$, which his an elliptic curve. The $r\ll0$ phase is an isomorphic elliptic curve characterized by the condition rank $A(p)=2$.

The singular loci in the moduli space are encoded in the critical locus of the effective potential of the Coulomb branch. This is given by
\begin{equation}
\label{weff}
  \mathcal{W}_{eff}=-\langle {t},\sigma\rangle-\sum_i\langle Q_i,\sigma\rangle\left(\ln(\langle Q_i,\sigma\rangle) -1\right)+i\pi\sum_{\alpha>0}\langle\alpha,\sigma\rangle, 
\end{equation}
with $t_i=r_i-i\theta_i$, $\sigma$ parametrizing the maximal torus $\mathfrak{t}\subset G$, and $\alpha$ the positive roots of $G$. By $\langle\cdot,\cdot\rangle$ we denote the pairing between $\mathfrak{t}$ and its dual $\mathfrak{t}^{\ast}$. For this example we get, taking into account the non-zero theta angle,
\begin{eqnarray}
  \mathcal{W}_{eff} & = &
-t\left(\sigma_1+\sigma_2\right)
+\pi i\left(\sigma_1-\sigma_2\right)-5\sigma_1\left(\ln\sigma_1-1\right)
-5\sigma_2\left(\ln\sigma_2-1\right)
\nonumber \\
& & \hspace*{1.5in}
 + 5\left(\sigma_1+\sigma_2\right)\left(\ln\left(-\sigma_1-\sigma_2\right)-1\right).
\end{eqnarray}
With $z=\sigma_2/\sigma_1$ the critical set is determined by
\begin{equation}
  e^{-t}=-\frac{1}{(1+z)^5}=-\frac{1}{\left(1+\frac{1}{z}\right)^5},\qquad z^5=1.
  \end{equation}
There is a Coulomb branch at $\exp(-t)=-(1/2)(11\pm 5\sqrt{5})$. The Picard-Fuchs operator associated to the codimension $5$ complete intersection in $G(2,5)$ is
\begin{equation}
  \mathcal{L}=\theta^2-z(11\theta^2+11\theta+3)-z^2(\theta+1)^2.
\end{equation}
The discriminant obtained from $\mathcal{L}$ coicides with the location of the Coulomb branch of the GLSM upon identification $z=e^{-t}$. The fundamental period is
\begin{equation}
  \varpi_0=\sum_{n=0}^{\infty}\sum_{k=0}^n\left(\begin{array}{c}n\\k\end{array}\right)^2\left(\begin{array}{c}n+k\\k\end{array}\right)z^n .
\end{equation}

Now, let us come back to the join. The Coulomb branch analysis for this model has been done in \cite{Caldararu:2017usq}. Indeed, the three singular points can be obtained by taking products of the singular points of the elliptic curve. The Picard-Fuchs operator is (AESZ 101 in \cite{vanstraten2}) 
\begin{eqnarray}
  \mathcal{L}^{U(2)}*\mathcal{L}^{U(2)}&=&\theta^4-z(124 \theta ^4 + 242 \theta ^3 + 187 \theta ^2 + 66 \theta + 9)\nonumber\\
  &&+z^2(123 \theta ^4 - 246 \theta ^3 - 787 \theta
  ^2 - 554 \theta - 124)\nonumber\\
  &&+z^3(123 \theta ^4 + 738 \theta ^3 + 689 \theta ^2 + 210 \theta + 12)\nonumber\\
  &&-z^4(124 \theta ^4 + 254 \theta ^3 + 205 \theta ^2 + 78 \theta
  +12)+z^5(\theta+1)^4.
\end{eqnarray}
The fundamental period factorizes as expected:
\begin{equation}
  \varpi_0=\sum_{n=0}^{\infty}\left\{\sum_{k=0}^n\left(\begin{array}{c}n\\k\end{array}\right)^2\left(\begin{array}{c}n+k\\k\end{array}\right)\right\}^2z^n.
\end{equation}
 The discriminant can be obtained by considering the coefficient of $\theta^4$. This factorizes as
\begin{equation}
  (-1 + z)^2 (1 - 122 z - 122 z^2 + z^3).
\end{equation}
The relevant\footnote{The other point, $z=1$, has Riemann symbol $\{0,1,3,4\}$ and thus does not have the characteristic behavior of a conifold point that has Riemann symbol $\{0,1,1,2\}$.} component comes from the cubic polynomial. Its zeros coincide with the loci of the Coulomb branch in the GLSM and the result is consistent with \cite{inoue}.

\subsection{Example:  GLSMs with gauge group factor $O_+(2)$}
\label{sect:ex:o+2}

In this section we will consider another model of joins of elliptic
curves, where the elliptic curves are described differently.
This will involve an $O_+(2)$ gauge group\footnote{We use the notation and definitions of \cite{Hori:2011pd}.} rather than $U(2)$,
but in other respects, formally the results will be very similar to
the example of the previous section. In passing, we should mention that
recently other mathematics work \cite{prince} has appeared which also discusses
 this example and the results of \cite{inoue}.

\subsubsection{Elliptic curve}
\label{sect:ell-curve}

We will begin by describing an $O_+(2)$ GLSM for the elliptic curve
itself in this language, then we shall describe joins.
This model is also discussed in \cite{Hori:2013gga}[section 5.5.2],
and a closely related model is in \cite{Hori:2011pd}[section 6.1], giving a GLSM realization of \cite{Hosono:2011np},
though we quickly review the details here.

Specifically, consider a GLSM with gauge group
\begin{equation}
\frac{ U(1) \times O_+(2) }{ {\mathbb Z}_2 }
\: = \:
(U(1) \times U(1))\rtimes \mathbb{Z}_2,
\end{equation}
fields $x_i$, $p^k$, $i, j, k \in \{1, 2, 3\}$, in representations
\begin{center}
\begin{tabular}{c|cr|c}
& $x_i$ & $p^k$& FI \\ \hline
$O_+(2)$ & $\Box$ & ${\bf 1}$&- \\
$U(1)$ & $+1$ & $-2$&$2r$
\end{tabular}
\end{center}
and with superpotential
\begin{equation}
W \: = \: \sum_{ij} S^{ij}(p) x_i \cdot x_j,
\end{equation}
where
\begin{equation}
x_i \cdot x_j \: = \: x_i^a x_j^b \delta_{ab},
\end{equation}
and $S^{ij} = S^{ji}$ is a symmetric $3 \times 3$ matrix with entries linear
in $p$.
Alternatively,
as a $(U(1) \times U(1)) \rtimes {\mathbb Z}_2$ gauge theory, 
\begin{center}
\begin{tabular}{c|ccc|c}
& $u_i$ &  $v_i$ & $p^k$&FI \\ \hline
$U(1)$ & $0$ & $1$ & $-1$&$r$ \\
$U(1)$ & $1$ & $0$ & $-1$&$r$
\end{tabular}
\end{center}
where
\begin{eqnarray}
u_i & = & x_i^1 + i x_i^2,
\\
v_i & = & x_i^1 - i x_i^2.
\end{eqnarray}
Furthermore, we analyze this model in a regime where the two Fayet-Iliopoulos
parameters are taken to match:  $r_1 = r_2 = r$.

For $r \gg 0$, the $\{ u_i \}$ are not all zero, and also the
$\{ v_i \}$ are not all zero.  They act as homogeneous coordinates
on ${\mathbb P}^2 \times {\mathbb P}^2$.  Writing the symmetric matrix
\begin{equation}
S^{ij}(p) \: = \: s^{ij}_k p^k,
\end{equation}
the superpotential can be usefully rewritten as
\begin{equation}
W \: = \: \sum_k p^k \left( s^{ij}_k x_i \cdot x_j \right)
\: = \: \frac{1}{2}  \sum_k p^k s^{ij}_k \left( u_i v_j +
v_i u_j \right),
\end{equation}
and so we see that this phase can be interpreted as a complete
intersection of three hypersurfaces of degree $(1,1)$ in
a free $\mathbb{Z}_2$-quotient of ${\mathbb P}^2 \times {\mathbb P}^2$, which is an elliptic curve.

The ${\mathbb Z}_2$ acts by exchanging the two ${\mathbb P}^2$
factors, and we assume that the curve is sufficiently generic to
not intersect the fixed-point loci of that involution.
(This is also condition C in \cite{Hori:2011pd}[section 6.1].)

For $r \ll 0$, the $p^k$ are not all zero,
and act as homogeneous coordinates on ${\mathbb P}^2$.
Here, we interpret the superpotential as a mass matrix for the $x$ fields.
Generically, one expects that the (symmetric) mass matrix has no zero
eigenvalues, so the $x$s are all massive. 
As noted in \cite{Hori:2011pd}[section 4.4], an $SO(k)$ gauge
theory with $N \leq k-2$ doublets has no supersymmetric vacua.
On the other hand, from \cite{Hori:2011pd}[section 4.5],
an $SO(k)$ gauge theory with $N = k-1$ doublets does have vacua
(corresponding to $(1/2) k (k-1)$ free mesons).  Here, for
an $SO(2)$ theory, this means that if we have 1 massless doublet,
we get a unique vacuum, hence we restrict to the subvariety of
${\mathbb P}^2$ over which the mass matrix has rank 2.
This is the locus $\{ \det S^{ij}(p) = 0 \}$, and since
$S^{ij}$ is a $3 \times 3$ matrix, this is a degree 3 hypersurface
in ${\mathbb P}^2$, which is an elliptic curve.

The effective potential on the Coulomb branch is
\begin{eqnarray}
  \mathcal{W}_{eff} & = & -t\left(\sigma_1+\sigma_2\right)-3\sigma_1\left(\ln\sigma_1-1\right)-3\sigma_2\left(\ln\sigma_2-1\right)
\nonumber \\
 & & \hspace*{1.5in}
+ 3\left(\sigma_1+\sigma_2\right)\left(\ln\left(-\sigma_1-\sigma_2\right)-1\right).
\end{eqnarray}
Defining $z=\sigma_2/\sigma_1$, the critical locus is at
\begin{equation}
  e^{-t}=\frac{1}{(1+z)^3}=\frac{1}{\left(1+\frac{1}{z}\right)^3},\qquad z^3=1.
\end{equation}
The singular points are thus at $\exp(-t)=\{-1, 1/8 \}$.

The Picard-Fuchs operator associated to this elliptic curve is
\begin{equation}
  \mathcal{L}=\theta^2-z(7\theta^2+7\theta+2)-8z^2(\theta+1)^2.
\end{equation}
The discriminant is at $z=\{-1,1/8\}$, which is consistent with the GLSM result. The fundamental period is
\begin{equation}
  \varpi_0=\sum_{n=0}^{\infty}\sum_{k=0}^n\left(\begin{array}{c}n\\k\end{array}\right)^3z^n.
\end{equation}
This can also be obtained from the sphere partition function of the GLSM \cite{Hori:2013gga}.

For completeness, let us also describe a dual version of this
theory, using a duality described in
\cite{Hori:2011pd}[section 4.6] to dualize the
$O_+(2)$ part of the gauge theory, with $N=3$ doublets
$x_1, \cdots, x_3$, to an $SO(2)$ gauge theory with
$N=3$ doublets $\varphi^i_a$ ($i \in \{1, \cdots, 3\}$,
$a \in \{1, 2 \}$) with $(1/2) N (N+1) = 3$ singlets $m_{ij}$
and a superpotential
\begin{equation}
W \: = \: \sum_{ij} m_{ij} \varphi^i \cdot \varphi^j.
\end{equation}

In the present case, this results in a GLSM with gauge group
\begin{equation}
\frac{
U(1) \times SO(2)
}{
{\mathbb Z}_2
},
\end{equation}
fields
\begin{center}
\begin{tabular}{c|crr}
& $\varphi^i$ & $p^k$ & $m_{ij}$ \\ \hline
$SO(2)$ & $\Box$ & ${\bf 1}$ & ${\bf 1}$ \\ 
$U(1)$ & $-1$ & $-2$ & $+2$
\end{tabular}
\end{center}
with superpotential
\begin{equation}
W \: = \:
\sum_{ij} S^{ij}(p) m_{ij} \: + \:
\sum_{ij} m_{ij} \varphi^i \cdot \varphi^j.
\end{equation}

Note that if we define
\begin{eqnarray}
\tilde{u}^i & = & \varphi^i_1 + i \varphi^i_2,
\nonumber \\
\tilde{v}^i & = & \varphi^i_1 - i \varphi^i_2,
\end{eqnarray}
then we can take the gauge group to be 
$(U(1) \times U(1)) \rtimes {\mathbb Z}_2$ 
under which
\begin{center}
\begin{tabular}{c|cccc}
& $\tilde{u}^i$ & $\tilde{v}^i$ & $p^k$ & $m_{ij}$ \\ \hline
$U(1)$ & $1$ & $0$ & $-1$ & $+1$ \\
$U(1)$ & $0$ & $1$ & $-1$ & $+1$,
\end{tabular}
\end{center}
with superpotential
\begin{equation}
W \: = \:
\sum_{ij} S^{ij}(p) m_{ij} \: + \:
\frac{1}{2} \sum_{ij} m_{ij} \left( \tilde{u}^i \tilde{v}^j
+ \tilde{v}^i \tilde{u}^j \right).
\end{equation}

For $r \gg 0$, D-terms imply that $\{ \tilde{u}^i, m_{ij} \}$
are not all zero, and separately $\{ \tilde{v}^j, m_{ij} \}$ are
not all zero.  We can interpret the second superpotential term
\begin{displaymath}
\sum_{ij} m_{ij} \varphi^i \cdot \varphi^j
\end{displaymath}
as implying that we should restrict to the rank 2 locus of the
space of matrix elements $m_{ij}$.  (This is because, as pointed
out in \cite{Hori:2011pd}[section 4.4], an $SO(2)$ gauge theory with
no doublets\footnote{
To be clear, this is a statement about pure $SO(2)$ gauge theory in
two dimensions with suitable (discrete) theta angle.
For related statements for other pure gauge theories, see for
example \cite{Aharony:2016jki,Gu:2018fpm,Chen:2018wep}.
} has no supersymmetric vacua, whereas \cite{Hori:2011pd}[section 4.5]
an $SO(2)$ gauge theory
with one doublet has a unique vacuum.)
On this locus, we can write
\begin{displaymath}
m_{ij} \: \propto \: a_i b_j \: + \: a_j b_i,
\end{displaymath}
where we can take the $a_i$, $b_j$ to have equal and opposite charges,
so (glossing over excluded loci) this means the space of $m_{ij}$
reproduces ${\mathbb P}^2 \times {\mathbb P}^2$.  The first
term in the superpotential,
\begin{displaymath}
\sum_{ij} S^{ij}(p) m_{ij},
\end{displaymath}
then gives three hypersurfaces in the space of $m_{ij}$, reproducing
the complete intersection in ${\mathbb P}^2 \times {\mathbb P}^2$
that gives an elliptic curve.
This is a nonperturbative of geometry, in a phase which in the previous duality
frame was a perturbatively-understood geometry.

For $r \ll 0$, D-terms imply that the $p^k$ are not all zero.
The superpotential can be helpfully rewritten as
\begin{equation}
W \: = \: \sum_{ij} m_{ij} \left( s^{ij}_k p^i \: + \: \frac{1}{2}\left(
 \tilde{u}^i \tilde{v}^j
+ \tilde{v}^i \tilde{u}^j \right) \right).
\end{equation}
In this phase, the $m_{ij}$ act analogously to a Lagrange multiplier,
giving the constraint that
\begin{displaymath}
S^{ij}(p) \: \propto \: \tilde{u}^i \tilde{v}^j
+ \tilde{v}^i \tilde{u}^j,
\end{displaymath}
or more simply that the $3 \times 3$ matrix $S^{ij}(p)$ have rank 2,
restricting the allowed $p$.  In fact, this is essentially the
PAXY model of \cite{Jockers:2012zr}[section 3.5]
for a symmetric determinantal representation. 
In any event, as discussed previously, the locus on which $S^{ij}$ has
rank no more than 2 is the locus $\{ \det S^{ij} = 0 \}$,
a degree 3 hypersurface in ${\mathbb P}^2$, which is an elliptic curve.
Here, in this dual GLSM, we see this geometry realized perturbatively.

\subsubsection{Ambient join}

Now, let us describe what is morally the join of the ambient space of the
model above with itself, in GLSM language.  Here, we describe the
'ambient space' by a GLSM with gauge group
\begin{equation}
\frac{
U(1) \times O_+(2) 
}{
{\mathbb Z}_2
},
\end{equation}
and three doublets $x_i$, $i \in \{1, 2, 3\}$, in representations
\begin{center}
\begin{tabular}{c|c}
& $x_i$ \\ \hline
$O_+(2)$ & $\Box$ \\
$U(1)$ & $1$
\end{tabular}
\end{center}

Morally, the join of the ambient space with itself is then described
by a GLSM with gauge group
\begin{equation}
\left(
\frac{
U(1) \times O_+(2) 
}{
{\mathbb Z}_2
}
\right)^2 \times U(1)_3,
\end{equation}
with fields in representations
\begin{center}
\begin{tabular}{c|ccrr}
& $x_i$ & $\tilde{x}_i$ & $z_1$ & $z_2$ \\ \hline
$O_+(2)$ & $\Box$ & ${\bf 1}$ & ${\bf 1}$ & ${\bf 1}$ \\
$U(1)$ & $1$ & $0$ & $-2$ & $0$ \\
$O_+(2)$ & ${\bf 1}$ & $\Box$ & ${\bf 1}$ & ${\bf 1}$ \\
$U(1)$ & $0$ & $1$ & $0$ & $-2$ \\
$U(1)_3$ & $0$ & $0$ & $1$ & $1$ 
\end{tabular}
\end{center}

Proceeding as before, if we integrate out the divisor
$\{ z_1 = 0\}$ and remove $U(1)_3$, we get
\begin{center}
\begin{tabular}{c|ccr}
& $x_i$ & $\tilde{x}_i$ & $z_2$ \\ \hline
$O_+(2)$ & $\Box$ & ${\bf 1}$ & ${\bf 1}$ \\
$U(1)$ & $1$ & $0$ & $-2$ \\
$O_+(2)$ & ${\bf 1}$ & $\Box$ & ${\bf 1}$ \\
$U(1)$ & $0$ & $1$ & $-2$
\end{tabular}
\end{center}
and then blowing down $\{ z_2 = 0 \}$ and removing the second $U(1)$,
this becomes our final result for the join of the two
'ambient spaces,'
\begin{center}
\begin{tabular}{c|cc}
& $x_i$ & $\tilde{x}_i$ \\ \hline
$O_+(2)$ & $\Box$ & ${\bf 1}$ \\
$U(1)$ & $1$ & $1$ \\
$O_+(2)$ & ${\bf 1}$ & $\Box$ 
\end{tabular}
\end{center}
To be clear, this is now a 
\begin{equation}
\frac{
O_+(2) \times O_+(2) \times U(1)
}{
{\mathbb Z}_2 \times {\mathbb Z}_2
}
\end{equation}
gauge theory.

In passing, this GLSM, for the `join of the ambient spaces,'
also includes as a subset the GLSM for the self-intersection of 
two elliptic curves, described in the same fashion.
The GLSM for the intersection, following \cite{Caldararu:2017usq},
has gauge group
\begin{equation}
\frac{
O_+(2) \times O_+(2) \times U(1)
}{
{\mathbb Z}_2 \times {\mathbb Z}_2
},
\end{equation}
with fields
\begin{center}
\begin{tabular}{c|ccrrr}
& $x_i$ & $\tilde{x}_i$ & $p^k$ & $\tilde{p}^k$ & $q^{ij}$ \\ \hline
$O_+(2)$ & $\Box$ & ${\bf 1}$ & ${\bf 1}$ & ${\bf 1}$ & ${\bf 1}$ \\
$U(1)$ & $+1$ &  $+1$ & $-2$ & $-2$ &  $-2$ \\
$O_+(2)$ & ${\bf 1}$ & $\Box$ & ${\bf 1}$ & ${\bf 1}$ & ${\bf 1}$
\end{tabular}
\end{center}
with superpotential
\begin{equation}
W \: = \: \sum_{ij} S^{ij}(p) x_i \cdot x_j \: + \:
\sum_{ij} \tilde{S}^{ij}(\tilde{p}) \tilde{x}_i \cdot \tilde{x}_j
\: + \:
\sum_{ij} q^{ij} \left( f_{ij}(B) - \tilde{B}_{ij} \right),
\end{equation}
where
\begin{equation}
B_{ij} \: = \: x_i^a x_j^b \delta_{ab}, 
\: \: \:
\tilde{B}_{ij} \: = \: \tilde{x}^a_i \tilde{x}^b_j \delta_{ab},
\end{equation}
and $f_{ij}$ encodes a linear rotation of the symmetry
group $GL(3)$.
Clearly, if we removed the superpotential and the fields
$p^k$, $\tilde{p}^k$, and $q^{ij}$, we would recover the GLSM
for the `ambient join' discussed above.

\subsubsection{Calabi-Yau complete intersection}

Next, we will describe the join of two elliptic curves described
in this fashion, or equivalently a complete intersection in the
`ambient join' of the previous subsection.

Our starting point is then a GLSM with gauge group
\begin{equation}
\left(
\frac{
U(1) \times O_+(2) 
}{
{\mathbb Z}_2
}
\right)^2 \times U(1)_3,
\end{equation}
with fields in representations
\begin{center}
\begin{tabular}{c|ccrrrr}
& $x_i$ & $\tilde{x}_i$ & $p^k$ & $\tilde{p}^k$ & $z_1$ & $z_2$ \\ \hline
$O_+(2)$ & $\Box$ & ${\bf 1}$ & ${\bf 1}$ & ${\bf 1}$ 
& ${\bf 1}$ & ${\bf 1}$ \\
$U(1)$ & $1$ & $0$ & $-2$ & $0$ & $-2$ & $0$ \\
$O_+(2)$ & ${\bf 1}$ & $\Box$ & ${\bf 1}$ & ${\bf 1}$
& ${\bf 1}$ & ${\bf 1}$  \\
$U(1)$ & $0$ & $1$ & $0$ & $-2$ & $0$ & $-2$ \\
$U(1)_3$ & $0$ & $0$ & $0$ & $0$ & $1$ & $1$
\end{tabular}
\end{center}
with superpotential
\begin{equation}
W \: = \: \sum_{ij} S^{ij}(p) x_i \cdot x_j \: + \:
\sum_{ij} \tilde{S}^{ij}(\tilde{p}) \tilde{x}_i \cdot \tilde{x}_j.
\end{equation}

Now, as written, this GLSM does not sit at an RG fixed point:
for example, the sum of the $U(1)_3$ charges is nonzero.
We proceed as in the previous example, by successive blowdowns to
a different GLSM where the Calabi-Yau condition can be seen on the
ambient space, without using any relations (as in 
appendix~\ref{app:nonambientcy}).

Blowing down $\{ z_1 = 0 \}$ and eliminating $U(1)_3$, we get the
fields
\begin{center}
\begin{tabular}{c|ccrrr}
& $x_i$ & $\tilde{x}_i$ & $p^k$ & $\tilde{p}^k$ &  $z_2$ \\ \hline
$O_+(2)$ & $\Box$ & ${\bf 1}$ & ${\bf 1}$ & ${\bf 1}$
 & ${\bf 1}$ \\
$U(1)$ & $1$ & $0$ & $-2$ & $0$ & $-2$  \\
$O_+(2)$ & ${\bf 1}$ & $\Box$ & ${\bf 1}$ & ${\bf 1}$
 & ${\bf 1}$  \\ 
$U(1)$ & $0$ & $1$ & $0$ & $-2$  & $-2$ 
\end{tabular}
\end{center}

Blowing down $\{ z_2 = 0 \}$ and eliminating the second $U(1)$,
we get our final GLSM for the join of the two elliptic curves, 
described by the fields
\begin{center}
\begin{tabular}{c|ccrr}
& $x_i$ & $\tilde{x}_i$ & $p^k$ & $\tilde{p}^k$  \\ \hline
$O_+(2)$ & $\Box$ & ${\bf 1}$ & ${\bf 1}$ & ${\bf 1}$ \\
$U(1)$ & $1$ & $1$ & $-2$ &  $-2$  \\
$O_+(2)$ & ${\bf 1}$ & $\Box$ & ${\bf 1}$ & ${\bf 1}$
\end{tabular}
\end{center}
with gauge group
\begin{equation}
\frac{
O_+(2) \times U(1) \times O_+(2)
}{
{\mathbb Z}_2 \times {\mathbb Z}_2
},
\end{equation}
and superpotential
\begin{equation}
W \: = \: \sum_{ij} S^{ij}(p) x_i \cdot x_j \: + \:
\sum_{ij} \tilde{S}^{ij}(\tilde{p}) \tilde{x}_i \cdot \tilde{x}_j.
\end{equation}
Here, for example, the sum of the $U(1)$ charges vanishes, so we see
that this GLSM describes a Calabi-Yau.

We can change variables as before from $x_i$, $\tilde{x}_i$
to $u_i$, $v_i$, $\tilde{u}_i$, $\tilde{v}_i$. Before taking the
${\mathbb Z}_2$ quotient, the charges of the fields, and FI parameters 
corresponding to $U(1)$ factors, are given by
\begin{equation*}
  \begin{array}{c|rrrrrr|c}
    & u_i&v_i&\tilde{u}_i&\tilde{v}_i&p_{\alpha}&\tilde{p}_{\beta}&\text{FI}\\
    \hline
    U(1)&1&1&1&1&-2&-2&2r\\
    O(2)_1&1&-1&0&0&0&0&0\\
    O(2)_2&0&0&1&-1&0&0&0
    \end{array}
\end{equation*}
To carry out the $\mathbb{Z}_2$-quotient at the level of the charges make a change of basis by adding the second and third row to the first. Then we end up with
\begin{equation*}
  \begin{array}{c|rrrrrr|c}
    & u_i&v_i&\tilde{u}_i&\tilde{v}_i&p_{\alpha}&\tilde{p}_{\beta}&\text{FI}\\
    \hline
    a&2&0&2&0&-2&-2&2r\\
    b&1&-1&0&0&0&0&0\\
    c&0&0&1&-1&0&0&0
    \end{array}
\end{equation*}
Now we see that the action of the first $U(1)$, denoted by $a$ acts non-minimally. We mod out by the $\mathbb{Z}_2$ by dividing the first line by $2$. From this we obtain the charges for the fields in the free quotient
\begin{equation}
  \label{o2o2charges}
  \begin{array}{c|rrrrrr|c}
    & u_i&v_i&\tilde{u}_i&\tilde{v}_i&p_{\alpha}&\tilde{p}_{\beta}&\text{FI}\\
    \hline
    \tilde{a}&1&0&1&0&-1&-1&r\\
    b&1&-1&0&0&0&0&0\\
    c&0&0&1&-1&0&0&0
    \end{array}
\end{equation}
To reveal the geometry in the $r\gg0$ phase we can make further manipulations
\begin{equation}
  \label{o2o2charges2}
  \begin{array}{c|rrrrrr|c}
    & u_i&v_i&\tilde{u}_i&\tilde{v}_i&p_{\alpha}&\tilde{p}_{\beta}&\text{FI}\\
    \hline
    \tilde{a}&1&0&1&0&-1&-1&r\\
    \tilde{a}b^{-1}&0&1&1&0&-1&-1&r\\
    \tilde{a}c^{-1}&1&0&0&1&-1&-1&r
    \end{array}
\end{equation}
In this latter form, the geometric interpretation in terms of joins is more clear. To reveal the freely acting $\mathbb{Z}_2$ it is also useful to further rewrite this as
\begin{equation*}
  \begin{array}{c|rrrrrr|c}
    & u_i&v_i&\tilde{u}_i&\tilde{v}_i&p_{\alpha}&\tilde{p}_{\beta}&\text{FI}\\
    \hline
    \tilde{a}&1&0&1&0&-1&-1&r\\
    \tilde{a}^2b^{-1}c^{-1}&1&1&1&1&-2&-2&2r\\
    \tilde{a}c^{-1}&1&0&0&1&-1&-1&r
    \end{array}
\end{equation*}
The ambient geometry corresponds to a $\mathbb{P}^{11}$ with two $\mathbb{P}^5$s blown up. The freely acting $\mathbb{Z}_2$ exchanges the two $\mathbb{P}^5$s. 

In the $r\gg0$ phase, this describes a (singular) Calabi-Yau,
given by a codimension $6$ complete intersection in the ambient geometry defined by $u_i,v_i\tilde{u}_i,\tilde{v}_i$.
The first $U(1)$ describes an ambient ${\mathbb P}^5$ with homogeneous
coordinates
$\{ u_i, \tilde{u}_i \}$; the next $U(1)$ describes a blowup in which
one inserts a ${\mathbb P}^2$ along the locus where all the $u_i$ vanish;
the third $U(1)$ describes a blowup in which one inserts a ${\mathbb P}^2$
along the locus where all of the $\tilde{u}_i$ vanish.
We then take a complete intersection of six hyperplanes of degrees $(1,1,1)$
in this toric variety.  
The resulting space (the join of two elliptic curves) is a Calabi-Yau
(from the fact that charges sum to zero), of dimension 3,
as expected.  

To be clear, as before, this join is singular, which is reflected
in the fact that the GLSM phase has noncompact branches we have omitted.
In the next section, we will deform this model to a smooth Calabi-Yau,
and for example will compute the 
Hodge numbers of the resulting space.

For $r \ll 0$, D-terms imply that
the $\{p^k, \tilde{p}^k \}$ are not all zero.
We can interpret each of the superpotential terms
\begin{displaymath}
\sum_{ij} S^{ij}(p) x_i \cdot x_j,
\: \: \:
\sum_{ij} \tilde{S}^{ij}(\tilde{p}) \tilde{x}_i \cdot \tilde{x}_j
\end{displaymath}
as mass matrices, one for the doublets $x_i$, the other for the
doublets $\tilde{x}_i$.  Just as in our previous analysis,
we only get vacua from $SO(2)$ gauge theories with one massless
doublet \cite{Hori:2011pd}[section 4], so the vacua
are defined by $p^k$ such
that the matrix $S^{ij}(p)$ has rank 2,
and $\tilde{p}^k$ such that the matrix $\tilde{S}^{ij}(\tilde{p})$
has rank 2.  Putting this together, we see that the geometry in this
phase is the join of the two determinantal varieties (one defined by
the locus where $S^{ij}$ has rank $2$, the other by the locus where
$\tilde{S}^{ij}$ has rank $2$).

\subsubsection{Deformation to self-intersection}

The space described by the GLSM of the previous section is
extremely singular, as expected for a join.  It contains loci where
each of two spaces shrinks to a point.  We can smooth the singularities
by deforming the superpotential to a more generic form compatible
with the symmetries of the theory:
\begin{equation}
W \: = \:
\sum_{ij} S^{ij}(p, \tilde{p}) x_i \cdot x_j \: + \:
\sum_{ij} \tilde{S}^{ij}(p, \tilde{p}) \tilde{x}_i \cdot \tilde{x}_j.
\end{equation}
In other words, we take the matrices $S^{ij}$, $\tilde{S}^{ij}$ to be
functions of both $p$ and $\tilde{p}$, rather than one set apiece.

In this section we will argue that the smooth deformation of the join
above is equivalent to a self-intersection of the ambient space,
closely analogous to the relation we saw in the previous section
between 
the Calabi-Yau complete
intersection in the join of $G(2,5)$ with itself, and the
self-intersection $G(2,5) \cap G(2,5)$ discussed in \cite{Caldararu:2017usq}.

Now, let us compare this to the self-intersection of the ambient space,
after a generic $GL(3)$ rotation, following 
\cite{Caldararu:2017usq}.  Following the same pattern as for
$G(2,5) \cap G(2,5)$ described there,
this self-intersection is described by a GLSM with gauge group
\begin{equation}
\frac{ U(1) \times O_+(2) \times O_+(2)
}{
{\mathbb Z}_2 \times {\mathbb Z}_2 
},
\end{equation}
with fields in representations
\begin{center}
\begin{tabular}{c|ccr}
& $x_i$ & $\tilde{x}_i$ & $q^{ij}$ \\ \hline
$O_+(2)$ & $\Box$ & ${\bf 1}$ & ${\bf 1}$ \\
$U(1)$ & $1$ & $1$ & $-2$ \\
$O_+(2)$ & ${\bf 1}$ & $\Box$ & ${\bf 1}$
\end{tabular}
\end{center}
with superpotential
\begin{equation}
W \: = \: q^{ij} \left( f_{ij}(x \cdot x) \: - \: \tilde{x}_i \cdot
\tilde{x}_j \right),
\end{equation}
where
\begin{equation}
f_{ij}(x \cdot x) \: = \: f^{k \ell}_{ij} x_k \cdot x_{\ell},
\end{equation}
and $f^{k \ell}_{ij}$ are constants, encoding a generic $GL(3)$
rotation.

To relate this self-intersection of the ambient space to
the deformation of the join of the elliptic curves, 
first make the field redefinition
\begin{equation}
q^{ij} \: = \: - \tilde{S}^{ij}(p, \tilde{p} ).
\end{equation}
(As a consistency check, note that there are altogether
six $p$, $\tilde{p}$ fields, which matches the number of $q$ fields.)
Then, define the $f_{ij}^{k \ell}$ as the solutions to the equations
\begin{equation}
q^{ij} f_{ij}^{k \ell} x_k \cdot x_{\ell} \: = \: S^{k \ell}(p, \tilde{p})
x_k \cdot x_{\ell},
\end{equation}
or in other words,
\begin{equation}
- \tilde{S}^{ij}(p, \tilde{p} )  f_{ij}^{k \ell}  \: = \: S^{k \ell}(p, \tilde{p}).
\end{equation}
To see why these equations can be solved for the $f^{k \ell}_{ij}$, 
note that if we distinguish
coefficients of $p^k$ from those of $\tilde{p}^k$, then this is a system
of 12 equations (two for every entry in the matrix $S^{ij}$)
with $6^2 = 36$ unknowns (the entries of $f^{k \ell}_{ij}$).  Trivially,
for generic choices, (multiple) solutions for $f^{k \ell}_{ij}$ exist.

With these definitions, the superpotential for the self-intersection
of the ambient space can be written
\begin{equation}
W \: = \: S^{ij}(p, \tilde{p} ) x_i \cdot x_j \: + \:
\tilde{S}^{ij}(p, \tilde{p}) \tilde{x}_i \cdot \tilde{x}_j,
\end{equation}
and as the $q^{ij}$ are equivalent to the $p^k$, $\tilde{p}^k$,
we see then that the self-intersection of the ambient space, for generic
$GL(3)$ rotation, is equivalent to a deformation of the join of two copies
of the elliptic curve of the subsection~\ref{sect:ell-curve}.

Now, let us take a moment to discuss the phases of this new GLSM,
describing the deformation of the join, or equivalently, the 
self-intersection of the ambient theory.  The $r \gg 0$ phase is
precisely the geometry we have been describing,
namely the self-intersection defined by
\begin{equation}
f_{ij}(x \cdot x) \: - \: \tilde{x}_i \cdot \tilde{x}_j \: = \: 0
\end{equation}
in the space of doublets $\{ x_i, \tilde{x}_i \}$, or equivalently,
if we write 
\begin{eqnarray}
S^{ij}(p,\tilde{p}) & = &
\sum_k p^k (s_1)^{ij}_k \: + \:
\sum_k \tilde{p}^k (s_2)^{ij}_k,
\nonumber \\ 
\tilde{S}^{ij}(p,\tilde{p}) & = &
\sum_k p^k (\tilde{s}_1)^{ij}_k \: + \:
\sum_k \tilde{p}^k (\tilde{s}_2)^{ij}_k,
\end{eqnarray}
the complete intersection
\begin{equation}
(s_1)^{ij}_k x_i \cdot x_j \: + \:
(\tilde{s}_1)^{ij}_k \tilde{x}_i \cdot \tilde{x}_j 
\: = \: 0 \: = \:
(s_2)^{ij}_k x_i \cdot x_j \: + \:
(\tilde{s}_2)^{ij}_k \tilde{x}_i \cdot \tilde{x}_j .
\end{equation} 
For $r \ll 0$, D-terms imply that $\{ p^k, \tilde{p}^k \}$ are not all zero,
and as for the join, we interpret each of the superpotential terms
\begin{displaymath}
S^{ij}(p, \tilde{p} ) x_i \cdot x_j,
\: \: \:
\tilde{S}^{ij}(p, \tilde{p}) \tilde{x}_i \cdot \tilde{x}_j,
\end{displaymath}
as mass matrices, one for the doublets $x_i$, the other for the
doublets $\tilde{x}_i$.  As in our previous discussion,
one only gets vacua from $SO(2)$ gauge theories with
one massless doublet \cite{Hori:2011pd}[section 4], so the vacua are
defined by the intersection of the
loci on $\{ p^k, \tilde{p}^k \}$ over which $S^{ij}(p,\tilde{p})$ has
rank 2, and $\tilde{S}^{ij}(p,\tilde{p})$ has rank 2. The first locus
is (generically) the locus $\{ \det S = 0 \}$, a degree three hypersurface,
and similarly the second is $\{ \det \tilde{S} = 0 \}$, another
degree three hypersurface.  Thus, the $r \ll 0$ phase describes
the complete intersection of two degree 3 hypersurfaces
in ${\mathbb P}^5$ (with homogeneous coordinates $\{ p^k, \tilde{p}^k \}$),
which is another Calabi-Yau.

For the rest of this section, we will consider a generic deformation
of the join, or equivalently a self-intersection of the ambient space,
as we dualize GLSMs and interpret the phases.

Before going on, we will outline the computation of the Hodge numbers of
the Calabi-Yau obtained in the $r \gg 0$ phase. For this purpose we relate our model to a toric three-parameter model defined by $\{u_i,v_i,\tilde{u}_i,\tilde{v}_i\}$ with weights as in (\ref{o2o2charges2}). To compute the Hodge numbers of the three-parameter model we use the program cohomCalg \cite{Blumenhagen:2010pv}, which requires the Stanley-Reisner ideal as input. This is obtained by computing the maximal star triangulation of the N-lattice polytope associated to the toric variety computable via TOPCOM\cite{topcom}. Using the {\tt nef.x}-function of PALP \cite{Braun:2012vh}, we can determine the $12$ vertices of the polytope:
\begin{equation*}
  \begin{array}{rrrrrrrrrrrr}
    \nu_1&\nu_2&\nu_3&\nu_4&\nu_5&\nu_6&\nu_7&\nu_8&\nu_9&\nu_{10}&\nu_{11}&\nu_{12}\\
    \hline
    0&  0&  0&  0&  0&  0&  0&  0&  0&  1&  0& -1\\
    0&  0&  0&  0&  0&  0&  0&  0&  0&  0&  1& -1\\
    0&  0&  0&  0&  0&  0&  1&  0& -1&  0&  0&  0\\
    0&  0&  0&  0&  0&  0&  0&  1& -1&  0&  0&  0\\
    1&  0&  0&  0&  0&  0&  0&  0&  0&  0&  0&  0\\
    0&  1& -1&  0&  0&  0&  0&  0&  0&  0&  0&  0\\
    0&  0& -1&  0&  0&  0&  0&  0&  0&  0&  0&  0\\
    0&  0&  1&  1&  0&  0&  0&  0& -1&  0&  0& -1\\
    0&  0&  1&  0&  1&  0&  0&  0& -1&  0&  0& -1\\
    0&  0&  1&  0&  0&  1&  0&  0& -1&  0&  0& -1\\
    \hline
    1&  1&  1&  0&  0&  0&  1&  1&  1&  0&  0&  0\\
    0&  0&  0&  1&  1&  1&  1&  1&  1&  0&  0&  0\\
    1&  1&  1&  0&  0&  0&  0&  0&  0&  1&  1&  1
    \end{array}
  \end{equation*}
Adding the origin, there are four maximal triangulations, all of which yield the same result for the Mori generators and the Stanley-Reisner ideal via the {\tt mori.x}-function of PALP. Assigning toric divisors $D_{\nu_i}$ to the vertices, the Stanley-Reisner ideal is
\begin{equation}
  D_{\nu_1}D_{\nu_2}D_{\nu_3}=0,
\qquad  D_{\nu_7}D_{\nu_8}D_{\nu_9}=0,
\qquad  D_{\nu_4}D_{\nu_5}D_{\nu_6} D_{\nu_{10}}D_{\nu_{11}}D_{\nu_{12}}=0.
\end{equation}
Using this and the hypersurface degrees as input for cohomCalg, the complete intersection in the toric ambient space has Hodge numbers $(h^{1,1}(\tilde{X}),h^{2,1}(\tilde{X}))=(3,39)$ and the Euler number is $\chi(\tilde{X})=-72$. Now we have to take into account that our Calabi-Yau is a one-parameter complete intersection in a free $\mathbb{Z}_2$-quotient of this ambient space which means that $\chi(X)=\chi(\tilde{X})/2=-36$. After identifying the K\"ahler parameters we can this deduce that  $(h^{1,1}(\tilde{X}),h^{2,1}(\tilde{X}))=(1,19)$.

The form of the effective potential on the Coulomb branch can be inferred from (\ref{o2o2charges}):
\begin{eqnarray}
  \mathcal{W}_{eff}&=&-t\sigma_1-3\sigma_2\left[\ln\sigma_2-1\right]-3\left(\sigma_1+\sigma_2\right)\left[\ln\left(\sigma_1+\sigma_2\right)-1\right]
\nonumber\\
  &=&-3\left(\sigma_1+\sigma_3\right)\left[\ln\left(\sigma_1+\sigma_3\right)-1\right]-3\sigma_3\left[\ln\sigma_3-1\right]
\nonumber \\
& & \hspace*{1.5in} +6\sigma_1\left[\ln\left(-\sigma_1\right)-1\right].
\end{eqnarray}
Defining $z_1=\sigma_2/\sigma_2$ and $z_2=\sigma_3/\sigma_1$ the critical locus is at
\begin{equation}
  e^{-t}=(1+z_1)^3(1+z_2)^3,\qquad \left(\frac{1}{z_1}+1\right)^3=\left(\frac{1}{z_2}+1\right)^3=-1.
\end{equation}
Solving this, one finds a Coulomb branch at $\exp(-t)=\{1,-1/8,1/64\}$. Comparing with the result of the elliptic curve we see that the singularities of the join sit at points which are the products of the singular loci of the components elliptic curves. This is in agreement with statements in \cite{galkintalk}.

\subsubsection{Picard-Fuchs operator and Hadamard products}

Next, we analyze the Picard-Fuchs operator. Via the connection between joins and Hadamard products, the fundamental period is
\begin{equation}
  \sum_{n=0}^\infty\left\{\sum_{k=0}^n\left(\begin{array}{c}n\\k\end{array}\right)^3\right\}^2z^n .
\end{equation}
This is annihilated by the Picard-Fuchs operator \cite{vanstraten2} (AESZ 100)
\begin{eqnarray}
  \mathcal{L}^{O(2)}*\mathcal{L}^{O(2)}&=&\theta^4-z \left(73 \theta ^4+98 \theta ^3+77 \theta
  ^2+28 \theta +4\right)\nonumber\\
  &&+z^2\left(520 \theta ^4-1040 \theta ^3-2904 \theta
  ^2-2048 \theta -480\right)\nonumber\\
&&+64 z^3 \left(65 \theta ^4+390 \theta ^3+417 \theta
  ^2+180 \theta +28\right)\nonumber\\
 &&-512 z^4 \left(73 \theta ^4+194 \theta ^3+221 \theta
  ^2+124 \theta +28\right)+32768z^5 (\theta +1)^4 .\nonumber\\
\end{eqnarray}
To extract the discriminant we consider the coefficient of $\theta^4$. This factorizes as follows
\begin{equation}
  (1 - 8 z)^2 (1 - 57 z - 456 z^2 + 512 z^3).
\end{equation}
The component relevant to the discriminant is the cubic equation whose zeros coincide with the Coulomb branch analysis of the GLSM.

\subsubsection{Dualize one factor in complete intersection}

Next, we consider dualizing one of the $O_+(2)$ factors in the
gauge group.  From 
\cite{Hori:2011pd}[section 4.2],
an $O_+(2)$ gauge theory with $N = 3$ doublets $x_1^a, \cdots, x_3^a$ 
is dual to
an $SO(N-2+1) = SO(2)$ gauge theory with $N=3$ doublets
$\varphi^i_a$ ($i \in \{1, \cdots, 3\}$, $a \in \{1, 2\}$), 
$(1/2)N (N+1)$ singlets $m_{ij} = + m_{ji}$, and 
a superpotential
\begin{equation}
W \: = \: \sum_{ij} m_{ij} \varphi^i \cdot \varphi^j,
\end{equation}
where the singlets of the dual theory are related to the fundamental fields
of the original theory as
\begin{equation}
m_{ij} \: = \: x_i \cdot x_j.
\end{equation}

Here, if we apply the duality above to one of the $O_+(2)$ factors
of the final GLSM of the previous section, we produce a GLSM with
gauge group
\begin{equation}
\frac{
SO(2) \times U(1) \times O_+(2)
}{
{\mathbb Z}_2 \times {\mathbb Z}_2
},
\end{equation}
fields
\begin{center}
\begin{tabular}{c|ccrrr}
& $\varphi^i$ & $\tilde{x}_i$ & $p^k$ & $\tilde{p}^k$  & $m_{ij}$ \\ \hline
$SO(2)$ & $\Box$ & ${\bf 1}$ & ${\bf 1}$ & ${\bf 1}$ & ${\bf 1}$ \\
$U(1)$ & $-1$ & $1$ & $-2$ &  $-2$ & $+2$  \\
$O_+(2)$ & ${\bf 1}$ & $\Box$ & ${\bf 1}$ & ${\bf 1}$ & ${\bf 1}$
\end{tabular}
\end{center}
and a superpotential
\begin{equation}
W \: = \: \sum_{ij} S^{ij}(p,\tilde{p}) \, m_{ij} \: + \:
\sum_{ij} \tilde{S}^{ij}(p,\tilde{p}) \, \tilde{x}_i \cdot \tilde{x}_j
\: + \:
 \sum_{ij} m_{ij} \varphi^i \cdot \varphi^j.
\end{equation}

Next, we shall analyze the phases of this GLSM, and describe
how they reproduce the same geometries as we saw in the previous
subsection.

First, consider the phase $r \gg 0$.  In this phase,
D-terms imply that $\{ \tilde{x}_i, m_{ij} \}$ are not all zero.
The last term of the superpotential,
\begin{displaymath}
 \sum_{ij} m_{ij} \varphi^i \cdot \varphi^j,
\end{displaymath}
is a mass term for the doublets $\varphi^i$.
Just as in our previous analysis, from
\cite{Hori:2011pd}[section 4], there will be no vacua when there are
no massless doublets; to get a (unique) vacuum, we need one massless
doublet.  As a result, this superpotential term tells us that we
must restrict to the rank two locus of the matrices $m_{ij}$.
The first two superpotential terms,
\begin{eqnarray*}
\lefteqn{
\sum_{ij} S^{ij}(p,\tilde{p}) \, m_{ij} \: + \:
\sum_{ij} \tilde{S}^{ij}(p,\tilde{p}) \, \tilde{x}_i \cdot \tilde{x}_j
} \\
& = &
\sum_k p^k \left( (s_1)^{ij}_k m_{ij} \: + \:
(\tilde{s}_1)^{ij}_k \tilde{x}_i \cdot \tilde{x}_j \right)
\: + \:
\sum_k \tilde{p}^k \left( 
(s_2)^{ij}_k m_{ij} \: + \:
(\tilde{s}_2)^{ij}_k \tilde{x}_i \cdot \tilde{x}_j \right),
\end{eqnarray*}
where
\begin{eqnarray}
S^{ij}(p,\tilde{p}) & = &
\sum_k p^k (s_1)^{ij}_k \: + \:
\sum_k \tilde{p}^k (s_2)^{ij}_k,
\nonumber \\
\tilde{S}^{ij}(p,\tilde{p}) & = &
\sum_k p^k (\tilde{s}_1)^{ij}_k \: + \:
\sum_k \tilde{p}^k (\tilde{s}_2)^{ij}_k,
\end{eqnarray}
instruct us to restrict to the complete intersection
\begin{equation}  \label{eq:o+2:dual1:ci}
(s_1)^{ij}_k m_{ij} \: + \:
(\tilde{s}_1)^{ij}_k \tilde{x}_i \cdot \tilde{x}_j 
\: = \: 0 \: = \:
(s_2)^{ij}_k m_{ij} \: + \:
(\tilde{s}_2)^{ij}_k \tilde{x}_i \cdot \tilde{x}_j .
\end{equation}
We know from the duality that in terms of the dual variables
$x_i^a$,
\begin{displaymath}
m_{ij} \: = \: x_i \cdot x_j,
\end{displaymath}
(which correlates with the earlier requirement that we restrict
to the rank two locus of the $m_{ij}$, as each $x_i$ denotes a doublet).  
As a result, our complete
intersection~(\ref{eq:o+2:dual1:ci}) can be written as
\begin{equation}
(s_1)^{ij}_k x_i \cdot x_j \: + \:
(\tilde{s}_1)^{ij}_k \tilde{x}_i \cdot \tilde{x}_j 
\: = \: 0 \: = \:
(s_2)^{ij}_k x_i \cdot x_j \: + \:
(\tilde{s}_2)^{ij}_k \tilde{x}_i \cdot \tilde{x}_j .
\end{equation}
which is identical to the interpretation of the $r \gg 0$ phase of this
GLSM in the first duality frame.

Next, consider the phase $r \ll 0$.  In this phase,
D-terms imply that $\{ \varphi^i, p^k, \tilde{p}^k \}$ are not all
zero.
The first and third terms in the superpotential above, which can
be written
\begin{displaymath}
\sum_{ij} m_{ij} \left( S^{ij}(p,\tilde{p}) \: + \: \varphi^i \cdot \varphi^j
\right),
\end{displaymath}
imply that the $3 \times 3$ matrix $S^{ij}(p,\tilde{p})$ is of rank 2.
Specifically, this has the form of the PAXY model for a symmetric
determinantal variety \cite{Jockers:2012zr}[section 3.5].
The middle superpotential term,
\begin{displaymath}
\sum_{ij} \tilde{S}^{ij}(p,\tilde{p}) \tilde{x}_i \cdot \tilde{x}_j,
\end{displaymath}
is a mass matrix for the doublets $\tilde{x}_i$ over the space of
$\{ p^k, \tilde{p}^k \}$.  Just as in our previous analysis, from
\cite{Hori:2011pd}[section 4], there will be no vacua when there are
no massless doublets; to get a (unique) vacuum, we need one massless
doublet.  As a result, this superpotential term tells us that we
must restrict to $\{ p^k, \tilde{p}^k \}$ such that $S^{ij}(p, \tilde{p})$ 
has rank two.

Putting this together, we see that this phase describes the intersection
of two determinantal varieties:  one defined by the locus
where $S^{ij}$ has rank $2$, the other defined by the locus where
$\tilde{S}^{ij}$ has rank $2$.  This is the same geometry we obtained
for this phase of the GLSM in the previous duality frame.

Note that in both phases of the GLSM in this duality frame,
geometry emerges through a combination of perturbative considerations
(analogues of the PAXY model \cite{Jockers:2012zr}) and
strong-coupling effects.

\subsubsection{Dualize both factors in complete intersection}

In this subsection we apply the duality of \cite{Hori:2011pd}[section 4.2]
to both of the $O_+(2)$ factors in the gauge group of the GLSM,
and analyze the phases of the resulting GLSM, to verify that they
produce the same geometries as the original GLSM.

Dualizing twice gives us a GLSM with gauge group
\begin{equation}
\frac{
SO(2) \times U(1) \times SO(2)
}{
{\mathbb Z}_2 \times {\mathbb Z}_2
},
\end{equation}
fields
\begin{center}
\begin{tabular}{c|ccrrrr}
& $\varphi^i$ & $\tilde{\varphi}^i$ & $p^k$ & $\tilde{p}^k$  & $m_{ij}$ &
$\tilde{m}_{ij}$ \\ \hline
$SO(2)$ & $\Box$ & ${\bf 1}$ & ${\bf 1}$ & ${\bf 1}$ & ${\bf 1}$ 
& ${\bf 1}$ \\
$U(1)$ & $-1$ & $-1$ & $-2$ &  $-2$ & $+2$  & $+2$ \\
$SO(2)$ & ${\bf 1}$ & $\Box$ & ${\bf 1}$ & ${\bf 1}$ & ${\bf 1}$
& ${\bf 1}$
\end{tabular}
\end{center}
and a superpotential
\begin{equation}
W \: = \: \sum_{ij} S^{ij}(p, \tilde{p}) \, m_{ij} \: + \:
\sum_{ij} \tilde{S}^{ij}(p,\tilde{p}) \, \tilde{m}_{ij}
\: + \:
 \sum_{ij} m_{ij} \, \varphi^i \cdot \varphi^j
\: + \:
\sum_{ij} \tilde{m}_{ij} \, \tilde{\varphi}^i \cdot \tilde{\varphi}^j.
\end{equation}

Next, we shall analyze the phases of this GLSM, and describe
how they reproduce the same geometries as we saw in the previous
subsection.

First, we consider the phase $r \gg 0$.
From the D-terms we see that $\{ m_{ij}, \tilde{m}_{ij} \}$ are
not all zero.
From the superpotential terms
\begin{displaymath}
\sum_{ij} m_{ij} \, \varphi^i \cdot \varphi^j
\: + \:
\sum_{ij} \tilde{m}_{ij} \, \tilde{\varphi}^i \cdot \tilde{\varphi}^j,
\end{displaymath}
which act as mass matrices for the doublets $\varphi^i$ and
$\tilde{\varphi}^i$, we see that there will only be vacua where
$m_{ij}$ and $\tilde{m}_{ij}$ each have rank 2.
(For rank 3, there are no massless $SO(2)$ doublets, hence
no vacua \cite{Hori:2011pd}[section 4.4]; for rank 2,
each $SO(2)$ has one massless doublet, which leads to one
vacuum \cite{Hori:2011pd}[section 4.5], 
working locally in a Born-Oppenheimer approximation over
the space of $\{ m_{ij}, \tilde{m}_{ij} \}$.)

The remaining superpotential terms
\begin{displaymath}
\sum_{ij} S^{ij}(p,\tilde{p}) \, m_{ij} \: + \:
\sum_{ij} \tilde{S}^{ij}(p,\tilde{p}) \, \tilde{m}_{ij}
\end{displaymath}
define a set of 6 hyperplanes in the space of $\{ m_{ij},
\tilde{m}_{ij} \}$.  
Now, we also know from the duality that
\begin{equation}
m_{ij} \: = \: x_i \cdot x_j,
\: \: \:
\tilde{m}_{ij} \: = \: \tilde{x}_i \cdot \tilde{x}_j,
\end{equation}
in terms of the doublets $x_i$, $\tilde{x}_i$ of the first duality frame
(which correlates with the earlier observation that one must restrict
to the rank two locus of the $m_{ij}$, $\tilde{m}_{ij}$).
As a result, the complete intersection described above necessarily
matches that we derived for the geometry of the $r \gg 0$ phase of
this GLSM in the first duality frame.

Next, we consider the phase $r \ll 0$.
In this phase, D-terms imply that $\{ \varphi^i, \tilde{\varphi}^i,
p^k, \tilde{p}^k \}$ are not all zero.
We can rewrite the superpotential in this phase as
\begin{equation}
W \: = \: \sum_{ij} m_{ij} \left( S^{ij}(p,\tilde{p}) \: + \:
\varphi^i \cdot \varphi^j \right)
\: + \:
\sum_{ij} \tilde{m}_{ij} \left( \tilde{S}^{ij}(p,\tilde{p})
\: + \: \tilde{\varphi}^i \cdot \tilde{\varphi}^j \right).
\end{equation}
This is essentially two copies of the PAXY model for
symmetric determinantal varieties \cite{Jockers:2012zr}[section 3.5].
This tells us that this phase is the intersection of 
the rank 2 locus of $S^{ij}(p,\tilde{p})$,
and the rank 2 locus of $\tilde{S}^{ij}(p,\tilde{p})$.
This is the same geometry we derived for
this phase of this GLSM in the first duality frame.  (There, the 
geometry was derived utilizing strong-coupling effects; here,
in this duality frame, the geometry arises perturbatively from the
critical locus of a superpotential.)

Thus, we find the same geometry for the $r \gg 0$ phase of this
GLSM in all duality frames, and the same geometry for the
$r \ll 0$ phase of this GLSM in all duality frames, as expected.
In one duality frame, the geometry is realized perturbatively
as the critical locus of a superpotential; in another, via strong
coupling effects; and in the third, via a combination of 
perturbative and nonperturbative considerations.

\subsection{Joins of gauge theories}

Much of this paper is devoted to describing GLSMs realizing 
joins of geometries.  However, it is worth observing that in principle,
especially in section~\ref{sect:ex:o+2}, we have also implicitly defined a 
notion of
a join of gauge theories.  Given one (two-dimensional, (2,2) supersymmetric)
gauge theory with gauge group $U(1) \times G$ and chiral superfields
$\phi$ in some representation $R$, 
and another with gauge group $U(1) \times \tilde{G}$ and chiral
superfields $\tilde{\phi}$ in some representation $\tilde{R}$,
we can define a gauge-theoretic analogue of a join,
which is now a $U(1)^3 \times G \times \tilde{G}$ gauge theory
with fields
\begin{center}
\begin{tabular}{c|ccrr}
& $\phi$ & $\tilde{\phi}$ & $z_1$ & $z_2$ \\ \hline
$G$ & $R_G$ & ${\bf 1}$ & ${\bf 1}$ & ${\bf 1}$ \\
$U(1)_1$ & $R_{U(1)}$ & $0$ & $-a$ & $0$ \\
$\tilde{G}$ & ${\bf 1}$ & $\tilde{R}_{\tilde{G}}$ & ${\bf 1}$ & ${\bf 1}$ \\
$U(1)_2$ & $0$ & $\tilde{R}_{U(1)}$ & $0$ & $-b$ \\
$U(1)_3$ & $0$ & $0$ & $1$ & $1$
\end{tabular}
\end{center}
where $a$, $b$ define the `embeddings' of the two gauge theories,
and in general $R_{U(1)}$, $\tilde{R}_{U(1)}$ will be vectors
of integers, with as many components as irreducible components of
the representations $R_G$, $\tilde{R}_{\tilde{G}}$.

We can `blowdown' the $z_i$ in the same pattern as for geometric cases.
Eliminating $z_1$ and $U(1)_3$, we get a $U(1)^2 \times G \times \tilde{G}$
gauge theory with fields
\begin{center}
\begin{tabular}{c|ccr}
& $\phi$ & $\tilde{\phi}$ &  $z_2$ \\ \hline
$G$ & $R_G$ & ${\bf 1}$ & ${\bf 1}$  \\
$U(1)_1$ & $R_{U(1)}$ & $0$ & $+a$ \\
$\tilde{G}$ & ${\bf 1}$ & $\tilde{R}_{\tilde{G}}$ & ${\bf 1}$  \\
$U(1)_2$ & $0$ & $\tilde{R}_{U(1)}$  & $-b$ 
\end{tabular}
\end{center}
Eliminating $z_2$ and $U(1)_2$, we have a $U(1) \times G \times \tilde{G}$
gauge theory with fields
\begin{center}
\begin{tabular}{c|ccr}
& $\phi$ & $\tilde{\phi}$ \\ \hline
$G$ & $R_G$ & ${\bf 1}$   \\
$U(1)_1$ & $R_{U(1)}$ & $(a/b) \tilde{R}_{U(1)}$  \\
$\tilde{G}$ & ${\bf 1}$ & $\tilde{R}_{\tilde{G}}$
\end{tabular}
\end{center}

We can slightly clean up the description of this last theory by rescaling
the $U(1)_1$ charges (glossing over potential subtleties involving
e.g. nonminimal charges as in
\cite{Pantev:2005rh,Pantev:2005zs,Hellerman:2006zs,Caldararu:2007tc}),
which leads us to
\begin{center}
\begin{tabular}{c|ccr}
& $\phi$ & $\tilde{\phi}$ \\ \hline
$G$ & $R_G$ & ${\bf 1}$   \\
$U(1)_1$ & $b R_{U(1)}$ & $a \tilde{R}_{U(1)}$  \\
$\tilde{G}$ & ${\bf 1}$ & $\tilde{R}_{\tilde{G}}$ 
\end{tabular}
\end{center}
This last $U(1) \times G \times \tilde{G}$ gauge theory is our
gauge-theoretic analogue of a classical join of two varieties.

In passing, note that if in each of the original gauge theories,
the sum of the $U(1)$ charges vanishes, then the same is true of the
classical join of the gauge theories.  In other words, if
\begin{equation}
\sum_i \left( R_{U(1)} \right)_i \: = \: 0 \: = \:
\sum_i \left( \tilde{R}_{U(1)} \right)_i,
\end{equation}
where the $i$ index counts the components of the vectors of $U(1)$ charges,
then the sum of the $U(1)$ charges in the classical join also vanishes.
This reflects the fact that the classical join of two Calabi-Yau's is another
Calabi-Yau.

We shall not use this notion of joins of gauge theories,
beyond the obvious application as a realization in GLSMs of
joins of geometries, but we thought it important to observe
that such a definition does exist.

\section{Multi-parameter examples and homological projective duality}
\label{sect:multiparameter}

In this section, we will give GLSMs whose phases realize the
various examples of homological projective duality discussed
in \cite{inoue}.  

The first set of examples of homological projective
duals we realize in GLSMs,
in section~\ref{sect:firstexs}, do not involve joins,
but will be used in subsequent join constructions.
Understanding these examples in GLSMs also turns out to be an
exercise in utilizing physical realizations of the
various embeddings described in our previous
work \cite{Caldararu:2017usq}.

In the next subsections, for each of the bundles ${\cal E}$ of 
subsection~\ref{sect:firstexs}, we will construct GLSMs describing complete
intersections of the form
\begin{displaymath}
J \times_{ {\mathbb P}( \wedge^2 V_5 \oplus V_N ) }
{\mathbb P} W
\end{displaymath}
in joins
\begin{displaymath}
J \: = \: {\rm Join}\left(G(2,5), {\mathbb P} {\cal E} \right),
\end{displaymath}
where $V_N$ is the vector space of sections of the relative hyperplane
class on ${\mathbb P} {\cal E}$.

After performing basic consistency tests (such as verifying that the
Calabi-Yau complete intersection behaves like a Calabi-Yau in GLSM
language), we will argue that the phases of each GLSM include a phase
describing a complete intersection of the form
\begin{displaymath}
J' \times_{ {\mathbb P} ( \wedge^2 V_5^* \oplus V_N^* ) }
{\mathbb P} W^{\perp},
\end{displaymath}
in joins 
\begin{displaymath}
J' \: = \: {\rm Join}\left( G(2,V_5^*), {\mathbb P} {\cal E}^{\perp} \right).
\end{displaymath}

In each case, we will also construct a dual GLSM, and verify the
same geometric interpretation in each phase.  This will be an
exercise in strong-coupling physics of $SU(2)$ gauge theories:
whenever a phase of one GLSM can be interpreted perturbatively,
understanding
the corresponding phase of the dual GLSM will require the 
methods of \cite{Hori:2006dk}.

\subsection{First examples: ${\mathbb P} {\cal E} \times_{ {\mathbb P} V }
{\mathbb P} W$  }
\label{sect:firstexs}

In this section we will describe GLSMs whose phases realize the first set of
examples of homological projective duality in \cite{inoue}.

Let ${\cal E}$ be a rank $r$ bundle on $Z$ satisfying the conditions in
\cite{inoue}[section 2.2], including that $\det {\cal E} \cong K_Z$.
Then, from \cite{inoue}[prop. 2.2], the intersection of a projective embedding
of
${\mathbb P} {\cal E}$ with $r$ hyperplanes should be
homologically projective dual to the intersection of
${\mathbb P} {\cal E}^{\perp}$ with a corresponding set of hyperplanes.

In each of the three examples discussed in
\cite{inoue}[section 2.2], we will construct a GLSM for
${\mathbb P} {\cal E} \times_{ {\mathbb P} V } {\mathbb P} W$,
and then observe that a different phase of the same GLSM describes
${\mathbb P} {\cal E}^{\perp} \times_{ {\mathbb P} V^*} {\mathbb P} W^{\perp}$.

\subsubsection{${\mathbb P}^2 \times {\mathbb P}^2$ }
\label{sect:first:ex1}

In this section, we will describe the GLSM for the first example,
and show explicitly that phases of the GLSM correspond to the
homologically-projective-dual varieties.  Before doing so, however,
we will quickly review the mathematical description of the
two spaces.

Consider the special case $Z = {\mathbb P}^2$,
${\cal E} = {\cal O}(-1)^3$,
$V = H^0(Z, {\cal E}^*)^* = {\mathbb C}^9$.
In this case, the map ${\mathbb P} {\cal E} = {\mathbb P}^2 \times
{\mathbb P}^2 \rightarrow {\mathbb P}^8$ proceeds via the Segre embedding:
\begin{displaymath}
[x_1, x_2, x_3] \times [y_1, y_2, y_3] \: \mapsto \: [ x_i y_j ].
\end{displaymath}
In the notation of \cite{inoue}[section 2.2], the vector subspace $W \subset V$
has codimension $r=3$, dimension $9-3=6$,
defined by three hyperplanes, and its
orthogonal complement $W^{\perp}$ has dimension $r=3$.
The intersection of three hyperplanes in ${\mathbb P} V = {\mathbb P}^8$
with the image of ${\mathbb P} {\cal E}$
can then be described as
\begin{equation}  \label{eq:p2xp2:rpos}
{\mathbb P} {\cal E} \times_{ {\mathbb P} V } {\mathbb P} W.
\end{equation}

Following the notation of \cite{inoue}[section 2.2],
the orthogonal bundle ${\cal E}^{\perp}$ is defined by
\begin{displaymath}
0 \: \longrightarrow \: {\cal E}^{\perp} \: \longrightarrow \:
{\cal O}^9 \: \stackrel{ x_{i} }{\longrightarrow} \: {\cal O}(+1)^3 \:
\longrightarrow \: 0,
\end{displaymath}
where the $(x_{\alpha})$ are understood as sections of ${\cal E}^*$.
Given that
\begin{displaymath}
0 \: \longrightarrow \: {\cal O} \: \longrightarrow \:
{\cal O}(1)^3 \: \longrightarrow \: T
\: \longrightarrow \: 0,
\end{displaymath}
one has
\begin{displaymath}
0 \: \longrightarrow \: \Omega^1(1) \: \longrightarrow \: 
{\cal O}^3 \: \longrightarrow \: {\cal O}(1) \: \longrightarrow \: 0,
\end{displaymath}
hence it is natural to suspect in this case that
${\cal E}^{\perp} = \oplus_3 \Omega^1(1)$.
The homologically projective dual space, which we will see arising
in another phase of the GLSM for the space~(\ref{eq:p2xp2:rpos}), is then
\begin{equation}
{\mathbb P} {\cal E}^{\perp} \times_{ {\mathbb P} V^* } {\mathbb P} W^{\perp}.
\end{equation}

Next, we shall construct the corresponding GLSM.
As discussed in \cite{Caldararu:2017usq},
the Segre embedding of ${\mathbb P}^2 \times {\mathbb P}^2$ can
be realized by a $U(1)^2$ gauge theory with matter:
\begin{itemize}
\item 3 chiral superfields $x_i$ of charge $(1,0)$,
\item 3 chiral superfields $y_j$ of charge $(0,1)$,
\item 9 chiral superfields $z_{ij}$ of charge $(1,1)$,
\item 9 chiral superfields $p^{ij}$ of charge $(-1,-1)$,
\end{itemize}
as summarized in the table below:
\begin{center}
\begin{tabular}{ccccr}
& $x_i$ & $y_j$ & $z_{ij}$ & $p^{ij}$ \\ \hline
$U(1)$ & $1$ & $0$ & $1$ & $-1$ \\
$U(1)$ & $0$ & $1$ & $1$ & $-1$
\end{tabular}
\end{center}
together with a superpotential
\begin{equation}
W \: = \: \sum_{ij} p^{ij}\left( z_{ij} - x_i y_j \right).
\end{equation}

To describe a Calabi-Yau complete intersection of 3 hyperplanes
in the image of the Segre embedding, we add
\begin{itemize}
\item 3 chiral superfields $q_m$ of charge $(-1,-1)$,
\end{itemize}
and a term to the superpotential, so that the complete superpotential
now reads
\begin{equation}
W \: = \: \sum_{ij} p^{ij}\left( z_{ij} - x_i y_j \right)
\: + \: \sum_m q_m G_m(z),
\end{equation}
where the hyperplanes are defined by $\{ G_m = 0 \}$.
The sums of the $U(1)$ charges of the fields can easily be checked
to vanish, so this is a Calabi-Yau.

This gives the GLSM realization of ${\mathbb P} {\cal E}
\times_{ {\mathbb P} V } {\mathbb P}W$,
in the notation of \cite{inoue}.

Now, we will show that this same GLSM contains another phase
which describes 
\begin{displaymath}
{\mathbb P} {\cal E}^{\perp} \times_{ {\mathbb P} V^*}
{\mathbb P} W^{\perp},
\end{displaymath}
in the notation of \cite{inoue}.

\begin{figure}
\begin{center}
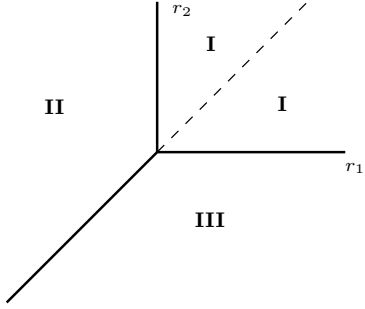
\end{center}
\caption{Phase diagram.}\label{fig-classicalp2p2}
\end{figure}

The $U(1)$ charges imply that there are four phases,
as illustrated in figure~\ref{fig-classicalp2p2}.
The D-terms are
\begin{eqnarray}
  |x_i|^2+|z_{ij}|^2-|p^{ij}|^2-|q_m|^2&=&r_1,\nonumber \\
  |y_j|^2+|z_{ij}|^2-|p^{ij}|^2-|q_m|^2&=&r_2.
\end{eqnarray}
In particular we note that
\begin{equation}
  |x_i|^2-|y_j|^2=r_1-r_2.
  \end{equation}
So, depending on whether $r_1>r_2$ or $r_1<r_2$, 
not all of the $x_i$ or $y_i$ are allowed to vanish, respectively. This implies in particular that there should be a phase boundary, spanned by the charges of $z_{ij}$ dividing the quadrant $r_{1,2}>0$. Let us consider the F-terms:
\begin{eqnarray}
  G_m(z)&=&0,\nonumber\\
  (z_{ij}-x_iy_j)&=&0,\nonumber\\
  p^{ij}+q_m\frac{\partial G}{\partial z_{ij}}&=&0,\nonumber\\
  p^{ij}x_i&=&0,\nonumber\\
   p^{ij}y_j&=&0.
\end{eqnarray}

Phase I corresponds to the geometry we constructed the GLSM to describe,
a complete intersection of hyperplanes in ${\mathbb P}^2 \times 
{\mathbb P}^2$ which is denoted ${\mathbb P} {\cal E}
\times_{ {\mathbb P} V } {\mathbb P}W$.
In this phase, where $r_1 \gg 0$ and $r_2 \gg 0$, neither all the $x_i$ nor
all the $y_i$ can vanish, as the second F-term would imply that $z_{ij}=0$
for generic hyperplanes,
which is forbidden by the D-terms.  The $x_i$ and $y_i$ then act as
homogeneous coordinates on either ${\mathbb P}^2$ factor.

As a further result, since for generic hyperplanes the superpotential 
prohibits the vanishing of all the $x_i$ and all the $y_i$, 
there is no phase boundary in the first quadrant along the line $r_1 = r_2$,
unlike on the ambient space.  This phase boundary of the ambient space
is lifted by the F terms.  Such liftings of phase boundaries occur commonly
in multi-parameter GLSMs, and we will see this again in later examples in
this paper.

Next we turn to phase III.  
In this phase, $r_2 \ll 0$ and $r_1 \gg r_2$.  The D terms then imply that
the $\{ p^{ij}, q_m \}$ are not all zero,
and also that the $x_i$ are not all zero.
In this phase, we 
interpret $p^{ij}$ as coordinates on
${\mathbb P}^8$, and $z_{ij}$ as Lagrange multipliers,
then the superpotential becomes
\begin{equation}
W \: = \: \sum_{ij} z_{ij} \left( p^{ij} +  \sum_m q_m G_m^{ij} \right)
\: - \: \sum_{ij} p^{ij} x_i y_j,
\end{equation}
where
\begin{displaymath}
G_m(z) \: = \: \sum_{ij} G_m^{ij} z_{ij}.
\end{displaymath}

We can interpret the second term of the superpotential,
\begin{displaymath} 
\sum_{ij} p^{ij} x_i y_j,
\end{displaymath}
as encoding the fact that in this phase, the $p^{ij}$ couple to
${\cal E}^{\perp}$.  To this end, note that there are nine $p^{ij}$,
and we can interpret the $y$'s as (analogues of) Lagrange multipliers
enforcing the condition that the $p^{ij}$ are in the kernel of the map
defined by the $x_i$.  (It is important for this interpretation that the
$x_i$ cannot all vanish, which is why this arises in phase III.)
If we label the two $U(1)$ gauge symmetries by
$(\lambda,\mu)$, then note that in the linear combination
$\lambda \mu^{-1}$, the $p^{ij}$ have zero charge, the $x_i$ have
charge $1$, and the $y_j$ have charge $-1$, consistent with an interpretation
of the $p^{ij}$ in terms of the bundle ${\cal O}^9$.

We can interpret the first terms of the superpotential,
\begin{displaymath}
 \sum_{ij} z_{ij} \left( p^{ij} +  \sum_m q_m G_m^{ij} \right),
\end{displaymath}
as saying that the $p^{ij}$ intersect $W^{\perp}$.  In the other phase,
the subspace $W \subset V$ was specified by the hyperplanes $\{ G_m(z) = 0\}$.
In this phase, the $q_m$ act as parameters on the complementary
hypersurface $W^{\perp}$:  the codimension of $W$ matches the dimension
of $W^{\perp}$.  In these terms, the $z_{ij}$ act as (analogues of)
Lagrange mulipliers, forcing the $p^{ij}$ to match the
(images of the) $q_m$.

Putting this together, we can interpret this second phase as the
geometry
\begin{displaymath}
{\mathbb P} {\cal E}^{\perp}
\times_{ {\mathbb P} V^* } {\mathbb P}W^{\perp},
\end{displaymath}
in the notation of \cite{inoue}.

Phase II has a nearly identical interpretation to phase III.
In phase II, one essentially flips the interpretations of the $x_i$ and
$y_j$.  The resulting geometry has the same description as in phase III.

\subsubsection{ Bl$_{\rm pt} {\mathbb P}^3$ }
\label{sect:first:ex2}

Next, consider the case that $Z = {\mathbb P}^2$,
${\cal E} = {\cal O}(-2) \oplus {\cal O}(-1)$,
$V = H^0( Z, {\cal E}^*)^* = {\mathbb C}^9$.
In this case,
${\mathbb P} {\cal E} = {\rm Bl}_{\rm pt} {\mathbb P}^3$
(as will be more clear from the GLSM description).

We can write a GLSM for ${\mathbb P} {\cal E}$ in terms of
chiral fields $x_1$, $x_2$, $x_3$, $y_1$, $y_2$,
charged under $U(1)^2$, as follows:
\begin{center}
\begin{tabular}{c|cccrr}
$U(1)$ & $x_1$ & $x_2$ & $x_3$ & $y_1$ & $y_2$ \\ \hline
$\lambda$ & $1$ & $1$ & $1$ & $-2$ & $-1$ \\
$\mu$ & $0$ & $0$ & $0$ & $1$ & $1$
\end{tabular}
\end{center}
which explicitly realizes ${\mathbb P} \left( {\cal O}(1) \oplus
{\cal O} \right)$.
After replacing $\lambda$ with $\lambda \mu^2$, we have the
equivalent description
\begin{center}
\begin{tabular}{c|cccrr}
$U(1)$ & $x_1$ & $x_2$ & $x_3$ & $y_1$ & $y_2$ \\ \hline
$\lambda \mu^2$ & $1$ & $1$ & $1$ & $0$ & $1$ \\
$\mu$ & $0$ & $0$ & $0$ & $1$ & $1$
\end{tabular}
\end{center}
which makes it clear that ${\mathbb P} {\cal E}$ is the blowup of
${\mathbb P}^3$ at a point.

The embedding ${\mathbb P} {\cal E} \rightarrow {\mathbb P} V$,
defined implicitly by $H^0(Z, {\cal E}^*)$, is realized as
\begin{equation}
[ x_1, x_2, x_3, y_1, y_2 ] \: \mapsto \:
[ y_1 x_1^2, y_1 x_2^2, y_1 x_3^2, y_1 x_1 x_2, y_1 x_1 x_3, y_1 x_2 x_3,
y_2 x_1, y_2 x_2, y_2 x_3 ],
\end{equation}
where all of the image coordinates have charge
$(\lambda, \mu) = (0,1)$.
For simplicity, define
\begin{equation}
\left(f_a(x,y) \right) \: = \: \left( y_1 x_1^2, y_1 x_2^2, y_1 x_3^2, y_1 x_1 x_2, y_1 x_1 x_3, y_1 x_2 x_3,
y_2 x_1, y_2 x_2, y_2 x_3 \right),
\end{equation}
e.g. $f_1(x,y) = y_1 x_1^2$.
Then, a GLSM for the embedding \cite{Caldararu:2017usq}
is described by adding a set of nine
pairs of fields $p^a$, $z_a$, with charges as below
\begin{center}
\begin{tabular}{c|cccrrrr}
$U(1)$ & $x_1$ & $x_2$ & $x_3$ & $y_1$ & $y_2$ & $z_a$ & $p^a$\\ \hline
$\lambda$ & $1$ & $1$ & $1$ & $-2$ & $-1$ & $0$ & $0$ \\
$\mu$ & $0$ & $0$ & $0$ & $1$ & $1$ & $1$ & $-1$
\end{tabular}
\end{center}
with superpotential
\begin{equation}
W \: = \: \sum_{a=1}^9 p^a \left( z_a - f_a(x,y) \right).
\end{equation}

\begin{figure}
\begin{center}
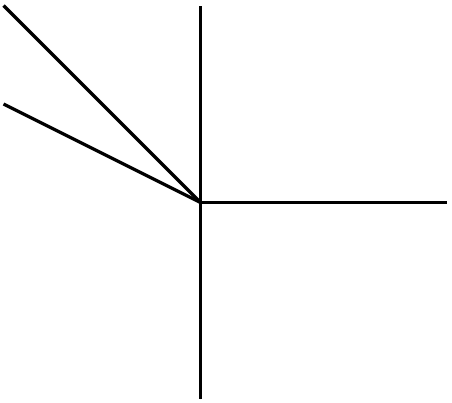
\end{center}
\caption{Phase diagram.}\label{fig-classicalp3}
\end{figure}

We can build a Calabi-Yau complete intersection in this
theory as a complete intersection of two hyperplanes of
degree $(\lambda,\mu) = (0,1)$, i.e., linear in $z_a$.
Let $G_m(z)$ denote the two hyperplanes, then a GLSM for
this intersection is defined by the fields
\begin{center}
\begin{tabular}{c|cccrrrrr|c}
$U(1)$ & $x_1$ & $x_2$ & $x_3$ & $y_1$ & $y_2$ & $z_a$ & $p^a$ & $q_m$&FI\\ \hline
$\lambda$ & $1$ & $1$ & $1$ & $-2$ & $-1$ & $0$ & $0$ & $0$&$r_{\lambda}$ \\
$\mu$ & $0$ & $0$ & $0$ & $1$ & $1$ & $1$ & $-1$ & $-1$&$r_{\mu}$
\end{tabular}
\end{center}
with superpotential
\begin{equation}
W \: = \: \sum_{a=1}^9 p^a \left( z_a - f_a(x,y) \right)
\: + \:
\sum_{m=1}^2 q_m G_m(z).
\end{equation}
This is the GLSM for
${\mathbb P} {\cal E} \times_{ {\mathbb P} V } {\mathbb P} W$. Let us study the phases in more detail. The charges of the fields indicate five phases as depicted in figure \ref{fig-classicalp3}. 
The D-terms are
\begin{eqnarray}
  |x_{\alpha}|^2-2|y_1|^2-|y_2|^2&=&r_{\lambda},\nonumber\\
  |y_1|^2+|y_2|^2+|z_a|^2-|p^a|^2-|q_m|^2&=&r_{\mu}.
\end{eqnarray}
To distinguish between phases II, III, and IV is is convenient to rewrite this as
\begin{eqnarray}
  |x_{\alpha}|^2-|y_1|^2+|z_a|^2-|p^a|^2-|q_m|^2&=&r_{\lambda}+r_{\mu},\nonumber\\
  |x_{\alpha}|^2+|y_2|^2+2|z_a|^2-2|p^a|^2-|q_m|^2&=&r_{\lambda}+2r_{\mu}.
  \end{eqnarray}
The F-terms are
\begin{eqnarray}
  z_a-f_a(x,y)&=&0,\nonumber\\
  G_m&=&0,\nonumber\\
  p^a+q_m\frac{\partial G_m}{\partial z_a}&=&0,\nonumber\\
  p^a\frac{\partial f_a}{\partial y_i}&=&0,\nonumber\\
  p^a\frac{\partial f_a}{\partial x_{\alpha}}&=&0.
\end{eqnarray}

In phase I, where $r_{\lambda} \gg 0$ and $r_{\mu} \gg 0$,
we have that the $x_{\alpha}$ are not all zero, and $\{ y_1, y_2, z_a \}$
are not all zero.
From the F-terms, in this phase we recover the complete intersection ${\mathbb P} {\cal E} \times_{ {\mathbb P} V } {\mathbb P} W$.

In phase II we have $r_{\lambda} \ll 0,$ $r_{\mu} \gg 0,$
$r_{\lambda}+r_{\mu} \gg 0$. This implies that $y_{1,2}$ are not allowed to 
vanish simultaneously, and also the $x_{\alpha}$ and $z_a$.

Phase III is characterized by $r_{\lambda} \ll 0,$
$r_{\mu} \gg 0,$
$r_{\lambda}+r_{\mu} \ll 0,$
$r_{\lambda}+2r_{\mu} \gg 0$. The D-terms imply, among other things,
that $\{ x_{\alpha}, y_2, z_a \}$ are not allowed to vanish simultaneously, 
and neither are $\{ y_1, p^a, q_m \}$ or $\{ y_1, y_2 \}$.

In phase IV we have $r_{\lambda} \ll 0$ and
$r_{\lambda}+2r_{\mu} \ll 0$.  In this phase,
 $\{ p^a, q_m \}$ and $\{ y_{1,2} \}$ are not allowed to vanish 
simultaneously.

In phase V with $r_{\lambda} \gg 0,$
$r_{\mu} \ll 0$, the D-terms imply that $\{ x_{\alpha} \}$
are not allowed to vanish simultaneously, and
$\{ p^a, q_m\}$ are not allowed to vanish simultaneously.
We will argue that this phase, phase V, describes
${\mathbb P} {\cal E}^{\perp} \times_{ {\mathbb P} V^* }
{\mathbb P} W^{\perp}$.

Write
\begin{displaymath}
G_m(z) \: = \: \sum_{a=1}^9 G_m^a z_a,
\: \: \:
f_a(x,y) \: = \: \sum_{\beta} y_{\beta} f_a^{\beta}(x)
\end{displaymath}
where for example the $f_a^{\beta}(x)$ are sections of ${\cal E}^*$,
then we can rewrite the superpotential as
\begin{equation}
W \: = \: \sum_a z_a \left( p^a + \sum_m q_m G_m^a \right)
\: - \: \sum_{a, \beta} y_{\beta} p^a f_a^{\beta}(x) .
\end{equation}

We can interpret the first term as saying that the $p$'s lie along
the hyperplanes defined by the $q_m$ -- in other words, that we are
intersecting with ${\mathbb P} W^{\perp}$, as before, with the
$z_a$ acting as analogues of Lagrange multipliers.
(Note that this requires that the $\{ p^a, q_m \}$ not all vanish.)

We can interpret the second term as saying that the $p$'s live in the
kernel of the map defined by the $f_a^{\beta}(x)$, with $y$'s acting
analogously to Lagrange multipliers.  The kernel of that map
is precisely ${\cal E}^{\perp}$ in the notation of \cite{inoue}[equ'n (2.6)]:
\begin{equation}
0 \: \longrightarrow \: {\cal E}^{\perp} \: \longrightarrow \:
H^0(Z, {\cal E}^*) \otimes {\cal O}_Z \: \longrightarrow \:
{\cal E}^* \: \longrightarrow \: 0,
\end{equation}
where the map to ${\cal E}^*$ is defined by the sections of ${\cal E}^*$.

The reader should 
note that this interpretation requires that the $\{x_{\alpha} \}$ not
all vanish.  Combined with the requirement that not all the
$\{p^a, q_m\}$ vanish, we see that only phase V can be described
in this fashion. 

Putting this together, we see that this phase (V) describes the
geometry
\begin{equation}
{\mathbb P} {\cal E}^{\perp} \times_{ {\mathbb P} V^* } 
{\mathbb P} W^{\perp},
\end{equation}
the homological projective dual of the first phase,
${\mathbb P} {\cal E} \times_{ {\mathbb P} V} {\mathbb P} W$,
as expected for phases of a GLSM.

\subsubsection{${\mathbb P}^1 \times {\mathbb P}^1 \times {\mathbb P}^1$}
\label{sect:first:ex3}

Next, consider the case that $Z = {\mathbb P}^1 \times {\mathbb P}^1$,
${\cal E} = {\cal O}(-1,-1)^{\oplus 2}$.

We can write a GLSM for ${\mathbb P} {\cal E}$ in terms of
chiral fields $x_1$, $x_2$, $y_1$, $y_2$, $w_1$, $w_2$, as
\begin{center}
\begin{tabular}{c|ccccrr}
$U(1)$ & $x_1$ & $x_2$ & $y_1$ & $y_2$ & $w_1$ & $w_2$ \\ \hline
$\lambda$ & $1$ & $1$ & $0$ & $0$ & $-1$ & $-1$ \\
$\mu$ & $0$ & $0$ & $1$ & $1$ & $-1$ & $-1$ \\
$\nu$ & $0$ & $0$ & $0$ & $0$ & $1$ & $1$
\end{tabular}
\end{center}
Picking a different basis for the $U(1)$ charges, this can be described as
\begin{center}
\begin{tabular}{c|ccccrr}
$U(1)$ & $x_1$ & $x_2$ & $y_1$ & $y_2$ & $w_1$ & $w_2$ \\ \hline
$\lambda \nu$ & $1$ & $1$ & $0$ & $0$ & $0$ & $0$ \\
$\mu \nu$ & $0$ & $0$ & $1$ & $1$ & $0$ & $0$ \\
$\nu$ & $0$ & $0$ & $0$ & $0$ & $1$ & $1$
\end{tabular}
\end{center}
which makes it clear that this is ${\mathbb P}^1 \times {\mathbb P}^1
\times {\mathbb P}^1$.

The embedding ${\mathbb P} {\cal E} \rightarrow {\mathbb P} V$,
$V \cong {\mathbb C}^8$,
defined implicitly by $H^0(Z,{\cal E}^*)$, is realized as
\begin{equation}
[x_1, x_2, y_1, y_2, w_1, w_2] \: \mapsto \:
[ x_i y_j w_k ],
\end{equation}
for all $i, j, k \in \{1, 2\}$.    Note that each of the
image coordinates has charge $(\lambda, \mu, \nu) = (0,0,1)$.

We can realize that embedding in GLSMs \cite{Caldararu:2017usq}
by adding fields $z_{ijk}$, $p^{ijk}$,
with charges
\begin{center} 
\begin{tabular}{c|ccccrrrr}
$U(1)$ & $x_1$ & $x_2$ & $y_1$ & $y_2$ & $w_1$ & $w_2$ & $z_{ijk}$ & $p^{ijk}$ \\ \hline
$\lambda$ & $1$ & $1$ & $0$ & $0$ & $-1$ & $-1$ & $0$ & $0$ \\
$\mu$ & $0$ & $0$ & $1$ & $1$ & $-1$ & $-1$ & $0$ & $0$ \\ 
$\nu$ & $0$ & $0$ & $0$ & $0$ & $1$ & $1$ & $1$ & $-1$
\end{tabular}
\end{center}
with superpotential
\begin{equation}
W \: = \: \sum_{ijk} p^{ijk} \left( z_{ijk} - x_i y_j w_k \right).
\end{equation}

Finally, we can construct a Calabi-Yau as a complete intersection of
the image of ${\mathbb P} {\cal E}$ with
a pair of hyperplanes in ${\mathbb P}V$.  
We can describe that complete intersection
with the GLSM with fields and charges as below
\begin{center}
\begin{tabular}{c|ccccrrrrr|c}
$U(1)$ & $x_1$ & $x_2$ & $y_1$ & $y_2$ & $w_1$ & $w_2$ & $z_{ijk}$ & $p^{ijk}$ & $q_m$&FI \\ \hline
$\lambda$ & $1$ & $1$ & $0$ & $0$ & $-1$ & $-1$ & $0$ & $0$ & $0$ & $r_{\lambda}$\\
$\mu$ & $0$ & $0$ & $1$ & $1$ & $-1$ & $-1$ & $0$ & $0$ & $0$ & $r_{\mu}$\\
$\nu$ & $0$ & $0$ & $0$ & $0$ & $1$ & $1$ & $1$ & $-1$ & $-1$ & $r_{\nu}$
\end{tabular}
\end{center}
with superpotential
\begin{equation}
W \: = \: \sum_{ijk} p^{ijk} \left( z_{ijk} - x_i y_j w_k \right)
\: + \:
\sum_{m=1}^2 q_m G_m(z).
\end{equation}

With increasing number of FI parameters,
 the phase structure becomes increasingly complicated. 
For this model we will content ourselves with identifying the geometric phase and its HPD dual. 
In the notation of \cite{inoue}, and by construction of this GLSM,
we expect one phase to be
\begin{displaymath}
{\mathbb P} {\cal E} \times_{ {\mathbb P} V } {\mathbb P} W.
\end{displaymath}
To recover this phase in the moduli space of the GLSM,
 it is convenient to recombine the $U(1)$ charges as follows
\begin{center}
\begin{tabular}{c|ccccrrrrr|c}
$U(1)$ & $x_1$ & $x_2$ & $y_1$ & $y_2$ & $w_1$ & $w_2$ & $z_{ijk}$ & $p^{ijk}$ & $q_m$&FI \\ \hline
$\lambda\nu$ & $1$ & $1$ & $0$ & $0$ & $0$ & $0$ & $1$ & $-1$ & $-1$ & $r_{\lambda}+r_{\nu}$\\
$\mu\nu$ & $0$ & $0$ & $1$ & $1$ & $0$ & $0$ & $1$ & $-1$ & $-1$ & $r_{\mu}+r_{\nu}$\\
$\nu$ & $0$ & $0$ & $0$ & $0$ & $1$ & $1$ & $1$ & $-1$ & $-1$ & $r_{\nu}$
\end{tabular}
\end{center}
We immediately see that in a region inside $r_i\gg 0$,
the solutions of the D-term and F-term equations lead to the expected geometry.

There is also a phase that can be interpreted as a nonlinear sigma model on the Calabi-Yau
\begin{equation}
{\mathbb P} {\cal E}^{\perp} \times_{ {\mathbb P} V^* } 
{\mathbb P} W^{\perp}.
\end{equation}
To identify this phase, it is useful to rewrite the superpotential in the form
\begin{equation}
W \: = \: \sum_{ijk} z_{ijk} \left( p^{ijk} + \sum_{m=1}^2 q_m G_m^{ijk}
\right) \: - \: \sum_{ijk} p^{ijk} x_i y_j w_k,
\end{equation}
where
\begin{equation}
G_m(z) \: = \: \sum_{ijk} G_m^{ijk} z_{ijk}.
\end{equation}
We can interpret the first terms in the superpotential as saying that
the $p_{ijk}$ will lie along $W^{\perp}$, parametrized by the $q_m$,
as before, with the $z_{ijk}$ acting as analogues of Lagrange
multipliers.  Similarly, the second term can be interpreted as the
statement that the $p^{ijk}$ are in the kernel of the map
defined by the sections $x_i y_j$ of ${\cal E}^*$, with the $w_k$
acting as analogues of Lagrange multipliers.  This means that the
$p^{ijk}$ couple to ${\cal E}^{\perp}$, defined by
\begin{equation}
0 \: \longrightarrow \: {\cal E}^{\perp} \: \longrightarrow \:
H^0(Z, {\cal E}^*) \otimes {\cal O}_Z \: \longrightarrow \:
{\cal E}^* \: \longrightarrow \: 0,
\end{equation}
where the map to ${\cal E}^*$ is defined by the sections of ${\cal E}^*$.

In order for this phase to be realized we have to demand that $\{ p^{ijk}, q_m \}$ do not vanish simultaneously, and for the ${\mathbb P}^1 \times
{\mathbb P}^1$ base to exist in the Higgs branch, we also need the $\{ x_i \}$
and, separately, the $\{ y_i \}$ to not simultaneously vanish.  
On the other hand, the $w_i$ and $z_{ijk}$ are allowed to vanish.
As a result, this geometry appears in a phase in the region
\begin{equation}
r_{\nu} \ll 0, \: \: \:
r_{\mu} \gg 0, \: \: \:
r_{\lambda} \gg 0.
\end{equation}

\subsection{ Join$(G(2,5), {\mathbb P}^2 \times {\mathbb P}^2)$ }

\subsubsection{First duality frame}

In this section we will describe the GLSM for the
resolved join of $G(2,5)$ and
${\mathbb P}^2 \times {\mathbb P}^2$, the first example
in \cite{inoue}[section 3.2].

First, recall we can describe $G(2,5)$ as a $U(2)$ gauge theory with
five chiral multiplets $\phi^i_a$ in the fundamental representation
($i \in \{1, \cdots, 5\}$, $a \in \{1, 2\}$).  We will describe
${\mathbb P}^2$ with a $U(1)^2$ gauge theory and a pair of sets of
fields $x_{\alpha}$, $y_{\beta}$, $\alpha, \beta \in \{1, 2, 3\}$,
corresponding to homogeneous coordinates
on either factor.  The ${\mathbb P}^1$ bundle will be described by
homogeneous coordinates $z_1, z_2$.  The GLSM for the resolved join
is then a $U(2) \times U(1)^3$ gauge theory with matter
\begin{center}
\begin{tabular}{c|ccccr}
 & $\phi^i_a$ & $x_{\alpha}$ & $y_{\beta}$ & $z_1$ & $z_2$ \\ \hline
$U(2)$ & $\Box$ & ${\bf 1}$ & ${\bf 1}$ & det$^{-1}$ & ${\bf 1}$ \\
$U(1)_{\lambda}$ & $0$ & $1$ & $0$ & $0$ & $-1$ \\
$U(1)_{\mu}$ & $0$ & $0$ & $1$ & $0$ & $-1$ \\
$U(1)_{\nu}$ & $0$ & $0$ & $0$ & $1$ & $1$
\end{tabular}
\end{center}

The line bundle $L$ in \cite{inoue}[section 3] is then defined by
$U(1)^3$ charges $(0,0,1)$.  It is straightforward to compute that
sections of $L$ are of the form
\begin{displaymath}
\epsilon^{ab} \phi^i_a \phi^j_b z_1, \: \: \:
x_{\alpha} y_{\beta} z_2
\end{displaymath}
These sections are in one-to-one correspondence with elements of
\begin{displaymath}
H^0(G(2,5), {\cal O}(1)) \oplus H^0( {\mathbb P}^2 \times {\mathbb P}^2,
{\cal O}(1,1) ),
\end{displaymath}
precisely as expected for $L$ for $G(2,5)$ in the Pl\"ucker embedding
and ${\mathbb P}^2 \times {\mathbb P}^2$ in the Segre embedding.
These sections also define the projective embedding of the
classical join of $G(2,5)$ and ${\mathbb P}^2 \times {\mathbb P}^2$
(with these embeddings).

Now, on the face of it, for the GLSM we have constructed so far,
the Calabi-Yau condition requires a (collection of) hypersurfaces
of total charge $(2,2,2)$ under the $U(1)^3$, and degree four under
$\det U(2)$.  However, this is different from the Calabi-Yau condition
described in \cite{inoue}.  The essential difference is that the
Calabi-Yau condition described in \cite{inoue} uses relations between
divisors which only exist on the Calabi-Yau, and do not extend to the
ambient space, whereas the Calabi-Yau condition above is one inherited
from the ambient space.

We saw related matters previously in section~\ref{sect:g25:2:cyci}
in describing Calabi-Yau spaces inside other joins,
and we also again refer the reader to 
appendix~\ref{app:nonambientcy} for a discussion of
Calabi-Yau conditions
that utilize relations along the intersection that are not inherited from
the ambient space, as well as GLSM constructions to make the
pertinent Calabi-Yau condition more clear.  In the present case,
briefly, we can rewrite the GLSM (for generic intersections) so as to
make the pertinent Calabi-Yau condition more clear, 
by blowing down the resolved join to a classical join.

We will do this by rewriting the GLSM,
by eliminating $z_{1,2}$ and a pair of $U(1)$ gauge
symmetries.
Consider a linear combination of the $U(1)$ charges:
\begin{center}
\begin{tabular}{c|ccrcr}
 & $\phi^i_a$ & $x_{\alpha}$ & $y_{\beta}$ & $z_1$ & $z_2$ \\ \hline
$U(2)$ & $\Box$ & ${\bf 1}$ & ${\bf 1}$ & det$^{-1}$ & ${\bf 1}$ \\
$U(1)_{\lambda\mu}$ & $0$ & $1$ & $1$ & $0$ & $-2$ \\
$U(1)_{\lambda \mu^{-1}}$ & $0$ & $1$ & $-1$ & $0$ & $0$ \\
$U(1)_{\nu}$ & $0$ & $0$ & $0$ & $1$ & $1$
\end{tabular}
\end{center}
In terms of these linear combinations, we see that, schematically,
$D_{\nu} \equiv 2 D_{\lambda \mu}$, and $D_{\nu} \equiv D_{\rm det}$.

Before blowing down, let us write the data above in terms of
$SU(2) \times U(1)$ representations.  In other words, mechanically
we use the fact that
\begin{equation}
U(2) \: = \: \frac{
SU(2) \times U(1)_{\rm det} 
}{
{\mathbb Z}_2
},
\end{equation}
and then write the defining representations in terms of $SU(2) \times
U(1)_{\rm det}$ instead of $U(2)$.  In this language, the data above becomes
\begin{center}
\begin{tabular}{c|ccrrr}
 & $\phi^i_a$ & $x_{\alpha}$ & $y_{\beta}$ & $z_1$ & $z_2$ \\ \hline
$SU(2)$ & $\Box$ & ${\bf 1}$ & ${\bf 1}$ & ${\bf 1}$ & ${\bf 1}$ \\
$U(1)_{\rm det}$ & $1$ & $0$ & $0$ & $-2$ & $0$ \\
$U(1)_{\lambda\mu}$ & $0$ & $1$ & $1$ & $0$ & $-2$ \\
$U(1)_{\lambda \mu^{-1}}$ & $0$ & $1$ & $-1$ & $0$ & $0$ \\
$U(1)_{\nu}$ & $0$ & $0$ & $0$ & $1$ & $1$
\end{tabular}
\end{center}

If we blowdown the divisor $\{z_1=0\}$, eliminating $U(1)_{\nu}$, we get
\begin{center}
\begin{tabular}{c|ccrrr}
 & $\phi^i_a$ & $x_{\alpha}$ & $y_{\beta}$ &  $z_2$ \\ \hline
$SU(2)$ & $\Box$ & ${\bf 1}$ & ${\bf 1}$ &  ${\bf 1}$ \\
$U(1)_{\rm det}$ & $1$ & $0$ & $0$ & $+2$ \\
$U(1)_{\lambda\mu}$ & $0$ & $1$ & $1$ &  $-2$ \\
$U(1)_{\lambda \mu^{-1}}$ & $0$ & $1$ & $-1$ &  $0$ 
\end{tabular}
\end{center}

Blowing down $\{z_2=0\}$, or equivalently,
eliminating $z_1$, $z_2$, $U(1)_{\lambda \mu}$, and
$U(1)_{\nu}$ from the original set, the GLSM field and gauge content reduces to
\begin{center}
\begin{tabular}{c|crr}
&  $\phi^i_a$ & $x_{\alpha}$ & $y_{\beta}$ \\ \hline
$SU(2)$ & $\Box$ & ${\bf 1}$ & ${\bf 1}$ \\
$U(1)_{\rm det}$ & $1$ & $1$ & $1$ \\
$U(1)_{\lambda \mu^{-1}}$ & $0$ & $1$ & $-1$
\end{tabular}
\end{center}

Now, technically, the matter fields above are inconsistent with gauge group
$U(2) \times U(1)_{\lambda \mu^{-1}}$: the $x$ and $y$ fields are only
in well-defined representations of $SU(2) \times U(1)^2$.  In particular,
because they are invariant under $SU(2)$ but of charge $1$ under
$U(1)_{\rm det}$, they do not descend to $U(2)$ representations.
To fix this problem, we will take the gauge group of the GLSM for the
classical join to be $SU(2) \times U(1)^2$ instead of $U(2) \times U(1)$.

The sum of the field charges under $U(1)_{\lambda \mu^{-1}}$ vanishes,
so the Calabi-Yau condition in this new GLSM
is now completely determined by $U(1)_{\rm det}$.
Following standard procedures, we see that for a Calabi-Yau complete
intersection, the sum of the degrees of the hypersurfaces under
$U(1)_{\rm det}$ must be $5(2) + (3) + (3) = 16$, and hyperplanes would be
linear in the field combinations
\begin{displaymath}
\epsilon^{ab} \phi^i_a \phi^j_b, \: \: \:
x_{\alpha} y_{\beta},
\end{displaymath}
which both transform with charge $2$ under $U(1)_{\rm det}$.
An intersection of $8$ hyperplanes should therefore be Calabi-Yau, which
precisely duplicates the Calabi-Yau condition for this join
stated in \cite{inoue}.
Furthermore, for generic hyperplanes in the field combinations above,
the complete intersection will not intersect the singularities of the
join, much as we saw previously in discussions of e.g. $G(2,5) \cap
G(2,5)$.

To realize a complete intersection of eight hyperplanes,
we add eight fields $q_m$, $m \in \{1, \cdots, 8\}$, as
\begin{center}
\begin{tabular}{c|ccrr|c} 
&  $\phi^i_a$ & $x_{\alpha}$ & $y_{\beta}$ & $q_m$ & FI\\ \hline
$SU(2)$ & $\Box$ & ${\bf 1}$ & ${\bf 1}$ & ${\bf 1}$&-\\
$U(1)_{\rm det}$ & $1$ & $1$ & $1$ & $-2$&$r_1$ \\
$U(1)_{\lambda \mu^{-1}}$ & $0$ & $1$ & $-1$ & $0$&$r_2$
\end{tabular}
\end{center}
along with a superpotential
\begin{equation}
W \: = \: \sum_m q_m G_m\left( \epsilon^{ab} \phi^i_a \phi^j_b,
x_{\alpha} y_{\beta} \right),
\end{equation}
where the $G_m$'s are linear in the field combinations above,
i.e.
\begin{equation}
  G_{m}=\sum_{ij}a_{ij,m}\epsilon^{ab} \phi^i_a \phi^j_b+\sum_{\alpha,\beta}b^{\alpha\beta,m}x_{\alpha}y_{\beta}.
\end{equation}
(Mathematically, for generic hyperplanes mixing 
\begin{displaymath}
\epsilon^{ab} \phi^i_a \phi^j_b, \: \: \:
x_{\alpha} y_{\beta},
\end{displaymath}
we expect the hyperplanes will not intersect the singularities in the
ambient join.)
The D-term equations are
\begin{eqnarray}
  \phi\phi^{\dagger}&=&\frac{1}{2}|\phi^i_a|^2,\nonumber\\
  |\phi^i_a|^2+|x_{\alpha}|^2+|y_{\alpha}|^2-2|q_m|^2&=&r_1 \: \equiv \:
r_{\rm det},\nonumber\\
  |x_{\alpha}|^2-|y_{\beta}|^2&=&r_2 \: \equiv \: r_{\lambda \mu^{-1}}.
\end{eqnarray}
The F-term equations are
\begin{eqnarray}
  G_m(\epsilon^{ab} \phi^i_a \phi^j_b,x_{\alpha}y_{\beta})&=&0,\nonumber\\
  q_ma_{ij,m}\epsilon^{ab}\phi_j^b&=&0,\nonumber\\
  q_mb^{\alpha\beta,m}x_{\alpha}&=&0,\nonumber\\
  q_mb^{\alpha\beta,m}y_{\beta}&=&0.
\end{eqnarray}

The gauge charges encode the classical phase diagram that one obtains by solving the D-term equations. 
\begin{figure}
\begin{center}
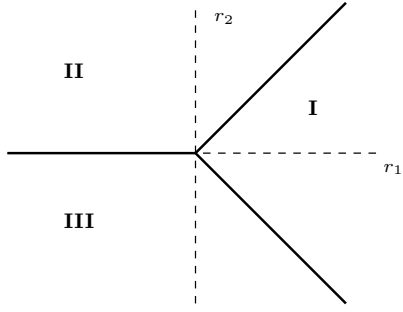
\end{center}
\caption{Phase diagram.}\label{fig-classical1}
\end{figure}
However, we assume that we have generic hypersurfaces, which do not
intersect the singularities of the join, and so in particular does not 
intersect a region where either all the $x_{\alpha} = 0$ or 
$y_{\beta} = 0$, and so the phase boundary along the positive $r_1$
axis is lifted by the superpotential.  (Indeed,
at any of this loci, the F-terms would describe a complete intersection of codimension $8$ in $G(2,5)$ which would imply $\phi^i=0$ which is forbidden by the $SU(2)$ D-term.) This lifts the phase boundary along $r_1>0$. 
The remaining boundaries are along
$r_1 = \pm r_2$, for $r_1 > 0$, and along the negative $r_1$ axis. The classical phase diagram is depicted in figure \ref{fig-classical1}. 

First we consider phase I, where $r_1 > |r_2| \gg 0$.
If $r_2 \gg 0$, then one D-term equation implies that not all
$x_{\alpha}$ can vanish, and in addition,
\begin{equation}
 |\phi_i^a|^2+2|y_{\alpha}|^2-2|q_m|^2
\: = \: r_1 - r_2,
\end{equation}
so we see that the $\phi$ and $y_{\beta}$ also cannot all vanish.
Similarly, if $r_2 \ll 0$, then not all the $y_{\beta}$ can vanish,
and in addition,
\begin{equation}
 |\phi_i^a|^2+2|x_{\alpha}|^2-2|q_m|^2
\: = \: r_1 + r_2,
\end{equation}
so we see that the $\phi$ and $x_{\alpha}$ also cannot all vanish.
This is, in any event, the geometric phase we built the GLSM to
describe, namely
\begin{displaymath}
J \times_{ {\mathbb P} ( \wedge^2 V_5 \oplus V_N ) } {\mathbb P} W,
\end{displaymath}
in the notation of \cite{inoue}, a complete intersection in the join
\begin{displaymath}
J \: = \: {\rm Join}\left( G(2,5), {\mathbb P} {\cal E} \right)
\: = \: {\rm Join}\left( G(2,5), {\mathbb P}^2 \times {\mathbb P}^2 \right).
\end{displaymath}

Now, let us turn our attention to phase II, in which $r_2 \gg 0$
but $r_1 < r_2$.
We will argue that this phase describes the geometry
\begin{displaymath}
J' \times_{ {\mathbb P} ( \wedge^2 V_5^* \oplus V_N^* ) } 
{\mathbb P} W^{\perp},
\end{displaymath}
a Calabi-Yau complete intersection in the join
\begin{displaymath}
J' \: = \: {\rm Join} \left( G(2,V_5^*), {\mathbb P} {\cal E}^{\perp} \right).
\end{displaymath}

In this phase, the D terms imply that not all the $x_{\alpha}$
vanish and that not all the $q_m$ vanish.
As a result, the $\{ q_m \}$ act like homogeneous coordinates on
${\mathbb P}^7$.
The superpotential can be rewritten in the form
\begin{equation}
W \: = \: \sum_{ij} \epsilon^{ab} \phi^i_a \phi^j_b \left(
\sum_m q_m G_{m, ij} \right) \: + \:
\sum_{\alpha, \beta} x_{\alpha} y_{\beta} \left(
\sum_m q_m G_m^{\alpha \beta} \right),
\end{equation}
where
\begin{equation}
G_m \: = \: G_{m, ij}  \epsilon^{ab} \phi^i_a \phi^j_b
\: + \:
G_m^{\alpha \beta} x_{\alpha} y_{\beta}.
\end{equation}

The second term in the superpotential,
\begin{equation}
\sum_{\alpha, \beta} x_{\alpha} y_{\beta} \left(
\sum_m q_m G_m^{\alpha \beta} \right),
\end{equation}
provides three constraints on the $q_m$.  Specifically, much as in
section~\ref{sect:first:ex1}, we can interpret the $y_{\beta}$ as analogues
of Lagrange multipliers, requiring
\begin{displaymath}
\sum_m q_m G_m^{\alpha \beta}
\end{displaymath}
to be in the kernel of the map defined by the $x_{\alpha}$,
and so couple to ${\cal E}^{\perp}$.

The first term in the superpotential is a mass matrix for the
$\phi^i_a$.  We can interpret
\begin{equation}
A_{ij} \: \equiv \: \sum_m q_m G_{m, ij}
\end{equation}
as an antisymmetric $5 \times 5$ matrix, encoding the masses of
the $\phi$ fields.  Our analysis of this term then closely follows
the analysis of the R{\o}dland example in \cite{Hori:2006dk}.
Since antisymmetric matrices have even rank, its rank must be one of
$\{ 4, 2, 0 \}$.  Over loci where it has rank 4, there is only one
massless doublet of $SU(2)$, so there are no vacua, following
\cite{Hori:2006dk}.  Over loci where it has rank 2, there are three
doublets, which correspond to a single vacuum, following
\cite{Hori:2006dk}.  Thus, due to this mass matrix, we are effectively
restricting to loci in the space of $q_m$ (${\mathbb P}^7$) where the
matrix $(A_{ij})$ has rank 2.  There are eight $q_m$, but, the low-energy
limit of the second superpotential term imposes three constraints,
so overall the $q_m$ represent five degrees of freedom.  
The locus over which $(A_{ij})$ has rank two, over a five-dimensional space,
is the space denoted $Pf(5)$ in \cite{Donagi:2007hi}[section 4.2.2],
which is (the dual of) the Grassmannian $G(2,5)$, which following
\cite{inoue} we will denote $G(2,V_5^*)$, where $V_5$ is a five-dimensional
vector space.

Letting $J'$ denote Join$(G(2,5), {\mathbb P} {\cal E}^{\perp})$,
since the $q_m$ parametrize $W^{\perp}$,
altogether the GLSM is describing in this phase
\begin{displaymath}
J' \times_{ {\mathbb P}( \wedge^2 V_5^* \oplus V_N^* ) } {\mathbb P} W^{\perp},
\end{displaymath}
where $V_{N'}^*$ denotes the vector space with monomial basis
$\{x_{\alpha} y_{\beta} \}$.

Phase III is very similar to phase II.  Here, since $r_2 \ll 0$,
the $y_{\beta}$ do not all vanish.  The analysis is largely identical
to that of phase II, but exchanging $x_{\alpha}$ with $y_{\beta}$.
In particular, the $y_{\beta}$ now serve as the base space of the
bundle ${\cal E}$.  The final resulting geometry is otherwise the
same as that of phase II. This is in agreement with \cite{inoue} where a discussion of the mirror reveals three large complex structure points, two of which correspond to the same geometry.

We now complete our analysis by discussing 
the Coulomb and mixed branches of the GLSM. The effective potential on the Coulomb branch is given by (\ref{weff}).
In our case we get
\begin{eqnarray}
  \mathcal{W}_{eff}&=&-t_1\sigma_1-t_2\sigma_2
-5\left(\sigma_1+\sigma_0\right)\left[\ln\left(\sigma_1+\sigma_0\right)-1\right]-5\left(\sigma_1-\sigma_0\right)\left[\ln\left(\sigma_1-\sigma_0\right)-1\right]
\nonumber\\
  &&-3\left(\sigma_1+\sigma_2\right)\left[\ln\left(\sigma_1+\sigma_2\right)-1\right]-3\left(\sigma_1-\sigma_2\right)\left[\ln\left(\sigma_1-\sigma_2\right)-1\right]
\nonumber\\
  &&-8\left(-2\sigma_1\right)\left[\ln\left(-2\sigma_1\right)-1\right].
\end{eqnarray}
The critical locus is at
\begin{eqnarray}
  e^{-t_1}&=&\frac{1}{2^{16}}\frac{(\sigma_1+\sigma_0)^5(\sigma_1-\sigma_0)^5(\sigma_1+\sigma_2)^3(\sigma_1-\sigma_2)^3}{\sigma_1^{16}},\nonumber\\
  e^{-t_2}&=&\frac{(\sigma_1+\sigma_2)^3}{(\sigma_1-\sigma_2)^3},\qquad \frac{(\sigma_1+\sigma_0)^5}{(\sigma_1-\sigma_0)^5}=1.
  \end{eqnarray}
Defining $z=\sigma_0/\sigma_1$ and $w=\sigma_2/\sigma_1$ this can be written as
\begin{eqnarray}
  e^{-t_1}&=&\frac{1}{2^{16}}(1+z)^5(1-z^5)(1+w)^3(1-w)^3,\nonumber\\
  e^{-t_2}&=&\frac{(1+w)^3}{(1-w)^3},\qquad \frac{(1+z)^5}{(1-z)^5}=1.
  \end{eqnarray}
The last equation can be solved explicitly and inserted into the remaining two equations which then only depend on $w$. Furthermore we note
\begin{equation}
e^{-t_1-t_2}=(1\pm z)^{10}(1+w)^6,
\qquad e^{-t_1+t_2}=(1\pm z)^{10}(1-w)^6.
  \end{equation}
Specific values of $w$ correspond to limiting regions which determine the legs of the amoeba that maps out the ``quantum'' phase diagram. At $w=1$ we recover the phase boundary at $r_1=-r_2>0$. To see this, we observe that at this locus
\begin{equation}
  e^{-t_1}=0,
\quad e^{-t_2}\rightarrow\infty,
\quad e^{-t_1-t_2}=\mathrm{const.},
\quad e^{-t_1+t_2}=0. 
\end{equation}
This implies that $r_1\rightarrow\infty$ and $r_2\rightarrow -\infty$ while $r_1\sim -r_2$.  Similarly, $w=-1$ corresponds to $r_1=r_2>0$. Indeed,
in this case
\begin{equation}
  e^{-t_1}=0,
\quad e^{-t_2}\rightarrow 0,
\quad e^{-t_1-t_2}=0,
\quad e^{-t_1+t_2}=\mathrm{const}. 
\end{equation}
Hence, we get $r_1\sim r_2\rightarrow\infty$. Finally, $w\rightarrow\infty$ gives the phase boundary at $r_1<0,r_2=0$, due to
\begin{equation}
  e^{t_1}\rightarrow\infty,
\quad e^{-t_2}\rightarrow \mathrm{const.},
\quad e^{-t_1-t_2}\rightarrow\infty,
\quad e^{-t_1+t_2}\rightarrow\infty. 
  \end{equation}
Let also show that the phase boundary at $r_1>0$ is lifted also at the quantum level. The extra boundary must be encoded in a mixed branch, where only a subgroup of the maximal torus of $G$ is preserved.  In more complicated examples it is non-trivial to see which combination of $U(1)$s gives a non-trivial contribution to the discriminant. In our case, the additional contribution hides in the case where $U(1)_{\lambda\mu^{-1}}$ is unbroken and $\sigma_2$ and all the fields that are not charged under $U(1)_{\lambda\mu^{-1}}$ are large. Following \cite{Morrison:1994fr} (see also \cite{Hori:2016txh} for the notation we are using), we write $\sigma_2\equiv\sigma_L$ for the $\sigma$-fields that take large values in the unbroken maximal torus of the gauge group. Next, we divide the matter fields into $(\dot{\phi},\hat{\phi})$, where $\hat{\phi}$ receive mass by $\sigma_L$ and the $\dot{\phi}$ do not. Also the remaining $\sigma$ fields are divided into $(\dot{\sigma},\hat{\sigma})$, depending on whether they receive mass by $\sigma_L$ or not. The massive fields can be integrated out. The low-energy theory has a scalar potential
\begin{equation}
  U_{eff}=\frac{1}{2}\left(|\dot{Q}(\dot{\sigma})\dot{\phi}|^2+|\dot{Q}(\dot{\bar{\sigma}})\dot{\phi}|^2\right)+\frac{e_{eff}^2}{2}\left(\mu_{eff}(\dot{\phi})-r_{eff}\right)^2+|dW(\dot{\phi})|^2,
  \end{equation}
where $W(\dot{\phi})$ means restriction to $\dot{\phi}$ and $r_{eff}(\dot{\sigma},\sigma_L)$ is the real part of $t_{eff}=-d\mathcal{W}_{eff}$, i.e.
\begin{equation}
  t_{eff,i}=t_i+\hat{Q_i}\ln\hat{Q}(\sigma),
  \end{equation}
where $\hat{Q}$ denotes the gauge charges of the massive fields $\hat{\phi}_i$.

\begin{figure}
  \begin{center}
    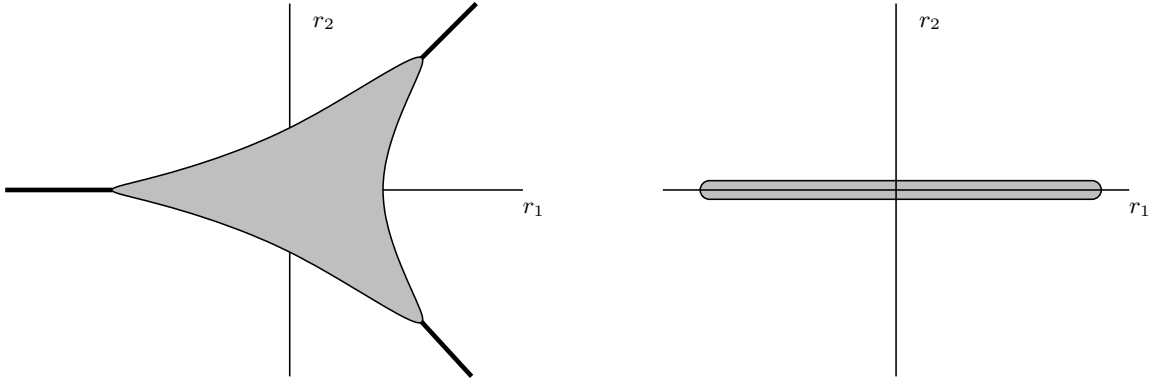
    \end{center}\caption{Components of the discriminant. The mixed branch (right) gets lifted by the superpotential.}\label{fig-amoeba1}
\end{figure}

In this particular example we have $\dot{\phi}=\{\phi^i_a,q_m\}$ and $\hat{\phi}=\{x_{\alpha},y_{\beta}\}$. Furthermore we have $\dot{\sigma}={\sigma_0,\sigma_2\equiv\sigma_{L}}$ and $\hat{\sigma}=\{\sigma_1\}$. We find
\begin{eqnarray}
  U_{eff}&=&\frac{1}{2}\left(|(\sigma_0+\sigma_1)\phi^i_1|^2+|(-\sigma_0+\sigma_1)\phi^i_2|^2+|-2\sigma_1 q_m|^2+\left(\sigma_i\rightarrow \bar{\sigma}_i\right)\right)\nonumber\\
  &&+\frac{1}{2}\sum_{a,b}e^2_{eff,a,b}(\mu_{eff,a}-r_{eff,a})(\mu_{eff,b}-r_{eff,b})+|dW(\phi^i_a,q_m)|^2.
\end{eqnarray}
The effective D-terms are
\begin{eqnarray}
  \mu_{eff,0}&=&\phi\phi^{\dagger}-\frac{1}{2}|\phi^i_a|^2,\nonumber\\
  \mu_{eff,1}&=&|\phi^i_a|^2-2|q_m|^2,\nonumber\\
   \mu_{eff,2}&=&0.
\end{eqnarray}
The effective couplings are
\begin{eqnarray}
  t_{eff,1}&=&t_1+3\ln(\sigma_1+\sigma_2)+3\ln(\sigma_1-\sigma_2),\nonumber\\
  t_{eff,2}&=&t_2+3\ln(\sigma_1+\sigma_2)-3\ln(\sigma_1-\sigma_2).
\end{eqnarray}
The minimum is at $\sigma_0=\sigma_1=0$.  The values for $r_1$ remain unconstrained whereas $r_2=0$ and $\theta_2=\pi$. This would imply that the mixed branch accounts for the phase boundaries along the $r_1$-axis. The situation is depicted in figure \ref{fig-amoeba1}.
However, in order to obtain $U_{eff}=0$ one also has to satisfy $dW(\dot{\phi})=0$. This in particular implies that $q_m=0$ which is in contradiction with the D-term $\mu_{eff,1}=t_{eff,1}$, since the left-hand side is positive definite while the right-hand side is not.

\subsubsection{Analysis of dual gauge theory}
\label{sect:joins:ex1:dual}

Now, let us consider a dual of this GLSM, dualizing the $SU(2)$
factor in the gauge group.  We will re-examine the geometric
interpretations and come to the same conclusions as in the
previous section, albeit with the twist that geometries that
arose via perturbative physics in the previous duality
frame will here arise via the strong-coupling physics of
\cite{Hori:2006dk}, and conversely.

From \cite{Hori:2011pd}[section 5.6],
for an odd number $N \geq k+3$
massless fundamentals, there is a duality between an $Sp(k)$ gauge theory
with $N$ fundamentals $\phi^i$, and an $Sp(N-k-1)$ gauge theory with
$N$ fundamentals $\varphi_i$, $(1/2) N (N-1)$ singlets $b^{ij} = -
b^{ji}$ (dual to the baryons of the original theory), with a superpotential
\begin{displaymath}
W \: = \: \sum_{ij} b^{ij} [ \varphi_i \varphi_j] \: = \:
\sum_{ij} \sum_{ab} b^{ij} \varphi_i^a \varphi_j^b J_{ab},
\end{displaymath}
where $J$ is the $(N-k-1) \times (N-k-1)$ symplectic form
\begin{displaymath}
J \: = \: \left[ \begin{array}{cc}
0 & 1 \\
-1 & 0 \end{array} \right].
\end{displaymath}
In the present case, we have an $SU(2) = Sp(2)$ gauge theory with five
fundamentals, so $k=2$, $N=5$, and $N-k-1 = 2$, so the dual gauge
theory is another $SU(2)$ gauge theory with five fundamentals
$\varphi_i$, plus ten $SU(2)$ singlets $b^{ij}$
and superpotential given above.  Also, for $SU(2)$, we can
identify $J_{ab} = \epsilon_{ab}$.

The dual theory can therefore be described by the following fields:
\begin{center}
\begin{tabular}{c|crrrr}
& $\varphi_i^a$ & $b^{ij}$ & $x_{\alpha}$ & $y_{\beta}$ & $q_m$ \\ \hline
$SU(2)$ & $\Box$ & ${\bf 1}$ & ${\bf 1}$ & ${\bf 1}$ & ${\bf 1}$ \\
$U(1)_1$ & $-1$ & $+2$ & $1$ & $1$ & $-2$ \\
$U(1)_2$ & $0$ & $0$ & $1$ & $-1$ & $0$
\end{tabular}
\end{center}
with the superpotential
\begin{equation}
W \: = \: \sum_m q_m G_m\left( b^{ij}, x_{\alpha} y_{\beta} \right)
\: + \:
\sum_{ij} \sum_{ab} b^{ij} \varphi_i^a \varphi_j^b \epsilon_{ab},
\end{equation}
where the $b^{ij}$ of the dual theory are related to the
$\phi^i_a$ of the original theory by
\begin{equation}
b^{ij} \: = \: \epsilon^{ab} \phi^i_a \phi^j_b,
\end{equation}
and for $SU(2)$, we have identified $J_{ab} = \epsilon_{ab}$.
We identify $U(1)_1$ with $\det U(2)$ of the original gauge theory.
The $U(1)_{1,2}$ charges of $\varphi_i^a$ were determined from
self-consistency of the last term in the superpotential.

First, let us consider phase I, in which $r_1 > |r_2| \gg 0$.
In this page, the $\{ b^{ij}, x_{\alpha}, y_{\beta} \}$ are not\footnote{
Strictly speaking, if $r_2 \gg 0$, then the $x_{\alpha}$ cannot
all vanish and $\{ b^{ij}, y_{\beta} \}$ cannot vanish,
and if $r_2 \ll 0$, then the $y_{\beta}$ cannot all vanish and
$\{ b^{ij}, x_{\alpha} \}$ cannot vanish.
} all zero. Since the superpotential forbids the locus $x=y=0$, there is no phase boundary at $r_1>0$, in agreement with the original model.

The first term of the superpotential,
\begin{equation}
 \sum_m q_m G_m\left( b^{ij}, x_{\alpha} y_{\beta} \right),
\end{equation}
restricts us to the locus where all the hyperplanes $G_m$ vanish,
as is typical for GLSMs for complete intersections.

Understanding the second term of the superpotential in this
phase in this duality frame will require us to use strong-coupling
physics from \cite{Hori:2006dk}.
Working in a Bohr-Oppenheimer approximation,
we interpret the second term of the superpotential,
\begin{equation}
\sum_{ij} \sum_{ab} b^{ij} \varphi_i^a \varphi_j^b \epsilon_{ab},
\end{equation}
in terms of a mass matrix for the $\varphi$ over the space of $b^{ij}$.
Our analysis then follows that of the R{\o}dland example in \cite{Hori:2006dk}.
This mass matrix, $(b^{ij})$, is (trivially) linear in the $b^{ij}$.
As it is an antisymmetric $5 \times 5$ matrix, it can have rank 4, 2, or 0.
If it has rank 4, then there is one massless $SU(2)$ doublet,
and hence no supersymmetric vacua, following \cite{Hori:2006dk}.
If it has rank 2, then there are three massless $SU(2)$ doublets,
and there is one supersymmetric vacuum, following \cite{Hori:2006dk}.

Now, one description of the Grassmannian $G(2,5)$ 
\cite{Donagi:2007hi}[section 4.2.2] is as the subset of
the space of
$(b^{ij})$ such that the skew-symmetric matrix $(b^{ij})$ has 
rank $2$.  Thus, we see that the second term of the superpotential
is restricting the $\{b^{ij}\}$ to a copy of $G(2,5)$.

The interpretation of the first term is now straightforward:
it is describing a complete intersection inside the product
of $G(2,5)$ and the space of $x_{\alpha} y_{\beta}$.
As a result, 
this phase of the GLSM should
be interpreted, in the notation of \cite{inoue}, as
\begin{displaymath}
J \times_{ {\mathbb P}\left( \wedge^2 V_5 \oplus V_N \right) }
{\mathbb P} W,
\end{displaymath}
where $J$ is the join
\begin{displaymath}
J \: = \: {\rm Join}\left( G(2,5), {\mathbb P}^2 \times {\mathbb P}^2 \right),
\end{displaymath}
as expected, matching the geometry of the corresponding phase of the
dual gauge theory.

Now, let us turn to phase II, where $r_2 \gg 0$ and $r_1 < r_2$.
From the D terms, not all of the $x_{\alpha}$ can vanish, and
separately, not all of 
the $q_m$ and
$\varphi_i^a$ can vanish.  If we expand
\begin{equation}
G_m\left( b^{ij}, x_{\alpha} y_{\beta} \right)
\: = \:
\sum_{ij} G_{m, ij} b^{ij} \: + \:
\sum_{\alpha, \beta} G_m^{\alpha \beta} x_{\alpha} y_{\beta},
\end{equation}
then we can write the superpotential as
\begin{equation}
W \: = \:
\sum_{ij} b^{ij} \left( \sum_m q_m G_{m,ij} \: + \: \sum_{ab}
\epsilon_{ab} \varphi_i^a \varphi_j^b \right) \: + \:
\sum_{\alpha, \beta} \sum_m q_m G_m^{\alpha \beta} x_{\alpha} y^{\beta}.
\end{equation}

From the second term in the superpotential,
\begin{equation}
\sum_{\alpha, \beta} \sum_m q_m G_m^{\alpha \beta} x_{\alpha} y^{\beta},
\end{equation}
we see that the $q_m$ couple to ${\cal E}^{\perp}$.
If $r_2 \gg 0$, then we can interpret the $y_{\beta}$ as analogues of
Lagrange multipliers, which force the linear combinations
\begin{displaymath}
\sum_m q_m G_m^{\alpha \beta} 
\end{displaymath}
to lie in the kernel of the map defined by $x_{\alpha}$,
hence couple to ${\cal E}^{\perp}$.  (If $r_2 \ll 0$, the same remarks
apply after swapping the interpretation of $x_{\alpha}$ and $y_{\beta}$.)

Similarly, the first term in the superpotential,
\begin{equation}
\sum_{ij} b^{ij} \left( \sum_m q_m G_{m,ij} \: + \: \sum_{ab}
\epsilon_{ab} \varphi_i^a \varphi_j^b \right),
\end{equation}
identifies the linear combinations
\begin{displaymath}
\sum_m q_m G_{m,ij}
\end{displaymath}
with the image of the Pl\"ucker embedding of $G(2,5)$.  Since this model
was obtained by dualizing, relative to the first duality frame we can
identify this $G(2,5)$ with $G(2,V_5^*)$, the `dual' Grassmannian to
that appearing earlier.

Altogether, this means that this phase of the GLSM can be identified with
\begin{displaymath}
J' \times_{ {\mathbb P}\left( \wedge^2 V_5^* \oplus V_N^* \right) }
{\mathbb P} W^{\perp},
\end{displaymath}
in the notation of \cite{inoue},
where $J'$ is the join
\begin{displaymath}
J' \: = \: {\rm Join}\left(G(2,V_5^*), {\mathbb P} {\cal E}^{\perp}\right).
\end{displaymath}

As before, the analysis of phase III is nearly identical to the analysis
of phase II, merely swapping the roles of $x_{\alpha}$ and $y_{\beta}$.
The geometric interpretation is the same.

\subsection{ Join$(G(2,5), {\rm Bl}_{\rm pt} {\mathbb P}^3)$ }

In this section we will build a GLSM description of the second of
the examples of joins discussed in \cite{inoue},
namely
\begin{displaymath}
{\rm Join}\left(G(2,5), {\rm Bl}_{\rm pt} {\mathbb P}^3\right)
\end{displaymath}
and Calabi-Yau complete intersections therein.

\subsubsection{First duality frame}

Proceeding as before, we can write down the GLSM for the resolved join
immediately, as a ${\mathbb P}^1$ bundle over the product.
This GLSM is a $U(2) \times U(1)^3$ gauge theory with fields as follows:
\begin{center}
\begin{tabular}{c|ccrrcr}
 & $\phi^i_a$ & $x_{\alpha}$ & $y_1$ & $y_2$ & $z_1$ & $z_2$ \\ \hline
$U(2)$ & $\Box$ & ${\bf 1}$ & ${\bf 1}$ & ${\bf 1}$ &
det$^{-1}$ & ${\bf 1}$ \\
$U(1)_{\lambda}$ & $0$ & $1$ & $-2$ & $-1$ & $0$ & $0$ \\
$U(1)_{\mu}$ & $0$ & $0$ & $1$ & $1$ & $0$ & $-1$ \\
$U(1)_{\nu}$ & $0$ & $0$ & $0$ & $0$ & $1$ & $1$
\end{tabular}
\end{center}
where $\alpha \in \{1, 2, 3\}$, $i\in,\{1,\ldots,5\}$, and $a\in\{1,2\}$.

The projective embedding of Bl$_{\rm pt} {\mathbb P}^3$ is 
defined by a line bundle $L$ whose sections have $U(1)^2$
charges $(\lambda,\mu) = (0,1)$, as discussed in section~\ref{sect:first:ex2},
which defines the divisor $H_2$ of \cite{inoue} (by the
same charge assignments) and the $U(1)$ charges
of $z_2$.

We can perform a blowdown of the divisor $\{z_1=0\}$,
eliminating $z_1$ and $U(1)_{\nu}$, to get
\begin{center}
\begin{tabular}{c|ccrrr}
 & $\phi^i_a$ & $x_{\alpha}$ & $y_1$ & $y_2$ &  $z_2$ \\ \hline
$U(2)$ & $\Box$ & ${\bf 1}$ & ${\bf 1}$ & ${\bf 1}$ &
 det \\
$U(1)_{\lambda}$ & $0$ & $1$ & $-2$ & $-1$ & $0$ \\
$U(1)_{\mu}$ & $0$ & $0$ & $1$ & $1$ &  $-1$ \\
\end{tabular} 
\end{center} 

If we next blowdown the divisor $\{z_2=0\}$, eliminating $U(1)_{\mu}$, we get
\begin{center}
\begin{tabular}{c|ccrr}
 & $\phi^i_a$ & $x_{\alpha}$ & $y_1$ & $y_2$  \\ \hline
$U(2)$ & $\Box$ & ${\bf 1}$ & det & det \\  
$U(1)_{\lambda}$ & $0$ & $1$ & $-2$ & $-1$ 
\end{tabular}  
\end{center} 

The sum of charges under $U(1)_{\lambda}$ vanishes, so the Calabi-Yau condition
in this new GLSM is completely determined by det $U(2)$.  Following standard
procedures, we see that for a Calabi-Yau complete intersection, the sum
of the degrees of the hypersurfaces under det $U(2)$ must be $5 + 1 + 1 = 7$,
with hyperplanes linear in the field combinations
\begin{displaymath}
\epsilon^{ab} \phi^i_a \phi^j_b, \: \: \: 
x_{\alpha_1} x_{\alpha_2} y_1, \: \: \:
x_{\alpha} y_2,
\end{displaymath}
all of which transform with charge $1$ under det $U(2)$.
This precisely duplicates the Calabi-Yau condition for the join
stated in \cite{inoue}.  Furthermore, for generic hyperplanes, 
the Calabi-Yau should not intersect the singularities of the join.

Putting this together, we can describe a complete intersection of seven
hyperplanes in the join above as a GLSM with fields
\begin{center}
\begin{tabular}{c|ccrrc|c}
 & $\phi^i_a$ & $x_{\alpha}$ & $y_1$ & $y_2$ & $q_m$&FI \\ \hline
$U(2)$ & $\Box$ & ${\bf 1}$ & det & det & det$^{-1}$&$r$ \\
$U(1)_{\lambda}$ & $0$ & $1$ & $-2$ & $-1$ & $0$&$r_3$
\end{tabular}  
\end{center}
with superpotential
\begin{equation}  \label{eq:join:ex2:sup}
W \: = \: \sum_m q_m G_m\left( \epsilon^{ab} \phi^i_a \phi^j_b,
x_{\alpha_1} x_{\alpha_2} y_1, x_{\alpha} y_2 \right),
\end{equation}
where $m \in \{1, \cdots, 7\}$ and
\begin{equation}
  G_m=\sum_{ij}a_{ij,m}\epsilon^{ab} \phi^i_a \phi^j_b+\sum_{\alpha_1\alpha_2}b_{\alpha_1\alpha_2,m}x_{\alpha_1}x_{\alpha_2}y_1+\sum_{\alpha}c_{\alpha,m}x_{\alpha}y_2.
\end{equation}
The D-term equations are
\begin{eqnarray}
  \sum_{\alpha}|x_{\alpha}|^2-2|y_1|^2-|y_2|^2&=&r_3, \nonumber\\
  \phi\phi^{\dagger}+|y_1|^2{\bf 1}+|y_2|^2{\bf 1}-\sum_{m}|q_m|^2{\bf 1}&=&r{\bf 1}.
\end{eqnarray}
The F-term equations are
\begin{eqnarray}
  G_m(\epsilon^{ab}\phi_a^i\phi_b^j,x_{\alpha_1}x_{\alpha_2}y_1,x_{\alpha}y_2)&=&0,\nonumber\\
  q_ma_{ij,m}\epsilon^{ab}\phi^j_b&=&0,\nonumber\\
  q_m(b_{\alpha_1\alpha_2,m}x_{\alpha_2}y_1+c_{\alpha_1,m}y_2)&=&0,\nonumber\\
  q_mb_{\alpha_1\alpha_2,m}x_{\alpha_1}x_{\alpha_2}&=&0,\nonumber\\
  q_mc_{\alpha,m}x_{\alpha}&=&0.
\end{eqnarray}

\begin{figure}
  \begin{center}
    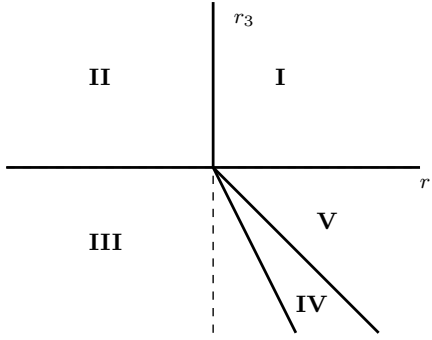
    \end{center}\caption{Phase diagram.}\label{fig-classical2}
\end{figure}

The gauge charges of the fields suggest a phase diagram with five phases as depicted in figure \ref{fig-classical2}. 

The phase boundaries are along the
positive $r$ and $r_3$ axes, along the negative $r$ axis, and along the
lines $r_3 = -2r$ and $r_3=-r$ in the fourth quadrant.
We will not try to systematically describe every
phase
here, but will instead outline highlights so that we can focus on the phases
pertinent to our physical realization of homological projective duality.

First we consider phase I, where $r \gg 0$ and $r_3 \gg 0$, the first
quadrant of the phase diagram.  Since $r_3 \gg 0$, the $x_{\alpha}$
are not all zero, and since $r \gg 0$, $\phi$ and the $y_i$ cannot all
vanish.
This phase describes
the space we constructed the
GLSM to describe, namely
\begin{displaymath}
J \times_{ {\mathbb P}\left( \wedge^2 V_5 \oplus V_N \right) } {\mathbb P} W
\end{displaymath}
in the join
\begin{displaymath}
J \: = \: {\rm Join}\left( G(2,5), {\mathbb P} {\cal E} \right) \: = \:
{\rm Join}\left( G(2,5), {\rm Bl}_{\rm pt} {\mathbb P}^3 \right),
\end{displaymath}
in the notation of \cite{inoue}.
The observant reader will note that for generic hypersurfaces, which
we assume, this will
not intersect the rank $0, 1$ loci of $\phi$.

Next, consider phase II, where $r \ll 0$ and $r_3 \gg 0$.  Since
$r_3 \gg 0$, the $x_{\alpha}$ are not all zero, and since
$r \ll 0$, the $q_m$ are not all zero.

In this phase, it will be useful to rewrite the
superpotential~(\ref{eq:join:ex2:sup}) in the form
\begin{equation}
W \: = \: \sum_{ij} \sum_m q_m G_{m,ij} \epsilon^{ab} \phi^i_a \phi^j_b
\: + \:
\sum_{\alpha_1 \alpha_2} x_{\alpha_1} x_{\alpha_2} y_1 \left(
\sum_m q_m G_m^{\alpha_1 \alpha_2} \right)
\: + \:
\sum_{\alpha} x_{\alpha} y_2 \left(
\sum_m q_m G_m^{\alpha} \right),
\end{equation}
where
\begin{equation}
G_m\left( \epsilon^{ab} \phi^i_a \phi^j_b,
x_{\alpha_1} x_{\alpha_2} y_1, x_{\alpha} y_2 \right)
\: = \:
\sum_{ij} G_{m,ij} \epsilon^{ab} \phi^i_a \phi^j_b
\: + \:
\sum_{\alpha_1 \alpha_2} G_m^{\alpha_1 \alpha_2}  x_{\alpha_1} x_{\alpha_2} y_1
\: + \:
\sum_{\alpha} G_m^{\alpha}  x_{\alpha} y_2.
\end{equation}

The second and third terms in the superpotential above,
\begin{equation}
\sum_{\alpha_1 \alpha_2} x_{\alpha_1} x_{\alpha_2} y_1 \left(
\sum_m q_m G_m^{\alpha_1 \alpha_2} \right)
\: + \:
\sum_{\alpha} x_{\alpha} y_2 \left(
\sum_m q_m G_m^{\alpha} \right),
\end{equation}
describe
\begin{displaymath}
\sum_m q_m G_m^{\alpha_1 \alpha_2}, \: \: \:
\sum_m q_m G_m^{\alpha}
\end{displaymath}
as lying in the kernel of the map defined by
$\{x_{\alpha_1} x_{\alpha_2}, x_{\alpha} \}$, and so they
couple to ${\cal E}^{\perp}$, as in section~\ref{sect:first:ex2}.
(The reader should note that this conclusion requires the
$x_{\alpha}$ to not all vanish, and so is only valid when
$r_3 \gg 0$.)

The first term in the superpotential,
\begin{equation}
\sum_{ij} \sum_m q_m G_{m,ij} \epsilon^{ab} \phi^i_a \phi^j_b,
\end{equation}
describes a mass matrix for the $\phi$ fields, with mass matrix given
specifically as the antisymmetric $5 \times 5$ matrix
with components
\begin{equation}
A_{ij} \: = \:  \sum_m q_m G_{m,ij}.
\end{equation}
Since this matrix is antisymmetric, its rank is one of
$\{4, 2, 0 \}$, corresponding to one, three, or five massless doublets,
respectively.  Working locally in a Born-Oppenheimer approximation
on the space of $q_m$, from the strong-coupling analysis of 
\cite{Hori:2006dk}, if there is one massless doublet, the theory has
no supersymmetric vacua.  If it has three massless doublets,
it will have one supersymmetric vacuum.  Thus, due to this mass matrix,
we are effectively restricting to loci in the
space of $q_m$ (${\mathbb P}^6$) where the matrix $(A_{ij})$ has rank
two.  There are seven $q_m$, but the low-energy limit of the
second and third terms in the
superpotential each impose a single constraint on the $q_m$ on vacua,
so overall the $q_m$ represent five degrees of freedom along vacua.
The locus over which $(A_{ij})$ has rank two, over a five-dimensional
space, is the space denoted $Pf(5)$ in \cite{Donagi:2007hi}[section 4.2.2],
which is (the dual of) the Grassmannian $G(2,5)$, which following
\cite{inoue} we will denote $G(2,V_5^*)$, where $V_5$ is a five-dimensional
vector space.

Letting $J'$ denote Join$(G(2,5),{\mathbb P} {\cal E}^{\perp})$, since the
$q_m$ parametrize $W^{\perp}$, altogether the GLSM is describing in this phase
\begin{displaymath}
J' \times_{ {\mathbb P}\left( \wedge^2 V_5^* \oplus V_N^* \right) }
{\mathbb P} W^{\perp},
\end{displaymath}
in the notation of \cite{inoue}, exactly as expected.

The other phases are not germane to this discussion, so we only summarize
a few highlights briefly.  First, note that the interpretation of
the second join requires both not all $x_{\alpha}$ to vanish and not all
$q_m$ to vanish, and so this description is only valid for phase II.
For example, in phases IV and V, the $y_i$ are not allowed to vanish
(since $r_3 \ll 0$), so as $r \gg 0$, the matrix $\phi\phi^{\dagger}$ 
need not necessarily have full rank.  This potentially implies a 
singularity, and suggests that phases IV and V describe a singular theory.
In phase III, 
one can set $x_{\alpha}=0$ and $\phi=0$, 
but one must keep at least one of the $y_i$ non-zero. 
The third F-term then gives three conditions on $q$ and $y_2$. 
This seems to be a valid vacuum manifold, whose details we will not
explore here.

\begin{figure}
  \begin{center}
    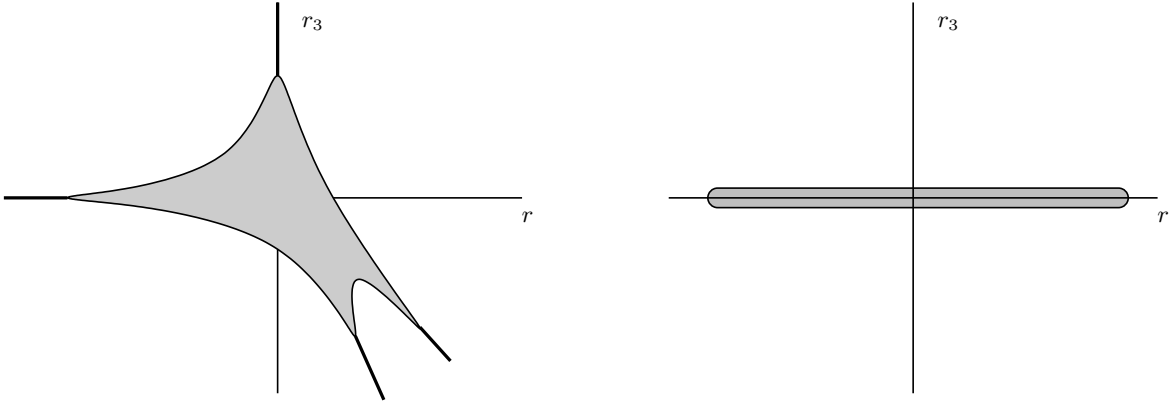
    \end{center}\caption{Components of the discriminant.}\label{fig-amoeba2}
\end{figure}

Finally, for completeness,
we discuss Coulomb and mixed branches. The effective potential on the Coulomb branch is
\begin{eqnarray}
  \mathcal{W}_{eff}&=&-t\left(\sigma_1+\sigma_2\right)
-t_3\sigma_3-5\sigma_1\left[\ln\sigma_1-1\right]
-5\sigma_2\left[\ln\sigma_2-1\right]
\nonumber\\
  &&-3\sigma_3\left[\ln\sigma_3-1\right]
-\left(\sigma_1+\sigma_2-2\sigma_3\right)\left[\ln\left(\sigma_1+\sigma_2-2\sigma_3\right)-1\right]
\nonumber\\
  &&-\left(\sigma_1+\sigma_2-\sigma_3\right)\left[\ln\left(\sigma_1+\sigma_2-\sigma_3\right)-1\right]
\nonumber\\
  &&+7\left(\sigma_1+\sigma_2\right)\left[\ln\left(-\sigma_1-\sigma_2\right)-1\right]+i\pi\left(\sigma_1+\sigma_2\right).
\end{eqnarray}
The critical locus is at
\begin{eqnarray}
  e^{-t}&=&\frac{\sigma_1^5(\sigma_1+\sigma_2-2\sigma_3)(\sigma_1+\sigma_2-\sigma_3)}{(\sigma_1+\sigma_2)^7}=\frac{\sigma_2^5(\sigma_1+\sigma_2-2\sigma_3)(\sigma_1+\sigma_2-\sigma_3)}{(\sigma_1+\sigma_2)^7},
\nonumber\\
  e^{-t_3}&=&\frac{\sigma_3^3}{(\sigma_1+\sigma_2-2\sigma_3)^2(\sigma_1+\sigma_2-\sigma_3)}.
\end{eqnarray}
Defining as before $z=\sigma_2/\sigma_1,
w=\sigma_3/\sigma_1$, we can write this as
\begin{eqnarray}
  e^{-t}&=&\frac{(1+z-2w)(1+z-w)}{(1+z)^7}=\frac{(\frac{1}{z}+1-\frac{2w}{z})(\frac{1}{z}+1-\frac{w}{z})}{(\frac{1}{z}+1)^7},
\nonumber\\
  e^{-t_3}&=&\frac{1}{(\frac{1}{w}+\frac{z}{w}-2)^2(\frac{1}{w}+\frac{z}{w}-1)}, \qquad z^5=1.
\end{eqnarray}
The following combinations are also of interest
\begin{equation}
  e^{-t-t_3}=\frac{w^3}{(1+z)^7(1+z-2w)}, 
\qquad
 e^{-2t-t_3}=\frac{w^3}{(1+z)^7(1+z-2w)}.
  \end{equation}
First we note that we have to exclude solutions with $z=1$ because this is fixed under the $U(2)$ Weyl group action. We consider $z$ to be any other fifth root of unity and discuss various values of $w$. The first case is $w=0$ where
\begin{equation}
  e^{-t}=\mathrm{const.},
 \qquad e^{-t_3}=0.
\end{equation}
The gives the positive $r_3$-axis and thus the phase boundary spanned by the charge vector of $x_{\alpha}$. Next, we consider $w=1+z$. Here we find
\begin{equation}
  e^{-t}=0, 
\quad e^{-t_3}\rightarrow-\infty,
\quad e^{-2t+t_3}=0,
\quad e^{-t-t_3}=\mathrm{const.}
\end{equation}
This gives the phase boundary along $r_3=-r$ spanned by the charges of $y_2$. For $w=\frac{1}{2}(1+z)$ one has
\begin{equation}
  e^{-t}=0,
\quad e^{-t_3}\rightarrow\infty,
\quad e^{-t-t_3}\rightarrow\infty,
\quad e^{-2t-t_2}=\mathrm{const.}
\end{equation}
This encodes the phase boundary $r_3=-2r$ spanned by the charge vector of $y_1$. Finally, we can consider $w\rightarrow\infty$, where we have
\begin{equation}
  e^{-t}\rightarrow\infty,
\qquad e^{t_3}=\mathrm{const.}
\end{equation}
This covers the phase boundary spanned by the charges of the $q_m$: $r_3=\mathrm{const.},r\rightarrow -\infty$. What is missing is the boundary on the positive $r$-axis. This must be encoded in a mixed branch. The calculation is very similar to the previous example. On the mixed branch $U(1)_{\lambda}$ remains unbroken. We have $\dot{\phi}=\{\phi^i_a,q_m\}$ and $\hat{\phi}=\{x_{\alpha},y_1,y_2\}$. Furthermore, $\dot{\sigma}=\sigma_3\equiv\sigma_L$ and $\hat{\sigma}=\{\sigma_1,\sigma_2\}$. The effective potential is 
\begin{eqnarray}
  U_{eff}&=&\frac{1}{2}\left(|\sigma_1\phi^i_1|^2+|\sigma_2\phi^i_2|^2+|(-\sigma_1-\sigma_2)q_m|^2+\left(\sigma_i\rightarrow \bar{\sigma}_i\right)\right)\nonumber\\
  &&+\frac{1}{2}\sum_{a,b}e^2_{eff,a,b}(\mu_{eff,a}-r_{eff,a})(\mu_{eff,b}-r_{eff,b})+|dW(\phi^i_a,q_m)|^2.
\end{eqnarray}
The effective D-terms are
\begin{eqnarray}
  \mu_{eff,U(2)}&=&\phi\phi^{\dagger}-\sum_{m}|q_m|^2{\bf 1}-r{\bf 1},
\nonumber\\
  \mu_{eff,3}&=&0.
\end{eqnarray}
The effective couplings are
\begin{eqnarray}
  t_{eff,U(2)}&=&t{\bf 1}+\ln\left(\sigma_1+\sigma_2-2\sigma_3\right){\bf 1}
+\ln\left(\sigma_1+\sigma_2-\sigma_3\right){\bf 1},\nonumber\\
  t_{eff,2}&=&t_3+3\ln\left(\sigma_3\right)
-2\ln\left(\sigma_1+\sigma_2-2\sigma_3\right)
-\ln\left(\sigma_1+\sigma_2-\sigma_3\right).
\end{eqnarray}
The minimum is at $\sigma_1=\sigma_2=0$, $r$ remains unconstrained, while $r_3=0$ and $\theta_3=\pi$. The two components of the discriminant are depicted in figure \ref{fig-amoeba2}. 

\subsubsection{Analysis of dual gauge theory}
\label{sect:joins:ex2:dual}

As mentioned previously, there is a duality \cite{Hori:2011pd}[section 5.6]
between an $SU(2)$ gauge theory with five fundamentals $\phi$,
and an $SU(2)$ gauge theory with five fundamentals $\varphi$, 
$(1/2)(5)(4) = 10$ singlets $b^{ij} = - b^{ji}$, and a superpotential
\begin{displaymath}
W \: = \: \sum_{ij} \sum_{ab} b^{ij} \varphi^a_i \varphi^b_j \epsilon_{ab}.
\end{displaymath}

In the present case, this means that the 
dual to the theory of the previous subsection is defined by fields
\begin{center}
\begin{tabular}{c|cccrrr}
 & $\varphi_i^a$ & $b^{ij}$ & $x_{\alpha}$ & $y_1$ & $y_2$ & $q_m$ \\ \hline
$SU(2)$ & $\Box$ & ${\bf 1}$ & ${\bf 1}$ & ${\bf 1}$ & ${\bf 1}$ & ${\bf 1}$ \\
$U(1)_1$ & $-1$ & $2$ & $0$ & $2$ & $2$ & $-2$ \\
$U(1)_{\lambda}$ & $0$ & $0$ & $1$ & $-2$ & $-1$ & $0$
\end{tabular}  
\end{center}
with the superpotential
\begin{equation} \label{eq:join:ex2:dual:sup}
W \: = \: \sum_m q_m G_m\left( b^{ij},
x_{\alpha_1} x_{\alpha_2} y_1, x_{\alpha} y_2 \right)
\: + \:
\sum_{ij} \sum_{ab} b^{ij} \varphi^a_i \varphi^b_j \epsilon_{ab},
\end{equation}
following the same pattern as in section~\ref{sect:joins:ex1:dual}.
The gauge group is
\begin{equation}
\frac{
SU(2) \times U(1)_1
}{
{\mathbb Z}_2
} \times U(1)_{\lambda}.
\end{equation}
The gauge factor $U(1)_1$ corresponds to det $U(2)$ in the previous
duality frame.  It is straightforward to check that the sums of
all charges for each $U(1)$ vanish, consistent with the statement
that this GLSM describes a Calabi-Yau.

Now, let us analyze the GLSM in phase I.

The first term of the superpotential,
\begin{equation}
\sum_m q_m G_m\left( b^{ij},
x_{\alpha_1} x_{\alpha_2} y_1, x_{\alpha} y_2 \right),
\end{equation}
restricts us to the intersection of the hyperplanes
$\{ G_m = 0 \}$, as is typical for GLSMs for complete intersections.

We can understand the second term of the superpotential,
\begin{equation}
\sum_{ij} \sum_{ab} b^{ij} \varphi^a_i \varphi^b_j \epsilon_{ab},
\end{equation}
as a mass matrix for the $\varphi$ over the space of $b^{ij}$.
As the mass matrix is an antisymmetric $5 \times 5$ matrix,
it can have rank $4$, $2$, or $0$.  If it has rank $4$,
then there is one massless $SU(2)$ doublet, which means no supersymmetric
vacua (working locally on the space of $b^{ij}$ in a Born-Oppenheimer
approximation), following \cite{Hori:2006dk}.  If it has rank two, then there
are three massless doublets, and one supersymmetric vacuum.
The subset of the space of $\{ b^{ij} \}$ over which the
skew-symmetric matrix $(b^{ij})$ has rank $2$ is precisely
$G(2,5)$ \cite{Donagi:2007hi}[section 4.2.2].

Putting this together, we see that the first superpotential
term is describing a complete intersection in a product of
$G(2,5)$ and the space of $x_{\alpha_1} x_{\alpha_2} y_1$, $x_{\alpha} y_2$.
This is precisely the complete intersection
\begin{displaymath}
J \times_{ {\mathbb P}\left( \wedge^2 V_5 \oplus V_N \right) } {\mathbb P} W
\end{displaymath}
in the join
\begin{displaymath}
J \: = \: {\rm Join}\left( G(2,5), {\mathbb P} {\cal E} \right) \: = \:
{\rm Join}\left( G(2,5), {\rm Bl}_{\rm pt} {\mathbb P}^3 \right),
\end{displaymath}
in the notation of \cite{inoue}.
As expected, this geometry matches that of the corresponding phase
of the dual gauge theory.

Now, let us turn to phase II.
In this phase, not all of the $q_m$ and $\varphi^a_i$ vanish.  If we
expand
\begin{equation}
G_m\left( b^{ij},
x_{\alpha_1} x_{\alpha_2} y_1, x_{\alpha} y_2 \right) \: = \:
\sum_{ij} G_{m,ij} b^{ij} \: + \:
\sum_{\alpha_1, \alpha_2} G_m^{\alpha_1 \alpha_2} x_{\alpha_1}
x_{\alpha_2} y_1 \: + \:
\sum_{\alpha} G_m^{\alpha} x_{\alpha} y_2,
\end{equation}
then we can write the superpotential~(\ref{eq:join:ex2:dual:sup}) as
\begin{equation}
W \: = \: \sum_{ij} b^{ij} \left( \sum_m q_m G_{m,ij} \: + \:
\sum_{ab} \epsilon_{ab} \varphi^a_i \varphi^b_j \right)
\: + \:
\sum_{\alpha_1, \alpha_2} \sum_m q_m G_m^{\alpha_1 \alpha_2} x_{\alpha_1}
x_{\alpha_2} y_1 
\: + \:
\sum_{\alpha} \sum_m q_m G_m^{\alpha} x_{\alpha} y_2.
\end{equation}

From the second and third terms in the superpotential,
\begin{equation}
\sum_{\alpha_1, \alpha_2} \sum_m q_m G_m^{\alpha_1 \alpha_2} x_{\alpha_1}
x_{\alpha_2} y_1 
\: + \:
\sum_{\alpha} \sum_m q_m G_m^{\alpha} x_{\alpha} y_2,
\end{equation}
the $y$ act analogously to Lagrange multipliers, forcing the $q_m$ to
be in the kernel of the map defined by $(x_{\alpha_1} x_{\alpha_2},
x_{\alpha})$, and hence couple to ${\cal E}^{\perp}$, in the notation
of \cite{inoue}.

Similarly, the first superpotential terms
\begin{equation}
\sum_{ij} b^{ij} \left( \sum_m q_m G_{m,ij} \: + \:
\sum_{ab} \epsilon_{ab} \varphi^a_i \varphi^b_j \right)
\end{equation}
can be interpreted as imposing a set of constraints on the low-energy
theory, in which the linear combinations
\begin{displaymath}
\sum_m q_m G_{m,ij}
\end{displaymath}
are identified with the image of the Pl\"ucker embedding of $G(2,5)$.
(The $b^{ij}$ act analogously to Lagrange multipliers.)
Since this model was obtained by dualizing relative to the first duality
frame, we can identify this $G(2,5)$ with $G(2,V_5^*)$, the
`dual' Grassmannian to that appearing earlier.

Altogether, this means that this phase of the GLSM can be identified with
\begin{displaymath}
J' \times_{ {\mathbb P}\left( \wedge^2 V_5^* \oplus V_N^* \right) }
{\mathbb P} W^{\perp},
\end{displaymath}
in the notation of \cite{inoue}, where $J'$ is the join
\begin{displaymath}
J' \: = \: {\rm Join}\left( G(2,V_5^*), {\mathbb P} {\cal E}^{\perp} \right).
\end{displaymath}

\subsection{ Join$(G(2,5), {\mathbb P}^1 \times {\mathbb P}^1 \times
{\mathbb P}^1 )$ }

In this section we will build a GLSM description of the third of
the examples of joins discussed in \cite{inoue},
namely
\begin{displaymath}
{\rm Join}\left(G(2,5), {\mathbb P}^1 \times {\mathbb P}^1 \times
{\mathbb P}^1 \right)
\end{displaymath}
and Calabi-Yau complete intersections therein.

\subsubsection{First duality frame}

Proceeding as before, we can write down the GLSM for the resolved join
immediately, as a ${\mathbb P}^1$ bundle over the product.
This GLSM is a $U(2) \times U(1)^4$ gauge theory with fields as follows:
\begin{center}
\begin{tabular}{c|cccrcr}
& $\phi^i_a$ & $x_{0,1}$ & $y_{0,1}$ & $w_{0,1}$ & $z_1$ & $z_2$ \\ \hline
$U(2)$ & $\Box$ & ${\bf 1}$ & ${\bf 1}$ & ${\bf 1}$ & 
det$^{-1}$ & ${\bf 1}$ \\
$U(1)_{\lambda}$ & $0$ & $1$ & $0$ & $-1$ & $0$ & $0$ \\
$U(1)_{\mu}$ & $0$ & $0$ & $1$ & $-1$ & $0$ & $0$ \\
$U(1)_{\nu}$ & $0$ & $0$ & $0$ & $1$ & $0$ & $-1$ \\
$U(1)_{\rho}$ & $0$ & $0$ & $0$ & $0$ & $1$ & $1$
\end{tabular}
\end{center}
The line bundle $L$ in \cite{inoue} is then defined by $U(1)^4$ charges
$(0,0,0,1)$.  It is straightforward to compute that sections of $L$
are of the form
\begin{displaymath}
\epsilon^{ab} \phi^i_a \phi^j_b z_1, \: \: \:
x_{\alpha} y_{\beta} w_{\gamma} z_2 .
\end{displaymath}

Blowing down the divisor $\{z_1=0\}$, by removing $z_1$ and
the last $U(1)$, this becomes
\begin{center}
\begin{tabular}{c|cccrcr}
& $\phi^i_a$ & $x_{0,1}$ & $y_{0,1}$ & $w_{0,1}$ &  $z_2$ \\ \hline
$U(2)$ & $\Box$ & ${\bf 1}$ & ${\bf 1}$ & ${\bf 1}$ &
 det \\
$U(1)_{\lambda}$ & $0$ & $1$ & $0$ & $-1$ & $0$ \\
$U(1)_{\mu}$ & $0$ & $0$ & $1$ & $-1$ &  $0$ \\
$U(1)_{\nu}$ & $0$ & $0$ & $0$ & $1$ &  $-1$ 
\end{tabular}
\end{center}

Blowing down the divisor $\{z_2=0\}$, by removing $z_2$ and the third
$U(1)$, this becomes
\begin{center}
\begin{tabular}{c|cccrc}
& $\phi^i_a$ & $x_{0,1}$ & $y_{0,1}$ & $w_{0,1}$  \\ \hline
$U(2)$ & $\Box$ & ${\bf 1}$ & ${\bf 1}$ & det \\ 
$U(1)_{\lambda}$ & $0$ & $1$ & $0$ & $-1$  \\
$U(1)_{\mu}$ & $0$ & $0$ & $1$ & $-1$  
\end{tabular}
\end{center}

A Calabi-Yau complete intersection will contain hypersurfaces charged
only under det $U(2)$, as the sums of the charges under the other
two $U(1)$s vanish.  Specifically, a complete intersection
of seven hyperplanes, of charge $1$ under det $U(2)$, will be
Calabi-Yau, agreeing with the statements in \cite{inoue}.
As before, for generic hyperplanes, the resulting Calabi-Yau will not
intersect the singularities of the join.

Such a theory, describing a complete intersection of seven hyperplanes
in the join above, is described by the fields
\begin{center}
\begin{tabular}{c|cccrc|c}
& $\phi^i_a$ & $x_{0,1}$ & $y_{0,1}$ & $w_{0,1}$ & $q_m$ & FI\\ \hline
$U(2)$ & $\Box$ & ${\bf 1}$ & ${\bf 1}$ & det & det$^{-1}$&$r$\\ 
$U(1)_{\lambda}$ & $0$ & $1$ & $0$ & $-1$ & $0$&$r_{\lambda}$ \\
$U(1)_{\mu}$ & $0$ & $0$ & $1$ & $-1$  & $0$& $r_{\mu}$
\end{tabular}
\end{center}
with superpotential
\begin{equation}   \label{eq:join:fr1:ex3:sup}
W \: = \: \sum_m q_m G_m\left( \epsilon^{ab} \phi^i_a \phi^j_b, 
x_{\alpha} y_{\beta} w_{\gamma} \right)=\sum_mq_m(G_{m,ij}\epsilon^{ab} \phi^i_a \phi^j_b+G_{m}^{\alpha\beta\gamma}x_{\alpha}y_{\beta}w_{\gamma}).
\end{equation}

The D-terms are
    \begin{eqnarray}
  \sum_{\alpha}|x_{\alpha}|^2-|w_{\alpha}|^2&=&r_{\lambda},\nonumber\\
  \sum_{\alpha}|y_{\alpha}|^2-|w_{\alpha}|^2&=&r{\lambda},\nonumber\\
  \mathrm{Tr}\phi\phi^{\dagger}+\sum_{\alpha}|w_{\alpha}|^2{\bf 1}-\sum_{m}|q_m|^2{\bf 1}&=&r{\bf 1}.
\end{eqnarray}
The F-terms are
\begin{eqnarray}
  G_m(\epsilon^{ab}\phi_a^i\phi_b^j,x_{\alpha}y_{\beta}w_{\gamma})&=&0,\nonumber\\
  q_mG_{m,ij}\epsilon^{ab}\phi_a^i&=&0,\nonumber\\
  q_mG_{m}^{\alpha\beta\gamma}x_{\alpha}y_{\beta}&=&0,\nonumber\\
  q_mG_{m}^{\alpha\beta\gamma}x_{\alpha}w_{\gamma}&=&0,\nonumber\\
  q_mG_{m}^{\alpha\beta\gamma}y_{\beta}w_{\gamma}&=&0.
\end{eqnarray}

As the K\"ahler moduli space of this example has dimension greater than
two, we will not try to give a systematic accounting of the location
of every phase, but we will note the existence of regions containing
two phases realizing the homologically-projective-dual geometries
corresponding to the pertinent join.
   
First, 
there exists a phase in the region where $r \gg 0$, $r_{\lambda} \gg 0$,
and $r_{\mu} \gg 0$, in which the $\{ \phi^i_a, x_{0,1}, y_{0,1}, w_{0,1} \}$
do not simultaneously vanish, but the $q_m$ are allowed to simultaneously
vanish.
In this phase, by construction, the GLSM is describing a complete
intersection
\begin{displaymath}
J \times_{ {\mathbb P} \left( \wedge^2 V_5 \oplus V_N \right) } {\mathbb P} W
\end{displaymath}
(in the notation of \cite{inoue}) in the join
\begin{displaymath}
J \: = \: {\rm Join}\left( G(2,5), {\mathbb P} {\cal E} \right).
\end{displaymath}

The other phase of interest is in the region $r_{\rm det} \ll 0$,
where the $q_m$ are not all zero, and also $r_{\lambda} \gg 0$,
$r_{\mu} \gg 0$, so that the $x_{0,1}$ do not both vanish, and also
so that the $y_{0,1}$ do not both vanish.
In this phase, it will be useful to rewrite the 
superpotential~(\ref{eq:join:fr1:ex3:sup})
in the form
\begin{equation}
W \: = \: \sum_{ij} \sum_m q_m G_{m,ij} \epsilon^{ab} \phi^i_a \phi^j_b
\: + \: \sum_{\alpha \beta \gamma} \sum_m q_m G_m^{\alpha \beta \gamma}
x_{\alpha} y_{\beta} w_{\gamma},
\end{equation}
where
\begin{equation}
G_m\left( \epsilon^{ab} \phi^i_a \phi^j_b, 
x_{\alpha} y_{\beta} w_{\gamma} \right) \: = \:
G_{m,ij}  \epsilon^{ab} \phi^i_a \phi^j_b \: + \:
G_m^{\alpha \beta \gamma} x_{\alpha} y_{\beta} w_{\gamma}.
\end{equation}

The second term in the superpotential,
\begin{equation}
\sum_{\alpha \beta \gamma} 
x_{\alpha} y_{\beta} w_{\gamma}\left(  \sum_m q_m G_m^{\alpha \beta \gamma}
\right),
\end{equation}
provides two constraints on the $q_m$.  Specifically, much as in 
section~\ref{sect:first:ex3} we can interpret the $w_{\gamma}$ as analogues of
Lagrange multipliers, requiring
\begin{displaymath}
 \sum_m q_m G_m^{\alpha \beta \gamma}
\end{displaymath}
to be in the kernel of the map defined by the $x_{\alpha} y_{\beta}$,
and so couple to ${\cal E}^{\perp}$.

We interpret the first term in the superpotential above as
a mass matrix for the $\phi^i_a$.  Proceeding as before,
\begin{equation}
A_{ij} \: = \: \sum_m q_m G_{m,ij}
\end{equation}
are the components of an antisymmetric $5 \times 5$ matrix, encoding the
masses of the $\phi$ fields.  The possible ranks of this matrix are
$\{4, 2, 0 \}$.  Over the locus in the space of $q_m$s where it has rank $4$,
there is only one massless doublet of $SU(2)$, so from \cite{Hori:2006dk},
there are no vacua.  Over the locus where $(A_{ij})$ has rank $2$,
there are three massless doublets of $SU(2)$, which
corresponds to a single vacuum from \cite{Hori:2006dk}.  Thus, this mass
matrix effectively restricts us to the locus in the space of $q_m$s
where the matrix $A_{ij}$ has rank $2$.  There are seven $q_m$s,
defining ${\mathbb P}^6$, but the low-energy limit of the second 
superpotential term imposes two constraints
so overall the $q_m$ represent five degrees of freedom.  The locus over
which $A_{ij}$ has rank 2, over a five-dimensional space, is the
space denoted $Pf(5)$ in \cite{Donagi:2007hi}[section 4.2.2], which is 
(the dual of) the Grassmannian $G(2,5)$, which following \cite{inoue}
we denote $G(2,V_5^*)$, where $V_5$ is a five-dimensional vector space.

Since the $q_m$ parametrize $W^{\perp}$, together this phase of the GLSM
is therefore describing the complete intersection
\begin{displaymath}
J' \times_{ {\mathbb P}\left( \wedge^2 V_5^* \oplus V_N^* \right) }
{\mathbb P} W^{\perp},
\end{displaymath}
where
\begin{displaymath}
J' \: = \: {\rm Join} \left( G(2,V_5^*), {\mathbb P} {\cal E}^{\perp} \right).
\end{displaymath}

\subsubsection{Analysis of dual gauge theory}

Next, we shall dualize the GLSM of the previous subsection and
analyze its phases.  We will find the same geometric interpretations of
the phases in this duality frame.  Furthermore, phases that could
be understood perturbatively in the previous frame, will now require
a strong-coupling analysis, and conversely.

Proceeding as before, we can dualize the $SU(2)$ gauge theory of
the previous section, applying \cite{Hori:2011pd}[section 5.6].
We replace the $SU(2)$ gauge theory with five fundamentals $\phi$,
by an $SU(2)$ gauge theory with five fundamentals $\varphi$,
ten singlets $b^{ij} = - b^{ji}$, and a superpotential.
The new GLSM is defined by fields as
\begin{center}
\begin{tabular}{c|ccccrr}
& $\varphi^a_i$ & $b^{ij}$ & $x_{0,1}$ & $y_{0,1}$ & $w_{0,1}$ &
$q_m$ \\ \hline
$SU(2)$ & $\Box$ & ${\bf 1}$ & ${\bf 1}$ & ${\bf 1}$ & ${\bf 1}$ 
& ${\bf 1}$ \\
$U(1)_1$ & $-1$ & $2$ & $0$ & $0$ & $2$ & $-2$ \\
$U(1)_{\lambda}$ & $0$ & $0$  & $1$ & $0$ & $-1$ & $0$ \\
$U(1)_{\mu}$ & $0$ & $0$ & $0$ & $1$ & $-1$ & $0$
\end{tabular}
\end{center}
with superpotential
\begin{equation}  \label{eq:join:ex3:sup}
W \: = \: \sum_m q_m G_m\left( b^{ij}, x_{\alpha} y_{\beta} w_{\gamma} \right)
\: + \: 
\sum_{i j} \sum_{a b} b^{ij} \varphi_i^a \varphi_j^b \epsilon_{ab}.
\end{equation}
To be clear, the gauge group is
\begin{equation}
\frac{
SU(2) \times U(1)_1
}{
{\mathbb Z}_2
} \times U(1)_{\lambda} \times U(1)_{\mu}.
\end{equation}

Now, let us turn to the first phase of interest, as described
in the previous section.
In this phase, $\{ b^{ij}, w_{0,1} \}$ are not all zero.

We can interpret the second term in the superpotential,
\begin{equation}
\sum_{i j} \sum_{a b} b^{ij} \varphi_i^a \varphi_j^b \epsilon_{ab},
\end{equation}
in terms of a mass matrix for the $\varphi_i^a$.  Following the same analysis
as for the previous joins, the mass matrix is the skew-symmetric
$5 \times 5$ matrix $( b^{ij} )$, which (as it is skew-symmetric) can
have rank $4$, $2$, or $0$.  Over loci for which it has rank $4$, 
there is one
massless doublet, which from the analysis of 
\cite{Hori:2006dk}, leads to no supersymmetric vacua.
Over loci for which it has rank $2$, there are three massless doublets,
which from the analysis of \cite{Hori:2006dk} leads to a single supersymmetric
vacuum.  The locus in the space of $\{ b^{ij} \}$ over which
the skew-symmetric matrix $(b^{ij} )$ has rank $2$ is precisely
$G(2,5)$, as described in \cite{Donagi:2007hi}[section 4.2.2], so we see that
this superpotential term is telling us to localize on a $G(2,5)$ inside
the space of $\{ b^{ij} \}$.

The first superpotential term describes a complete intersection in
$G(2,5)$ and the space of $x_{\alpha} y_{\beta} w_{\gamma}$.
This is precisely the complete intersection
\begin{displaymath}
J \times_{ {\mathbb P} \left( \wedge^2 V_5 \oplus V_N \right) } {\mathbb P} W
\end{displaymath}
(in the notation of \cite{inoue}) in the join
\begin{displaymath}
J \: = \: {\rm Join}\left( G(2,5), {\mathbb P} {\cal E} \right).
\end{displaymath}
Thus, we have recovered the same geometry as in the corresponding
phase of the dual gauge theory.

Now, let us turn to the second phase of interest,
as described in the previous section.
In this phase $\{ \varphi_i^a, q_m\}$ are not all zero.

It will be useful to rewrite the superpotential~(\ref{eq:join:ex3:sup})
in the form
\begin{equation}
W \: = \: \sum_{ij} b^{ij} \left( \sum_m q_m G_{m,ij} \: + \:
\varphi^a_i \varphi^b_j \epsilon_{ab} \right) \: + \:
\sum_{\alpha \beta \gamma} x_{\alpha} y_{\beta} w_{\gamma}
\left( \sum_m q_m G_m^{\alpha \beta \gamma} \right),
\end{equation}
where we have written
\begin{equation}
G_m\left( b^{ij}, x_{\alpha} y_{\beta} w_{\gamma} \right)
\: = \:
\sum_{ij} G_{m,ij} \varphi^a_i \varphi^b_j \epsilon_{ab}
\: + \:
\sum_{\alpha \beta \gamma} G_m^{\alpha \beta \gamma} x_{\alpha} y_{\beta}
w_{\gamma}.
\end{equation}

The second term in the superpotential above,
\begin{equation}
\sum_{\alpha \beta \gamma} x_{\alpha} y_{\beta} w_{\gamma}
\left( \sum_m q_m G_m^{\alpha \beta \gamma} \right),
\end{equation}
tells us that the linear combination
\begin{displaymath}
 \sum_m q_m G_m^{\alpha \beta \gamma}
\end{displaymath}
lies in the kernel of the map defined by $x_{\alpha} y_{\beta}$,
and so couples to ${\cal E}^{\perp}$, just as in
section~\ref{sect:first:ex3}.

Similarly, the first term in the superpotential,
\begin{equation}
\sum_{ij} b^{ij} \left( \sum_m q_m G_{m,ij} \: + \:
\varphi^a_i \varphi^b_j \epsilon_{ab} \right) ,
\end{equation}
identifies the linear combinations
\begin{displaymath}
 \sum_m q_m G_{m,ij}
\end{displaymath}
with the image of the Pl\"ucker embedding of $G(2,5)$.  
Since this model was obtained by dualizing, relative to the first duality
frame we can identify this $G(2,5)$ with $G(2,V_5^*)$, the
`dual' Grassmannian to that appearing in the other phase.

Altogether, this means that this phase of the GLSM can be identified with
\begin{displaymath}
J' \times_{ {\mathbb P}\left( \wedge^2 V_5^* \oplus V_N^* \right) }
{\mathbb P} W^{\perp},
\end{displaymath}
in the notation of \cite{inoue}, where $J'$ is the join
\begin{displaymath}
J' \: = \: {\rm Join}\left( G(2,V_5^*), {\mathbb P} {\cal E}^{\perp} \right).
\end{displaymath}

\section{Conclusions}

In this paper we have given a physical realization of 
some new examples \cite{inoue} of homological projective duality
\cite{kuz1,kuz2,kuz3}, as phases of GLSMs.  Geometries are realized
in these theories both perturbatively (as the critical locus of
a superpotential), and nonperturbatively (through strong coupling
effects), and we check that we realize the same geometries after
dualizing the GLSMs.  We also discussed the physical realizations of
joins, which play an important role in the homological projective
duality examples in \cite{inoue}, and outlined notions of joins of
more general gauge theories.
We also discussed Calabi-Yau conditions
that utilize relations amongst divisors and do not necessarily lift
to the ambient space.  We also discussed Hadamard products and Picard-Fuchs
equations in this context.

Another possible example of a join construction appears in
\cite{kk}[theorem 3.8], also involving the secant variety of a 
Segre embedding.  We leave the physical realization of that example
in GLSMs to future work. 

Finally, let us remark that a categorfied version of the notion of a 
join has recently been introduced in the context of homological 
projective duality \cite{kp18}. It would be interesting to understand this 
from a physics perspective.

\section{Acknowledgements}

We would like to thank A.~C\u{a}ld\u{a}raru, S.~Galkin,
S.~Hosono, S.~Katz, T.~Pantev, and E.~Scheidegger for useful discussions.
E.S. was partially supported by NSF grant PHY-1720321. J.K. was supported by the Austrian Science Fund (FWF): [P30904-N27].

\appendix

\section{Non-ambient Calabi-Yau conditions}
\label{app:nonambientcy}

Ordinarily when deriving Calabi-Yau conditions from GLSMs,
we derive conditions that can be stated on the ambient space,
without using divisor relations that may exist along hypersurfaces.
In particular, divisors that are not equivalent on the ambient space,
and which are associated to different GLSM charges, may become
equivalent after restriction to a subvariety, and such relations may be
used when deriving Calabi-Yau conditions.

In this appendix, we will give a simple model of such relations,
and then illustrate in a prototypical GLSM how to modify the GLSM so as
to reveal hidden potential Calabi-Yau hypersurfaces.

First, let us describe this issue in more detail.
Let $V \rightarrow M$ be a vector bundle of rank $r$,
and define $A = \det V$.  Define $J = {\mathbb P}(V)$,
a projective-space bundle over $M$.  We will consider the condition
for a subvariety of $J$ to be Calabi-Yau.

Over the space $J$ there is a unique line bundle, call it ${\cal O}_V(1)$,
such that $\pi_* {\cal O}_V(1) = V^*$.

The question we wish to answer is, for what positive integer $c$
is the intersection $Z$ of $c$ divisors $D_i \in | {\cal O}_V(1) |$,
Calabi-Yau?

First, let us derive the relation between $K_J |_Z$ and powers of
${\cal O}_V(1)|_Z$.  Let $Z$ be the intersection of $c$ divisors
in $| {\cal O}_V(1) |$ as above, and recall
\begin{displaymath}
0 \: \longrightarrow \: T Z \: \longrightarrow \:
TJ|_Z \: \longrightarrow \: N_{Z/J} \: \longrightarrow \: 0,
\end{displaymath}
where $N_{Z/J} = {\cal O}_V(1)^c|_Z$.
This implies that
\begin{displaymath}
\det TJ|_Z \: = \: \left( \det TZ \right) \otimes {\cal O}_V(c)|_Z,
\end{displaymath}
or in other words,
\begin{displaymath}
K_Z \: = \: K_J|_Z \otimes {\cal O}_V(c)|_Z.
\end{displaymath}
In order for $Z$ to be Calabi-Yau, then
we must require
\begin{equation}   \label{eq:app:cy1}
K_J|_Z \: = \: {\cal O}_V(-c)|_Z.
\end{equation}

So far, we have recovered a Calabi-Yau condition of a fairly standard
form in the physics literature.  Next, we shall derive a
Calabi-Yau condition using relations that exist on divisors but not on
the ambient $J$.  The first step is to note that for sufficiently `nice'
${\cal O}_V(1)$, divisors will not contain the fiber, and so in such cases,
$D_i \in | {\cal O}_V(1) |$ have the property that $D_i \cong M$.

Now, using the relation above, we shall compare ${\cal O}_V(1)|_{D_i}$
to $K_J|_{D_i}$.  On $J$, ${\cal O}_V(1)$ are independent, but
on $D_i$, they are not.

Let $\pi$ denote the projection $\pi: J \rightarrow M$.
Then,
\begin{displaymath}
0 \: \longrightarrow \: {\cal O} \: \longrightarrow \:
(\pi^* V ) \otimes {\cal O}_V(1) \: \longrightarrow \:
T_{\pi} \: \longrightarrow \: 0,
\end{displaymath}
which implies
\begin{displaymath}
\det T_{\pi} \: = \: \left( \pi^* \det V \right) \otimes {\cal O}_V(r),
\end{displaymath}
where $r$ is the rank of $V$.
It is also true that
\begin{displaymath}
\det T_{\pi} \: = \: K_J^{-1} \otimes \pi^* K_M.
\end{displaymath}
Thus, using the fact that $A = \det V$, 
\begin{displaymath}
K_J^{-1} \: = \: \pi^* \left( K_M^{-1} \otimes A \right) \otimes {\cal O}_V
(r).
\end{displaymath}
Now, for $D \in | {\cal O}_V(1) |$,
\begin{displaymath}
{\cal O}_V(1) |_D \: = \: N_{D/J},
\end{displaymath}
so we can apply adjunction:
\begin{displaymath}
0 \: \longrightarrow \: T D \: \longrightarrow \: TJ|_D \:
\longrightarrow \: N_{D/J} \: \longrightarrow \: 0,
\end{displaymath}
hence
\begin{equation}  \label{eq:app1}
K_J^{-1} |_D \: = \: K_D^{-1} \otimes {\cal O}_V(1) |_D.
\end{equation}
Since $D \cong M$, $K_D \cong K_M$.  Thus,
\begin{eqnarray*}
{\cal O}_V(1) |_D & = & K_J^{-1} |_D \otimes K_M ,
\\
& = & K_M^{-1} \otimes A \otimes {\cal O}_V(r)|_D \otimes K_M
\: = \: A \otimes {\cal O}_V(r)|_D,
\end{eqnarray*}
hence
\begin{displaymath}
{\cal O}_V(-r+1)|_D \: = \: A.
\end{displaymath}

Now, for simplicity, we shall assume henceforward that $r=2$,
i.e., that $J$ is a ${\mathbb P}^1$ bundle on $M$.
Then, from the above, we see that ${\cal O}_V(1)|_D = A^{-1}$,
for any divisor $D \in | {\cal O}_V(1) |$,
hence
\begin{displaymath}
{\cal O}_V(-c)|_D \: = \: A^c,
\end{displaymath}
and so
\begin{displaymath}
{\cal O}_V(-c)|_Z \: = \: A^c|_Z
\end{displaymath}
where $Z$ is the intersection of $c$ divisors $D \in | {\cal O}_V(1) |$,
defining our Calabi-Yau.
From equation~(\ref{eq:app1}),
\begin{displaymath}
K_J|_D \: = \: K_M \otimes {\cal O}_V(-1)|_D \: = \: 
K_M \otimes A,
\end{displaymath}
and so the Calabi-Yau condition~(\ref{eq:app:cy1}) reduces to
\begin{displaymath}
K_M|_Z \otimes A|_Z \: = \: A^c|_Z,
\end{displaymath}
or more simply,
\begin{equation}   \label{eq:app:cy2}
K_M|_Z \: = \: A^{c-1}|_Z
\end{equation}
on $Z$.

For a simple example, let us consider the Calabi-Yau condition
on a ${\mathbb P}^1$ bundle over ${\mathbb P}^n$.
Specifically, define $V = {\cal O}(-1) \oplus {\cal O}$,
and consider the total space of ${\mathbb P}V \rightarrow
{\mathbb P}^n$.  The GLSM for this space is a $U(1)^2$ gauge
theory with chiral superfields and charges as follows:
\begin{center}
\begin{tabular}{c|cccrc}
& $x_0$ & $\cdots$ & $x_n$ & $z_1$ & $z_2$ \\ \hline
$U(1)_{\lambda}$ & $1$ & $\cdots$ & $1$ & $-1$ & $0$ \\
$U(1)_{\nu}$ & $0$ & $\cdots$ & $0$ & $1$ & $1$
\end{tabular}
\end{center}
The $U(1)_{\nu}$ is responsible for the projectivization of the fibers.
In the language used above, ${\cal O}_V(1) \sim {\cal O}_{\nu}$,
the line bundle whose sections have charges $(0,1)$ under
$U(1)_{\lambda} \times U(1)_{\nu}$.
As a consistency check, note that
\begin{displaymath}
H^0( {\cal O}_{\nu}) \: = \: z_1 H^0({\mathbb P}^n, {\cal O}(1)) \oplus
\{ z_2 \} \: \cong \:
H^0( {\mathbb P}^n, {\cal O}(1) ) \oplus
H^0( {\mathbb P}^n, {\cal O} ).
\end{displaymath}
In this GLSM, the Calabi-Yau condition (determined by the divisors on the
ambient space, without using any relations that exist on subvarieties)
is that for a complete intersection, the sum of the charges with respect
to the two $U(1)$s must be $(n,2)$.

Next, note that on the patch $z_1 \neq 0$,
the line bundles ${\cal O}_{\lambda} \cong {\cal O}_{\nu}$,
by virtue of the trivialization defined by the section $z_1$.
If our Calabi-Yau complete intersection lies entirely in that
patch, then we can replace the GLSM above with another GLSM that will
realize the Calabi-Yau, in which we eliminate the field $z_1$
and $U(1)_{\nu}$, to get a $U(1)$ GLSM defined by
\begin{center}
\begin{tabular}{ccccc}
 & $x_0$ & $\cdots$ & $x_n$ & $z_2$ \\ \hline
$U(1)_{\lambda \mu}$ & $1$ & $\cdots$ & $1$ & $1$
\end{tabular}
\end{center}
In this alternative GLSM, the Calabi-Yau condition is simply
that the hypersurface or complete intersection must have
degree $n+2$ with respect to the single $U(1)$.

Now, let us compare to the Calabi-Yau condition~(\ref{eq:app:cy2})
arising in our previous analysis.
Here, the second GLSM is predicting that an intersection of
$n+2$ divisors in ${\cal O}(1)$ is Calabi-Yau.
In this example, $A = \det V = {\cal O}(-1)$,
and $M = {\mathbb P}^n$, so $K_M = {\cal O}(-n-1)$.
Plugging into~(\ref{eq:app:cy2}),
clearly
\begin{equation}
{\cal O}(-n-1) \: = \: {\cal O}(-1)^{n+1},
\end{equation}
hence $c-1 = n+1$ or more simply, $c=n+2$, matching the Calabi-Yau condition
derived from the GLSM.


\begin{thebibliography}{199}

\addcontentsline{toc}{section}{References}

\bibitem{Witten:1993yc}
  E.~Witten,
  ``Phases of N=2 theories in two-dimensions,''
  Nucl.\ Phys.\ B {\bf 403} (1993) 159-222,
   AMS/IP Stud.\ Adv.\ Math.\  {\bf 1} (1996) 143-211,
  {\tt hep-th/9301042}.


\bibitem{Hori:2006dk}
  K.~Hori and D.~Tong,
  ``Aspects of non-abelian gauge dynamics in two-dimensional N=(2,2) theories,''
  JHEP {\bf 0705} (2007) 079,
  {\tt hep-th/0609032}. 
  
\bibitem{Caldararu:2007tc}
  A.~C\u{a}ld\u{a}raru, J.~Distler, S.~Hellerman, T.~Pantev and E.~Sharpe,
  ``Non-birational twisted derived equivalences in abelian GLSMs,''
  Commun.\ Math.\ Phys.\  {\bf 294} (2010) 605-645, 
  {\tt arXiv:0709.3855}.
  
\bibitem{Hellerman:2006zs}
  S.~Hellerman, A.~Henriques, T.~Pantev, E.~Sharpe and M.~Ando,
  ``Cluster decomposition, T-duality, and gerby CFT's,''
  Adv.\ Theor.\ Math.\ Phys.\  {\bf 11} (2007)  751-818,
  {\tt hep-th/0606034}.
 

\bibitem{Hori:2011pd}
  K.~Hori,
  ``Duality in two-dimensional (2,2) supersymmetric non-abelian gauge theories,''
  JHEP {\bf 1310} (2013) 121,
  {\tt arXiv:1104.2853}.
  



 
\bibitem{kuz1} A. Kuznetsov, ``Homological projective duality,''
Publ. Math. Inst. Hautes \'Etudes Sci. {\bf 105} (2007) 157-220,
{\tt math.AG/0507292}.

\bibitem{kuz2} A. Kuznetsov, ``Homological projective duality for
Grassmannians of lines,''
{\tt math.AG/0610957}.

\bibitem{kuz3} A. Kuznetsov, ``Derived categories of quadric fibrations and
intersections of quadrics,''
Adv. Math. {\bf 218} (2008) 1340-1369,
{\tt math.AG/0510670}.

\bibitem{inoue} D. Inoue, ``Calabi-Yau 3-folds from projective
joins of del Pezzo manifolds,''
{\tt arXiv:1902.10040}.


\bibitem{galkintalk} S. Galkin, ``Joins and Hadamard products,''
talk at Steklov Mathematical Institute, September 17, 2015,
as part of the conference ``Categorical and analytic invariants in algebraic
geometry,'' video available
online.

\bibitem{Caldararu:2017usq}
  A.~C\u{a}ld\u{a}raru, J.~Knapp and E.~Sharpe,
  ``GLSM realizations of maps and intersections of Grassmannians and Pfaffians,''
  JHEP {\bf 1804} (2018) 119,
  {\tt arXiv:1711.00047}.


  \bibitem{vanstraten1} G.~Almkvist, W.~Zudilin,
``Differential equations, mirror maps and zeta values,''
pp. 481-515 in {\it Mirror symmetry V}, AMS/IP Stud. Adv. Math. {\bf 38},
Amer. Math. Soc., Providence, Rhode Island, 2006,
{\tt math/0402386}.


\bibitem{vanstraten2} G.~Almkvist, C.~van~Enckevort, D.~van~Straten,
and W.~Zudilin, ``Tables of Calabi--Yau equations,''
{\tt math/0507430}.


\bibitem{Hori:2013gga}
  K.~Hori and J.~Knapp,
  ``Linear sigma models with strongly coupled phases - one parameter models,''
  JHEP {\bf 1311} (2013) 070,
  {\tt arXiv:1308.6265}.

\bibitem{dm} C. Doran, A. Malmendier, ``Calabi-Yau manifolds realizing
symplectically rigid monodromy tuples,''
{\tt arXiv:1503.07500}.


\bibitem{Kapustka:2017jyt}
  M.~Kapustka and M.~Rampazzo,
  ``Torelli problem for Calabi-Yau threefolds with GLSM description,''
  {\tt arXiv:1711.10231}.

\bibitem{Witten:1993xi}
  E.~Witten,
  ``The Verlinde algebra and the cohomology of the Grassmannian,''
  pp. 357-422 in
{\it Geometry, topology, and physics (Cambridge, 1993)},
Conf. Proc. Lecture Notes Geom. Topology, IV, Int. Press,
Cambridge, Massachusetts, 1995,
  {\tt hep-th/9312104}.


\bibitem{Donagi:2007hi}
  R.~Donagi and E.~Sharpe,
  ``GLSM's for partial flag manifolds,''
  J.\ Geom.\ Phys.\  {\bf 58} (2008) 1662-1692,
  {\tt arXiv:0704.1761}.


  \bibitem{prince} T. Prince, ``Smoothing Calabi-Yau toric hypersurfaces using
the Gross-Siebert algorithm,'' {\tt arXiv:1909.02140}.

\bibitem{Hosono:2011np} 
  S.~Hosono and H.~Takagi,
  ``Mirror symmetry and projective geometry of Reye congruences I,''
  J.\ Alg.\ Geom.\  {\bf 23} (2014) 279-312,
  {\tt arXiv:1101.2746}.

\bibitem{Jockers:2012zr}
  H.~Jockers, V.~Kumar, J.~M.~Lapan, D.~R.~Morrison and M.~Romo,
  ``Nonabelian 2D gauge theories for determinantal Calabi-Yau varieties,''
  JHEP {\bf 1211} (2012) 166,
  {\tt arXiv:1205.3192}.

\bibitem{Aharony:2016jki}
  O.~Aharony, S.~S.~Razamat, N.~Seiberg and B.~Willett,
  ``The long flow to freedom,''
  JHEP {\bf 1702} (2017) 056,
  {\tt arXiv:1611.02763}.

\bibitem{Gu:2018fpm}
  W.~Gu and E.~Sharpe,
  ``A proposal for nonabelian mirrors,''
  {\tt arXiv:1806.04678}.

\bibitem{Chen:2018wep}
  Z.~Chen, W.~Gu, H.~Parsian and E.~Sharpe,
  ``Two-dimensional supersymmetric gauge theories with exceptional gauge groups,''
  {\tt arXiv:1808.04070}.

  \bibitem{Blumenhagen:2010pv}
  R.~Blumenhagen, B.~Jurke, T.~Rahn and H.~Roschy,
  ``Cohomology of line bundles: a computational algorithm,''
  J.\ Math.\ Phys.\  {\bf 51} (2010) 103525,
  {\tt arXiv:1003.5217}.

  \bibitem{topcom}
    J.~Rambau,
    ``TOPCOM: Triangulations of Point Configurations and Oriented Matroids,''
    Mathematical Software - ICMS 2002, 330 - 340 (2002)
    {\tt http://www.zib.de/PaperWeb/abstracts/ZR-02-17}.

\bibitem{Braun:2012vh}
  A.~P.~Braun, J.~Knapp, E.~Scheidegger, H.~Skarke and N.~O.~Walliser,
  ``PALP - a user manual,''
pp. 461-550 in {\it Strings, gauge fields, and the geometry behind:  the
legacy of Maximillian Kreuzer}, World Scientific, 2012,
  {\tt arXiv:1205.4147}.
  

\bibitem{Pantev:2005rh}
  T.~Pantev and E.~Sharpe,
  ``Notes on gauging noneffective group actions,''
  {\tt hep-th/0502027}.

\bibitem{Pantev:2005zs}
  T.~Pantev and E.~Sharpe,
  ``GLSM's for gerbes (and other toric stacks),''
  Adv.\ Theor.\ Math.\ Phys.\  {\bf 10} (2006)  77-121,
  {\tt hep-th/0502053}.




\bibitem{Morrison:1994fr}
  D.~R.~Morrison and M.~R.~Plesser,
  ``Summing the instantons: Quantum cohomology and mirror symmetry in toric varieties,''
  Nucl.\ Phys.\ B {\bf 440} (1995) 279-354,
  {\tt hep-th/9412236}.

    \bibitem{Hori:2016txh}
  K.~Hori and J.~Knapp,
  ``A pair of Calabi-Yau manifolds from a two parameter non-Abelian gauged linear sigma model,''
  {\tt arXiv:1612.06214}.


\bibitem{kk} G. Kapustka and M. Kapustka,
``A cascade of determinantal Calabi-Yau threefolds,''
Math. Nachr. {\bf 283} (2010) 1795-1809,
  {\tt arXiv:0802.3669}.


\bibitem{kp18} A. Kuznetsov and A. Perry,
  ``Categorical joins,''
  {\tt arXiv:1804.00144}

\end{thebibliography}
\end{document}